\documentclass{aa}  

\usepackage{graphicx}
\usepackage{txfonts}
\usepackage{xcolor}
\newcommand*{\spr}{{\sc Spritz}}
\newcommand*{\hers}{\textit{Herschel}}
\defcitealias{Polletta2007}{P07}
\defcitealias{Gruppioni2013}{G13}
\usepackage[flushleft]{threeparttable}
\usepackage{scrextend}
\usepackage{soul}
\begin{document}

   \title{Simulating the infrared sky with a \spr{}}

   \author{L. Bisigello
          \inst{1}\thanks{laura.bisigello@inaf.it} \and
          C. Gruppioni\inst{1} \and 
          A. Feltre\inst{1} \and
          F. Calura\inst{1} \and
          F. Pozzi\inst{1,2}  \and
          C. Vignali\inst{2,1} \and
          L. Barchiesi\inst{1,2} \and
          G. Rodighiero\inst{3,4} \and
          M. Negrello\inst{5}
          }

   \institute{INAF Osservatorio di Astrofisica e Scienza dello Spazio, via Gobetti 93/3, I-40129,
Bologna, Italy 
\and
    Dipartimento di Fisica e Astronomia, Università di Bologna, Via Gobetti 93/2, I-40129 Bologna, Italy
\and
    Dipartimento di Fisica e Astronomia, Università di Padova, Vicolo dell’Osservatorio, 3, I-35122, Padova, Italy
\and
    INAF Osservatorio Astronomico di Padova, vicolo dell’Osservatorio 5, I-35122 Padova, Italy
\and
    School of Physics and Astronomy, Cardiff University, The Parade, Cardiff CF24 3AA, UK
}

   \date{Received ; accepted }

 
  \abstract
   {}
   {Current hydrodynamical and semi-empirical simulations of galaxy formation and evolution have difficulties in reproducing the number densities of IR-detected galaxies. Therefore, a versatile, phenomenological new simulation tool is necessary to reproduce current and predict future observations at IR wavelengths. }
   {In this work we generate simulated catalogues starting from the \hers{} infrared luminosity functions of different galaxy populations, in order to consider in a consistent way different populations of galaxies and active galactic nuclei. We associated a spectral energy
distribution and physical properties, such as stellar mass, star-formation-rate and AGN contribution, to each simulated galaxy using a broad set of empirical relations. We compare the resulting simulated galaxies, extracted up to z$=$10, with a broad set of observational relations.}
   {The \emph{s}pectro-\emph{p}hotometric \emph{r}ealisations of \emph{i}nfrared-selected \emph{t}argets at all-\emph{z} (\spr{}) simulation will allow us to obtain in a fully consistent way simulated observations for a broad set of current and future facilities with photometric capabilities as well as low-resolution IR spectroscopy, like the \textit{James Webb Space Telescope} (\textit{JWST}) or the Origin Space Telescope (OST). The derived simulated catalogue contains galaxies and active galactic nuclei that by construction reproduce the observed IR galaxy number density, but it is also in agreement with the observed number counts from UV to far-IR wavelengths, the observed stellar mass function, the star-formation-rate vs. stellar mass plane and the luminosity function from the radio to the X-ray. The proposed simulation is therefore ideal to make predictions for current and future facilities, in particular, but not limited to, those operating at IR wavelengths. The \spr{} simulation will be publicly available.}
   {}


   \maketitle
%

\section{Introduction}

In the last decades numerous infrared (IR) extra-galactic surveys have been carried out thanks to ground and space telescopes, such as the Infrared Astronomical Satelite \citep[IRAS,][]{Neugebauer1984}, \textit{Spitzer} \citep{Werner2004}, \hers{} \citep{Pilbratt2010}, the two generations of the Submillimetre Common-User Bolometer Array at the James Clerk Maxwell Telescope \citep{Holland1999,Holland2013} and the Atacama Large Millimiter Array \citep[ALMA,][]{Wootten2009}. These observations highlight the large amount of energy contained in the IR background \citep{Hauser2001} and the existence of a population of massive objects,  i.e. M$^{*}>10^{10}\,{\rm M}_{\odot}$ \citep{daCunha2010}, with extreme IR luminosity, i.e. L$_{IR}>$10$^{12}\,L_{\odot}$ \citep[e.g.,][]{Soifer1984,Aaronson1984}. Both findings pin-point the importance of the IR galaxy population and its strong redshift evolution. \par
Indeed, the Ultraluminous and Hyper-luminous galaxies, with IR luminosity above 10$^{12}\,L_{\odot}$ and 10$^{13}\,L_{\odot}$ respectively, are very rare in the Local Universe, but their importance increases with redshift becoming responsible for a significant fraction of the comoving infrared luminosity density at z$>$1 \citep[e.g.,][]{LeFloch2005,PerezGonzales2005,Caputi2007}. Their extreme IR luminosity is partially due to the presence of active galactic nuclei (AGN), but even considering the AGN contribution, confusion, blending and possible lensing effects their star-formation rates can exceed 1000 $M_{\odot}\,yr^{-1}$ \citep{Rowanrobinson2000,Ruiz2013,Rowanrobinson2018}. The number density of the most luminous IR galaxies is generally underestimated by semi-analytic models of galaxy formation as well as by cosmological hydrodinamical simulations \citep{Hayward2013,Dowell2014,Gruppioni2015,AlcaldePampliega2019,Baes2020}. This tension between models and observations is not limited to the total IR luminosity, but it is present also for the luminosity function of the CO line \citep[e.g,][]{Decarli2019,Riechers2019}, and the dust-mass density \citep{Magnelli2020,Pozzi2020}. Therefore, these models are not suited for providing predictions for future IR missions, such as the \textit{JWST} \citep[][]{Gardner2009} and OST \footnote{\url{http://origins.ipac.caltech.edu/}} \citep[][]{Leisawitz2019}. 
This generation of IR telescopes will open up the possibility of investigating in more detail this population of strongly evolving, massive and dusty objects that challenge our understanding of the processes governing galaxy formation and evolution. \par
Given the difficulties of semi-analytical models and hydro-dynamical simulations in reproducing the number density of the most luminous IR galaxies, it is evident that predictions for future IR telescopes, such as JWST and OST, need to be computed exploiting ad-hoc simulations. 
In the literature there are several tools available to derive simulated catalogues for current and future telescopes.
For example, the Empirical Galaxy Generator \citep[EGG][]{Schreiber2017}, starting from the stellar mass function of quiescent and star-forming galaxies and using a series of empirical relations and galaxy templates, predicts the expected number counts in different broad-band filters including IR wavelengths. However, at the moment of writing, AGN and nebular emission lines are not included in the simulation. 
A similar approach is presented also in \citet{Williams2018}, which include nebular emission lines. This work does not include AGN and is focused on producing JWST simulated catalogues with spectral range limited to UV and near-IR wavelengths. Similar simulated catalogues are also presented in \citet{Bisigello2016,Bisigello2017}, which include nebular emission lines also from metal-free galaxies, but focuses on JWST predictions and does not include AGN. 
\par
Overall, a tool to create simulated catalogues from UV to IR wavelengths and that reproduces the expected IR number counts including both nebular emission lines and AGN is still missing.
In this perspective, we propose a new suite of simulation, the \emph{s}pectro-\emph{p}hotometric \emph{r}ealisations of \emph{i}nfrared-selected \emph{t}argets at all-\emph{z} (\spr{}), that starts from our current knowledge of the IR Universe to account for the number density of IR galaxies at different redshift and the contribution of star-formation and AGN to their IR light.\par
The paper is organised as follows. In section \ref{sec:deus_description} we describe the main steps of the construction of the \spr{} simulation and in section \ref{Sec:mocks} we illustrate the creation of the simulated catalogues.  In section \ref{sec:validation} we compare the derived simulated catalogues with a broad-set of observations, used to validate our method. We finally show some predictions for future telescopes in section \ref{sec:App} and we summarise our work in section \ref{sec:conclusions}. Throughout the paper, we consider a Chabrier initial mass function  \citep[IMF,][]{Chabrier2003}, a $\Lambda$CDM cosmology with $H_0=70\,{\rm km}{\rm s}^{-1}{\rm Mpc}^{-1} $, $\Omega_{\rm m}=0.27$, $\Omega_\Lambda=0.73$ and all magnitudes are in the AB system \citep{Oke1983}.

\section{The \spr{} ingredients}\label{sec:deus_description}
\begin{figure}
    \centering
    \includegraphics[width=1\linewidth,keepaspectratio,trim=40 40 40 40, clip]{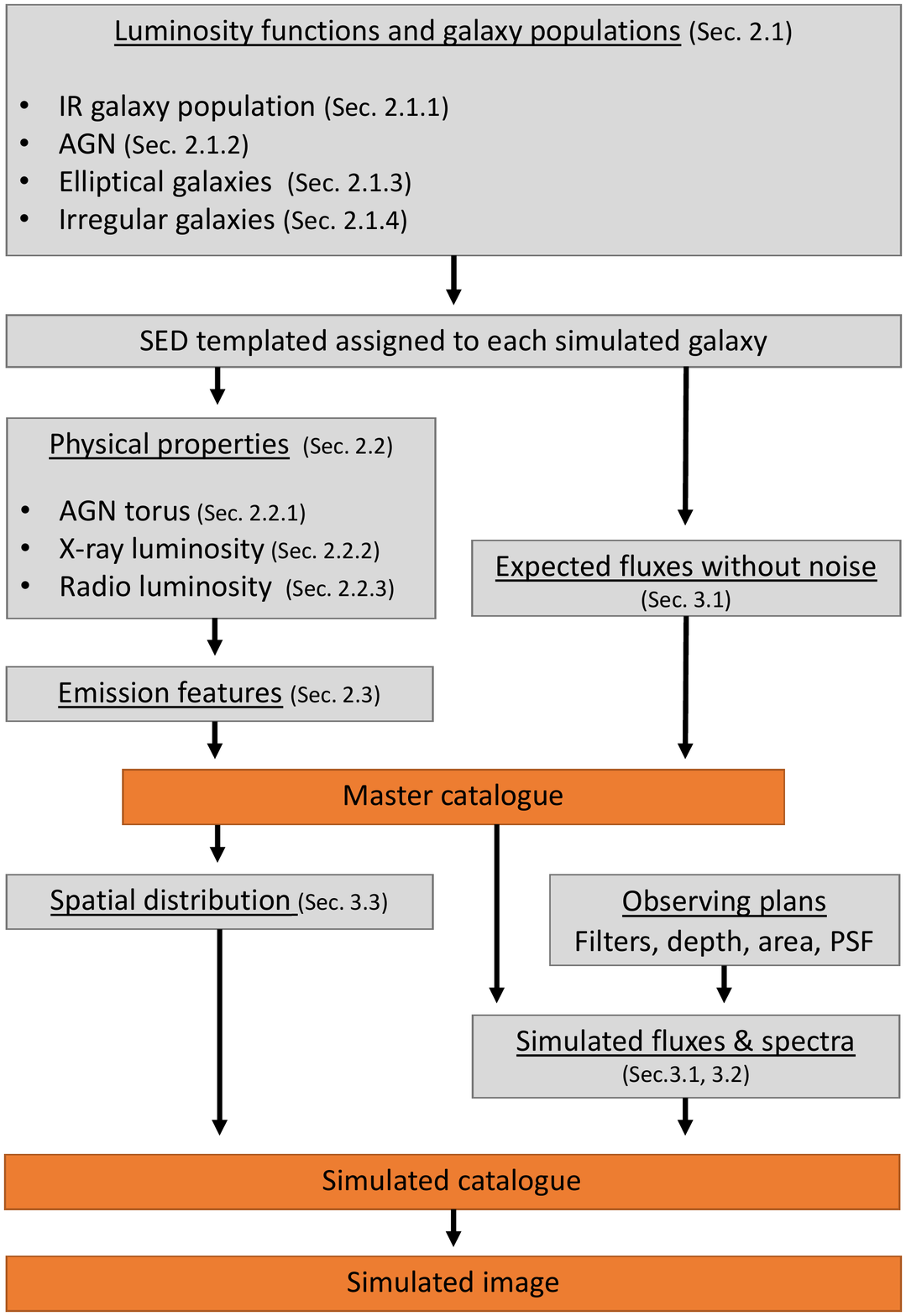}
    \caption{Workflow diagram summarising the main steps of \spr{}. Orange coloured boxes indicate the outputs of the simulation. }
    \label{fig:scheme}
\end{figure}

The \spr{} simulation is based on the previous works by \citet{Gruppioni2011,Gruppioni2013} and can be broadly applied to simulate spectro-photometric surveys for different current and future facilities. The main steps of \spr{} are summarised in Figure \ref{fig:scheme}. The main output consists of a master catalogue, which has no flux limits, from which simulated catalogues are created to mimic different spectro-photometric surveys. The master catalogue is generated starting from a set of seven observed luminosity functions (LFs) and a galaxy stellar mass function (GSMF), each of them representing a separate galaxy population. To each simulated galaxy we assigned a set of physical properties following various empirical relations. \par
We now proceed to describe the overall construction of the \spr{} master catalogue, which is made publicly available\footnote{\label{mywebsite}\url{https://sites.google.com/inaf.it/laurabisigello/spritz}}.  \par 

\subsection{Luminosity functions and galaxy populations}

\subsubsection{IR galaxy populations}

We started by extracting simulated galaxies from the \hers{} IR LF derived by \citet[][hereafter G13]{Gruppioni2013} for different galaxy populations, namely spiral galaxies, starburst (SB), `unobscured' type 1 AGN (AGN1), `obscured' type 2 AGN (AGN2) and two classes of composite systems (SB-AGN and SF-AGN). The latter two represent two galaxy populations without evident AGN activity, two populations dominated by the AGN activity and two mixed systems. The separation in these six galaxy populations was driven by the observational results presented in \citetalias{Gruppioni2013} and we maintained the same populations for consistency. We now describe the mentioned galaxy populations, their SED templates and their LFs.

\paragraph{Galaxy populations:}\label{sec:SEDmodels}
The population of spiral galaxies contains normal star-forming systems that are dominant in the Local Universe but their number density decreases at increasing redshift above z$\simeq$1. 
\begin{itemize}
    \item Spirals range from early bulge-dominated (S0) to late-type disky galaxies (Sdm), like M51. Their specific star-formation-rate (sSFR=SFR/$M^{*}$) spans the range from log$_{10}(sSFR/yr^{-1})=$-10.4 to -8.9 (see later the templates used to derive sSFR). 
    \item The SB population contains galaxies with intense episode of star-formation: their number increases with redshift and their sSFR varies from log$_{10}(sSFR/yr^{-1})=$-8.8 to -8.1. Examples of such objects are M82, NGC6090 and Arp220. Ultra-luminous galaxies (ULIRGs, L$_{IR}>10^{12}L_{\odot}$) are included in this galaxy population only if they do not show any AGN activity. 
    \item AGN1 and AGN2 populations contain very luminous AGN that dominate their galaxy energetic, compared to their star-formation activity. The difference between the two populations resides on  the optical/UV part of their spectrum being unobscured (AGN1) or obscured by dust (AGN2). Optically selected quasi stellar objects (QSO) are examples of AGN1, while optically obscured QSO, like  Mrk231 and IRAS19254, are part of the AGN2 population. These two populations, as the previous SB population, show an increase in their numbers with redshift and they dominate the expected galaxy population at z$>$2-3. 
    \item The two composite galaxy populations (SF-AGN and SB-AGN) represent objects that host an AGN that is not the dominant source of energy at any wavelength, apart from a small range in the mid-IR, where the AGN contribution is visible. \citetalias{Gruppioni2013} split them in two sub-populations as they show different redshift evolution and properties. In particular, SF-AGN have a spectrum similar to a spiral galaxy but host a low-luminosity AGN, while SB-AGN resemble starburst galaxies hosting an obscured AGN. SF-AGN dominate the galaxy population at z$=$1-2, while they are less numerous at lower redshift when spirals galaxies are more common. On the other hand, SB-AGN become more important at increasing redshift, i.e. z$>$2, and they can be interpreted as the heavily obscured phase of bright AGN. The star-formation of this galaxy population is very high (log$_{10}(sSFR/yr^{-1})\sim$-8.7 in our templates), but, following the interpretation of \citetalias{Gruppioni2013}, the system should follow a process of quenching becoming first a source with a dominant AGN and a lower level of star-formation, i.e. AGN1 or AGN2, and then an elliptical galaxy. Seyfert galaxies that show properties similar to star-forming galaxies \citep{GarciaGonzalez2016}, like Circinus or NGC1068, are good representatives of SF-AGN, while galaxies like IRAS20551, IRAS22491 and NGC6240, showing more intense star formation activity, are good examples of SB-AGN. \par
    \end{itemize}

\begin{figure}
    \centering
    \includegraphics[width=1\linewidth,keepaspectratio]{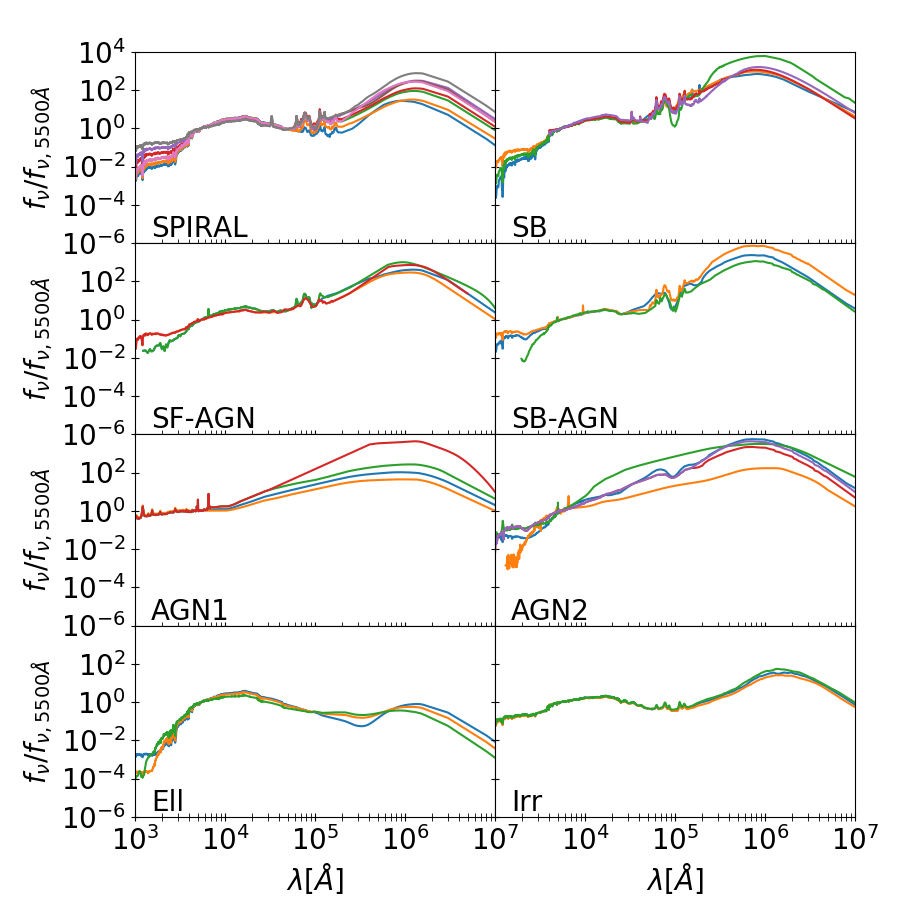}
    \caption{The SED templates used to derive \spr{} simulated galaxies. Each panel shows templates of a specific galaxy population and all templates are normalised to the flux density at 5500 \AA{}.}
    \label{fig:SED}
\end{figure}

\paragraph{SED templates:}
In \citetalias{Gruppioni2013} the \hers{} IR LFs have been derived by modelling the available optical-to-IR observations with the semi-empirical templates characteristic of each galaxy population. In particular, the templates are taken from the library by \citet[]{Polletta2007} with the addition of three starburst templates by \citet{Rieke2009} and some templates with a far-IR part modified to better match the \hers{} data \citep[see ][]{Gruppioni2010}. These adjustments on the original set of templates of \citet[]{Polletta2007} were performed by \citetalias{Gruppioni2013} to properly fit the SED of \hers{}-detected galaxies. In \spr{} we consider, for consistency, all 32 SED templates used in \citetalias{Gruppioni2013} (Fig. \ref{fig:SED}) with each galaxy population having from 3 to 8 SED templates associated with it. The relative importance of each template is dictated by the number of times that template is found to reproduce the photometry of the observed galaxies in \citetalias{Gruppioni2013} at each redshift and IR luminosity.

\begin{table*}[]
    \centering
    \caption{Parameters of the IR LFs used to derive the \spr{} simulated galaxies, as found by \citetalias{Gruppioni2013} for \hers{} galaxies or through the MCMC (for AGN1 and AGN2). From left to right: the name of the considered population, the faint-end slope $\alpha$, the bright-end slope cut $\sigma$, the two exponents of the power-law evolution of the characteristic density (k$_{\rho,1}$ and k$_{\rho,2}$) and the redshift at which the evolution changes (z$_{\rho}$), the two exponents of the power-law evolution of the characteristic luminosity (k$_{L,1}$ and k$_{L,2}$) and the redshift at which the evolution changes (z$_{L}$), the characteristic density ($\Phi^{*}_{0}$) and the characteristic luminosity ($L^{*}_{0}$) at z$=$0.  When the $z_{L}$ and k$_{L,2}$ ($z{\rho}$ and k$_{\rho,2}$) parameters are not listed, the corresponding populations show a single power-law evolution in the characteristic luminosity (density).}
    \label{tab:IRLF}
    \resizebox{1\textwidth}{!}{
    \begin{tabular}{c|cccccccccc}
       Population  & $\alpha$ & $\sigma$ & k$_{\rho,1}$ & k$_{\rho,2}$ & z$_{\rho}$ & k$_{L,1}$ & k$_{L,2}$ & z$_{L}$ & log$_{10}(\Phi^{*}_{0}/Mpc^{-3} dex^{-1})$ &  log$_{10}(L^{*}_{0}/L_{\odot})$ \\
       \hline
        SPIRALS & 1$\pm$0.05 & 0.5$\pm$0.01 & -0.54$\pm$0.12 & -7.13$\pm$0.24 & 0.53 & 4.49$\pm$0.15 & 0.00$\pm$0.46 & 1.1 & -2.12$\pm$0.01 & 9.78$\pm$0.04  \\
        SB & 1.2$\pm$0.20 & 0.35$\pm$0.10 & 3.79$\pm$0.21 & -1.06$\pm$0.05 & 1.1& 1.96$\pm$0.13 &  &  & -4.91$\pm$0.06 & 11.17$\pm$0.16  \\
        SF-AGN & 1.2$\pm$0.02 & 0.4$\pm$0.10 & 0.73$\pm$0.97 & -6.59$\pm$2.16 & 1.1& 3.59$\pm$0.40 &  &  & -2.95$\pm$0.23 & 10.16$\pm$0.13 \\
        SB-AGN & 1.2$\pm$0.02 & 0.4$\pm$0.10 & 1.81$\pm$0.68 &  &  & 1.51$\pm$0.55 &  &   & -4.59$\pm$0.24 & 11.22$\pm$0.19 \\
        AGN1-2\tablefootmark{a} & 1.31$^{+0.09}_{-0.09}$ & 0.48$^{+0.03}_{-0.03}$ & 1.33$^{+0.22}_{-0.22}$ & -2.75$^{+0.31}_{-0.30}$ & 1.88$^{+0.06}_{-0.04}$ & 2.85$^{+0.15}_{-0.16}$ & -0.06$^{+0.35}_{-0.34}$ & 2.75$^{+0.06}_{-0.03}$ & -5.42$^{+0.12}_{-0.13}$ & 10.93$^{+0.15}_{-0.16}$ \\
\end{tabular}}
    \tablefoot{
    \tablefoottext{a}{This LF has been derived applying an MCMC simultaneously to FUV and IR data (see Section \ref{sec:AGN1}), for both AGN1 and AGN2.}
    }
\end{table*}  

\begin{figure}
    \centering
    \includegraphics[width=1\linewidth,keepaspectratio]{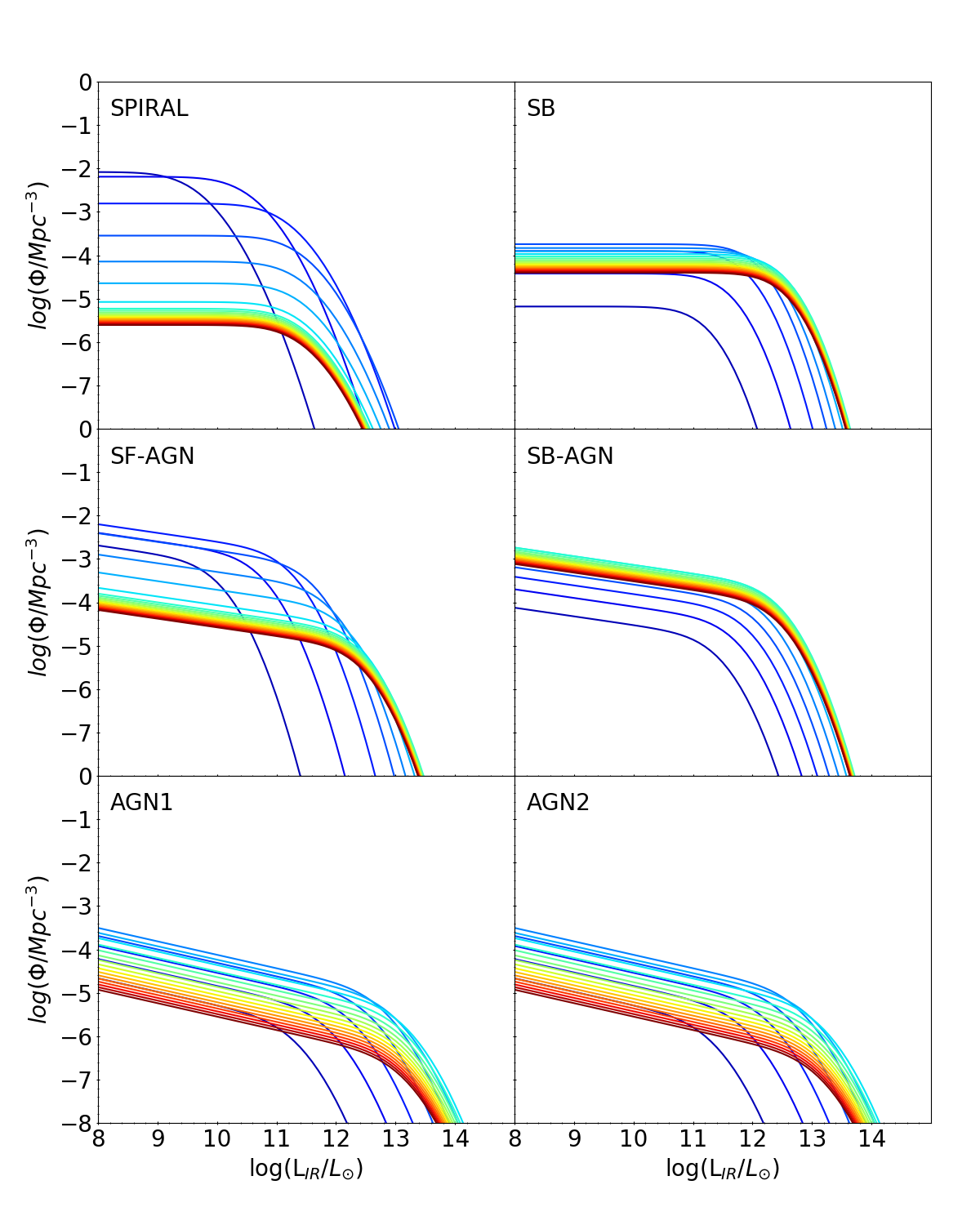}
    \caption{The IR LFs used in the \spr{} simulation to derive simulated galaxies. Each panel shows a different galaxy population, as indicated on the top left. The LFs are shown for different redshifts from 0.1 (\textit{dark blue solid line}) to 9.6 (\textit{dark red solid line}) with steps of 0.5. At z$>$3 we show the LFs derived assuming ${k_{\Phi}}=$-1 (see eq. \ref{eq:zevol}) for all the populations except for AGN1 and AGN2 for which observations up to z$=$5 are available.
    }
    \label{fig:IRLF}
\end{figure}

\paragraph{Luminosity functions:}
Briefly, the shape of the IR LF has been assumed to be well represented by the function proposed by \citet{Saunders1990} to describe the 60$\mu$m LF observed with IRAS at low-z:
\begin{equation}\label{eq:LF}
    \Phi(L)dlog_{10}L=\Phi^{*}\left(\frac{L}{L^{*}}\right)^{(1-\alpha)}exp\left[-\frac{1}{2\sigma^{2}}log^{2}_{10}\left(1+\frac{L}{L^{*}}\right)\right]dlog_{10}L
\end{equation}
where $L^{*}$ and  $\Phi^{*}$ are the luminosity and the number density at the knee of the LF, where the function changes from behaving as a power-law at low luminosities to a Gaussian at high luminosities. The parameter $\alpha$ is the faint-end slope of the LF, while $\sigma$ regulates the steepness of the bright-end slope. We assumed $\Phi^{*}$ and $L^{*}$ to evolve with redshift as :
\begin{equation}\label{eq:zevol}
\begin{array}{l}
    \Phi^{*} \propto \begin{cases} (1+z)^{k_{\rho,1}}, & \mbox{if } z\leq z_{\rho} \\ (1+z)^{k_{\rho,2}}, & \mbox{if } z_{\rho}<z<3 \\
    (1+z)^{k_{\Phi}}, & \mbox{if } z \geq 3 \end{cases} \\
    L^{*} \propto \begin{cases} (1+z)^{k_{L,1}}, & \mbox{if } z\leq z_{L} \\ (1+z)^{k_{L,2}}, & \mbox{if } z_{L}<z<3 \\
    \mbox{constant}, & \mbox{if } z_{L} \geq 3 \end{cases}
\end{array}
\end{equation}
 The full list of parameters is shown in Table \ref{tab:IRLF}, as derived by \citetalias{Gruppioni2013}, while the LF of each galaxy population is shown in Figure \ref{fig:IRLF}. The parameters $\Phi^{*}_{0}$ and $L^{*}_{0}$ present in the Table correspond to the number density and luminosity at the knee of the LF at z$=$0. Not all the considered populations show a broken power-law evolution, in these cases the $z_{L}$, k$_{L,2}$, $z{\rho}$ and $k_{\rho,2}$ parameters are not given. We note that the evolution of the IR LF at z$>$3 is an extrapolation, as \hers{} observations are generally limited to lower redshift. In particular, we assumed that the characteristic luminosity (L$^{*}$) remains constant while the galaxy number density decreases as $\Phi^{*}\propto(1+z)^{k_{\Phi}}$. We explored values for the coefficient ${k_{\Phi}}$ comprised between -1 and -4 to consider the effects of a steep and mild redshift evolution. For AGN1 and AGN2 we applied a different procedure described in the next section, because for the AGN1 population the direct use of the \hers{} LF brings to some discrepancies with far-ultraviolet (FUV) and X-ray observations.\par
 We populate the bright-end of the IR LF with galaxies with the SED template showing the highest L$_{IR}$-to-K ratio, i.e. those dominated by young stars with respect to the evolved population. When for a specific galaxy population there are multiple templates with similar L$_{IR}$-to-K ratio, we consider the template with the highest L$_{IR}$-to-FUV ratio, i.e. those populated by a relative small fraction of extremely young stars or with a larger amount of dust. 
 We found that this assumption is necessary to avoid an overestimation of the bright-end of the K-band and FUV LFs and it is applied to objects with log$_{10}(L_{IR}/L_{\odot})\geq$11.5 for spirals, log$_{10}(L_{IR}/L_{\odot})\geq$12 for SF-AGN and AGN2, log$_{10}(L_{IR}/L_{\odot})\geq$12.5 for SB and SB-AGN and log$_{10}(L_{IR}/L_{\odot})\geq$11 for AGN1. \par
 
\subsubsection{Type-1 and type-2 AGN}\label{sec:AGN1}
The large amount of available \hers{} data have allowed \citetalias{Gruppioni2013} to derive the LF for different galaxy populations. Achieving this result is not possible when the galaxy sample is limited in number, so a comparison with studies at other wavelengths is generally possible only with the LF of the total galaxy sample. However, in the literature there is a long list of works focused on studying the LF in the FUV of unobscured AGN-dominated galaxies, i.e. QSO, which dominate the bright-end of the FUV LF \citep[e.g.,][]{Croom2009,McGreer2013,Ross2013,Akiyama2018,Schindler2019}. These FUV observations can be used to check the conversion to these wavelengths of the AGN1 IR LF used in \spr{}. \par
To perform this comparison we converted the AGN L$_{IR}$ to FUV using the SED templates associated to this galaxy population (Fig. \ref{fig:SED}), but the AGN contribution to the FUV LF derived from \citetalias{Gruppioni2013} IR LF generally overestimates the observations, particularly at z$>$2 and at faint luminosities, where the IR LF was just extrapolated (Fig. \ref{fig:FUVLF_MCMC}). This showed the necessity to improve the AGN LF, but it also ensures that dust-free AGN observed in FUV and not in the IR are expected to be a minority. We therefore decided to improve the AGN1 LF by deriving a new IR LF for this galaxy population using a Monte-Carlo-Markov chain \citep[MCMC;][]{ForemanMackey2013} applied simultaneously to the FUV and \hers{} observations of AGN, the latter are taken for consistency from \citetalias{Gruppioni2013}. We consider the same function of the other IR galaxy populations (eq. \ref{eq:LF} and \ref{eq:zevol}), but instead of using the extrapolation at z$>$3, we exploited as additional constraints the FUV observations, which in the case of AGN1 are available up to z$=$5. Following the unification scheme of AGN \citep{Antonucci1993,Urry1995}, AGN1 and AGN2 should represent the same galaxy population observed from different viewing angles. For this reason, we do not expect extreme differences between the AGN1 and AGN2 IR LFs and, indeed, the \hers{} observed LFs derived by \citetalias{Gruppioni2013} are remarkably similar to each other, at least when both AGN populations are available. Given this similarity, we decided to include the observed \hers{} LF of both AGN1 and AGN2 to our MCMC run. The resulting IR LF well describes the \hers{} IR observations of both AGN1 and AGN2, but it is also consistent with the FUV QSO observations. The MCMC is described in detail in Appendix \ref{sec:MCMC} and the best result is listed in Table \ref{tab:IRLF}.  \par

\subsubsection{Elliptical galaxies}\label{sec:Ell}

\begin{figure}
    \centering
    \includegraphics[width=0.6\linewidth,keepaspectratio]{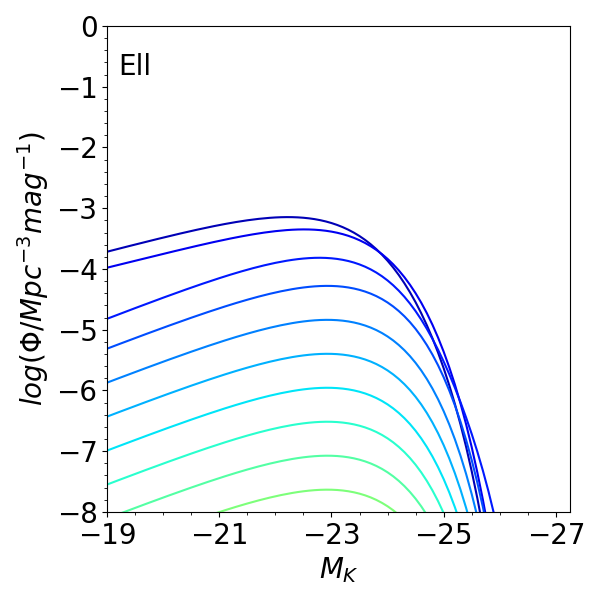}
    \caption{The K-band LF used to include simulated elliptical galaxies in the \spr{} simulation. The LFs are shown for different redshifts from 0.1 (\textit{dark blue solid line}) to 4.6 (\textit{green solid line}) with steps of 0.5, with the same colour code of Figure \ref{fig:IRLF}. At z$>$2 the LF has been extrapolated, starting from the \citet{Cirasuolo2007} K-band LF.}
    \label{fig:KLF}
\end{figure}

The great majority of elliptical galaxies (Ell) have low levels of SFR and small amounts of dust  \citep[e.g.][]{Knapp1989,Mazzei1994,Noeske2007,McDermid2015} and their IR luminosity originates, at least in local elliptical galaxies, from dust lanes or diffuse cirrus heated by the radiation field of old stars \citep[e.g.][]{Bertola1978,Goudfrooij1995,Kaviraj2012,Kokusho2019}. Given their faint IR fluxes, these galaxies are below the detection limit of most \hers{} observations and they are not present in the \citetalias{Gruppioni2013} LF study. However, future IR telescopes may be able to detect them at least at low-z. Therefore, we decide to include them in the \spr{} simulation by considering the average LFs derived by \citet{Arnouts2007}, \citet{Cirasuolo2007} and \citet{Beare2019} in the K-band. We consider these three LFs because they are obtained after removing star-forming systems, which are already included in the \hers{} LFs. We extrapolate the K-band LF by \citet{Cirasuolo2007}, which goes to higher redshift (z$=$2) than the other two, by maintaining constant the characteristic luminosity and varying the characteristic galaxy density $\Phi^{*}$ as $\propto(1+z)^{-1}$.
This extrapolation is not expected to have a crucial impact on the results, as the observed number density of elliptical galaxies at z$\sim$2 is already low, i.e. $\Phi^{*}=0.2\times 10^{-3}\,{\rm Mpc}^{-3}$. The resulting K-band LF at different redshift is shown in Figure \ref{fig:KLF}.\par
As done for the other galaxy populations, we associated to each galaxy extracted from the K-band LF one of the three empirical templates of elliptical galaxies by \citet[][Fig. \ref{fig:SED}]{Polletta2007}, considering each template with equal probability. The three templates corresponds to galaxies dominated by an old stellar population, but with a different amount of on-going star-formation, i.e. from log$_{10}(sSFR/yr^{-1})=$-12.5 to -11.3, which corresponds to different UV-slopes and IR dust bumps. We used these templates to derive physical parameters and to link the K-band LF to the IR LF. This is done trough the K-correction, which is different for each of the three considered templates, resulting in a non-rigid conversion between the K-band LF and the IR LF of elliptical galaxies. \par

\subsubsection{Irregular galaxies}\label{sec:Irr}

\begin{figure}
    \centering
    \includegraphics[width=1\linewidth,keepaspectratio]{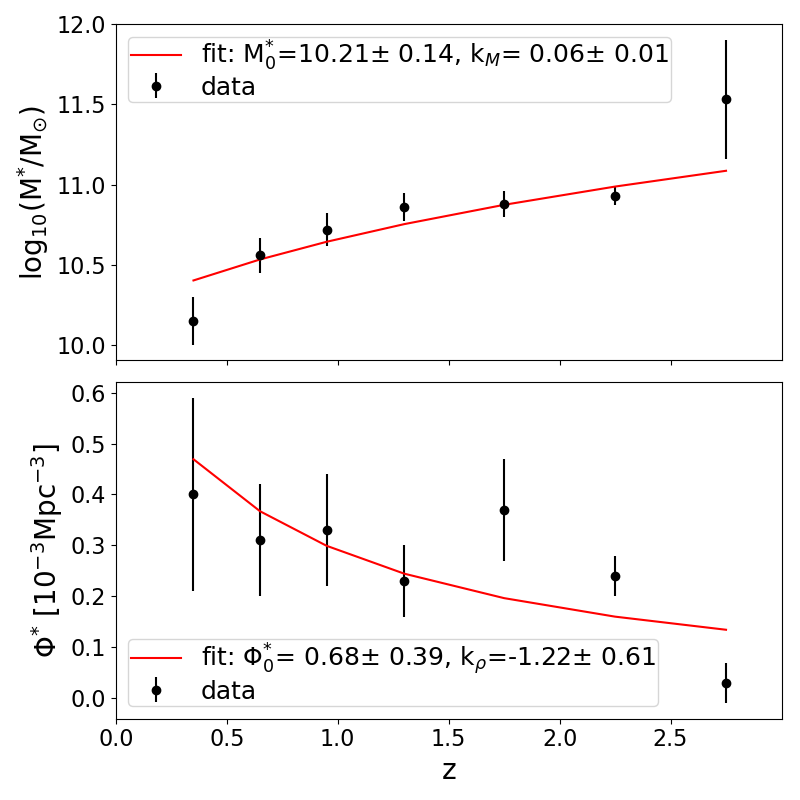}
    \caption{Redshift evolution of the characteristic stellar mass $M^{*}$  (\textit{top}) and characteristic density $\Phi^{*}$ (\textit{bottom}) of the GSMF of irregular galaxies. We show the data by \citet[][\textit{Black dots}]{Huertas-Company2016}  and our fit (\textit{red solid line}). The latter has been derived with the {\sc SciPy} package \citep{Virtanen2020} and the resulting parameters are reported in each panel. }
    \label{fig:Irrzev}
\end{figure}

Dwarf irregular galaxies make up another galaxy population that is missing from many \hers{} observations, excluding local objects. 
These galaxies are characterised by a relative low amount of dust, low metallicity and often high sSFR \citep{Hunter2010,Cigan2016,Bianchi2018}. They are therefore expected to have bluer optical spectra and fainter L$_{IR}$, with respect to the other galaxy populations analysed in this work.\par
We considered the GSMF of star-forming irregular galaxies by \citet{Huertas-Company2016} that is described by a single Schechter function and it has been derived for different redshift bins up to z$=$3. The faint-end slope $\alpha$ does not show a significant redshift evolution, therefore we assumed a constant value equal to the average weighted mean of the values found at different redshifts, i.e. $\alpha=-$1.58$\pm$0.19. On the other hand, we fitted the redshift evolution of the characteristic stellar mass (log$_{10}$(M$^{*})=$log$_{10}$(M$^{*}_{0})*(1+$z$)^{k_{M}}$) and density ($\Phi^{*}=\Phi^{*}_{0}*(1+$z$)^{k_{\Phi}}$) of the GSMF, similar to what has been done by \citetalias{Gruppioni2013} for the other galaxy populations. The fit is shown in Figure \ref{fig:Irrzev} and it results in log$_{10}($M$_{0}^{*}/$M$_{\odot})=$10.21$\pm$0.14, k$_{M}=$0.06$\pm$0.01, $\Phi_{0}^{*}=$0.68$\pm$0.39$\times$10$^3$ Mpc$^{-3}$ and k$_{\Phi}=$-1.22$\pm$0.61.\par
Similar to what has been done for the other galaxy populations, we associated a set of SED templates to simulated irregular galaxies extracted from the mentioned GSMF. 
In particular, we considered the SED corresponding to the median, the 16$\%$ and 84$\%$ of the template distribution of irregular galaxies by \citet{Bianchi2018}. These three templates are shown in Figure \ref{fig:SED}. The median template has been associated to 56$\%$ of the simulated galaxies, while the other two to 22$\%$ of the irregular galaxy population, in order to match the fraction of galaxies they represent in the observed sample \citep{Bianchi2018}. We made use of these three templates to derive the physical properties and each template is normalised to match the desired L$_{IR}$. \par
The LFs or GSMF of the different galaxy populations are sampled from z$=$0 to 10 and IR luminosity between $log_{10}(L_{IR}/L{\odot})=$5 and 15. The number density of each galaxy population in each luminosity-redshift bin is then used to derive the probability to observe each galaxy population in a desired area. This is fundamental for the creation of the simulated catalogues.

\begin{figure}
    \centering
    \includegraphics[width=0.6\linewidth,keepaspectratio]{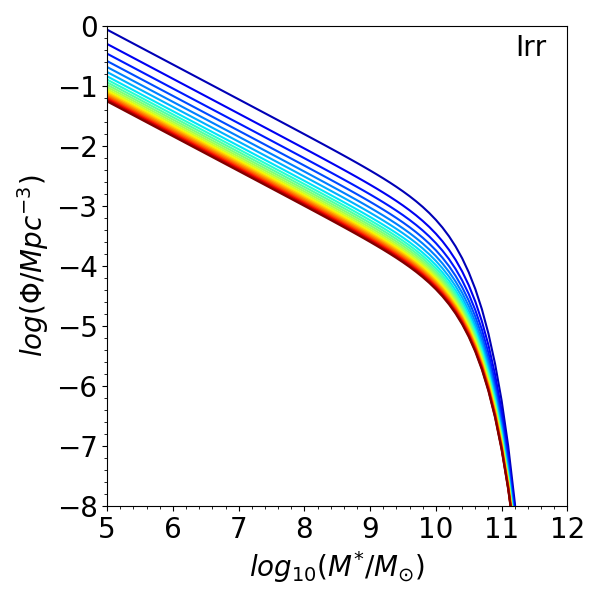}
    \caption{The GSMF used to include simulated irregular galaxies in the \spr{} simulation. The GSMFs are shown for different redshifts from 0.1 (\textit{dark blue solid line}) to 9.6 (\textit{dark red solid line}) with steps of 0.5, with the same colour code of Figure \ref{fig:IRLF}.}
    \label{fig:IrrMF}
\end{figure}

\subsection{Physical properties of simulated galaxies}\label{sec:PhysProp}
In this section we describe the method used to retrieve the various physical parameters associated to each L$_{IR}$-z bin of the master catalogue. For all templates, the total L$_{IR}$ is derived directly integrating the templates between 8 and 1000 $\mu$m. In order to derive the other main physical parameters, e.g. stellar mass and colour excess, for all simulated galaxies we applied the  {\sc Magphys} \citep[Multi-wavelength Analysis of Galaxy Physical Properties,][]{daCunha2008} package, or the software {\sc sed3fit} \citep{Berta2013} based on {\sc Magphys} when an AGN contribution is present, to the set of templates associated to each galaxy population. This SED fitting procedure serves to infer the physical properties associated to each SED template, as this information is not available a priori because of the empirical nature of the templates described in Sec. \ref{sec:SEDmodels}. To perform the fit, we created a large set of simulated observations of custom filters spanning from FUV to far-IR wavelengths and assigning S/N of 10 to each of them. For the AGN component, we considered a library with both smooth \citep{Fritz2006,Feltre2012} and clumpy torus \citep{Nenkova2008,Nenkova2008b}, as explained in more details in the next section. The analysis of the AGN component is necessary to disentangle the contribution of the AGN and star-formation to the total IR-luminosity and to take into account the contamination of the AGN when deriving the stellar mass. In the fits by both {\sc Magphys} and {\sc sed3fit} we consider \citet{Bruzual2003} stellar templates with \citet{Chabrier2003} IMF. For all considered empirical templates we retrieve the best fit value of the different physical parameters estimated by {\sc sed3fit} or {\sc Magphys}, as well as the 2.5$\%$, 16$\%$, 84$\%$ and 97.5$\%$ percentile when available\footnote{For some parameters the used codes extract only the best value.}. When assigning a physical property to a simulated galaxy, we randomise each parameter value following the corresponding probability distribution to take into account the uncertainties on the SED fitting procedure.  When a probability distribution is not available, we assumed a Gaussian distribution centred to the best value and an arbitrary $\sigma=$0.3 dex. In the master catalogue we considered only the median value of each parameter, as derived for each galaxy population and L$_{IR}$-z bin.  \par 
As new stars may be surrounded by dust, an estimation of the total amount of star-formation requires to account for both the direct light emitted by young stars, for example looking at UV wavelengths, and for the amount of light absorbed by dust, looking at the reprocessed light at IR wavelengths. In our case, we considered the contribution of the star-formation to the IR luminosity to estimate the obscured star-formation-rate (SFR) and the observed rest-frame UV at 1600 \AA{}, i.e. the luminosity at 1600 \AA{} computed on the template not corrected for dust attenuation, to estimate the unobscured component of SFR associated to each template. These two properties can be estimated following \citet{Kennicutt1998a,Kennicutt1998}:
\begin{equation}
\begin{aligned} 
    &SFR_{UV}[M_{\odot} yr^{-1}]=f_{IMF}\,4.5\times10^{-44}\,L_{\nu,1600\mu m}[erg s^{--1}],\\
    &SFR_{IR}[M_{\odot} yr^{-1}]=f_{IMF}\frac{L_{3-1000\mu m}[L_{\odot}]}{5.8\times10^9},
\end{aligned}
\end{equation}
where f$_{IMF}=$0.63 \citep{Murphy2011} represents the conversion from a \citet{Salpeter1955} to Chabrier IMF. The total SFR is the sum of the two components, SFR$_{tot}=$SFR$_{UV}+$SFR$_{IR}$.
\par

For each simulated galaxy we derived the gas-phase metallicity from the mass-metallicity relation by \citet{Wuyts2014}:
\begin{equation}\label{eq:met}
\begin{aligned} 
    &12 + {\rm log}_{10}(O/H) = Z_{0} + {\rm log}_{10} [1-exp(-(M^{*}/M_{0})^{\gamma} ], \\
    &{\rm log}_{10}(M_{0}/M_{\odot}) = (8.86 \pm 0.05) + (2.92 \pm 0.16)\,{\rm log}_{10}(1 + z).
\end{aligned}
\end{equation}
where the power-law slope at low metallicity is $\gamma=$0.40 and the asymptotic metallicity is $Z_{0}=$8.69. This mass-metallicity relation is in overall agreement with the relation proposed by \citet{Mannucci2010,Mannucci2011}, at least among intermediate and massive galaxies \citep[see ][]{Wuyts2014}. We did not consider the metallicity derived from the SED fitting, as otherwise the metallicity would be fixed to a specific value independently by the stellar mass.

\subsubsection{AGN torus library}\label{sec:torusmodels}
We made use of the {\sc sed3fit} code to derive the AGN contribution of all the galaxies with an AGN component, using two distinct torus libraries. The first one consists of the smooth torus models presented by \citet{Fritz2006} and \citet{Feltre2012}. The second one is the library of clumpy torus models presented in \citet{Nenkova2008,Nenkova2008b}. \par
In particular, we consider smooth torus models with an outer to inner radii ratio $R_{out}/R_{in}=$10 and 30. We limit the fit to models with $R_{out}/R_{in}\leq$30, as suggested by previous works and observations \citep[e.g.][]{Jaffe2004,Netzer2007,Hatziminaoglou2008,Pozzi2010,Mullaney2011,Garcia2019}. The considered torus models have torus amplitude angle $\Theta$, defined as the angular region occupied by the torus dust, from 60$^{\circ}$ to 140$^{\circ}$. The dust density distribution of the torus varies along the radial and vertical directions, as described in polar coordinates $\rho (r,\theta)=A\,r^{\beta}e^{-\gamma\,cos(\theta)}$. The spectrum of the central engine is instead modelled with a broken power-law \citep[eq. 1 of][]{Feltre2012}, with spectral index $\alpha$, $\lambda L(\lambda)\propto\lambda^{\alpha}$, at $0.125<\lambda<10.0\,\mu m$ fixed to $\alpha=$-0.5. Finally, the equatorial optical depth at 9.7 $\mu$m ($\tau_{9.7}$) is between 0.1 and 10 and we consider viewing angles ($\phi$) between 0$^{\circ}$ and 90$^{\circ}$.  \par

For the clumpy torus models we considered templates with an outer to inner radii ratio $R_{out}/R_{in}=$10 and 30, as done for the smooth torus models. The density profile of the dust clumps is described radially by a power-law and vertically by an exponential profile $\rho(r,\theta)\propto r^{-q}\,e^{-|\theta/\sigma|^{m}}$, with $m=$2 for smooth boundaries. We considered models with $q$ values between 0 and 3 and with torus width $\sigma$, i.e. Gaussian distribution of the clouds, between 20$^{\circ}$ and 70$^{\circ}$.  The selected models are also parametrized in terms of the average number of clouds along a radial direction on the equatorial plane, comprised between 1 and 15. \citet{RamosAlmeida2009} found a median number of clouds of 10, using mid-IR observations of Seyfert galaxies, but larger values are necessary to include very obscured AGN in the \spr{} simulation. For a given combination of parameters, all the clouds have the same optical depth $\tau_{v}$ and we considered values between 10 and 200.\par

\begin{figure*}[hbt!]
    \centering
    \includegraphics[width=1\linewidth,keepaspectratio]{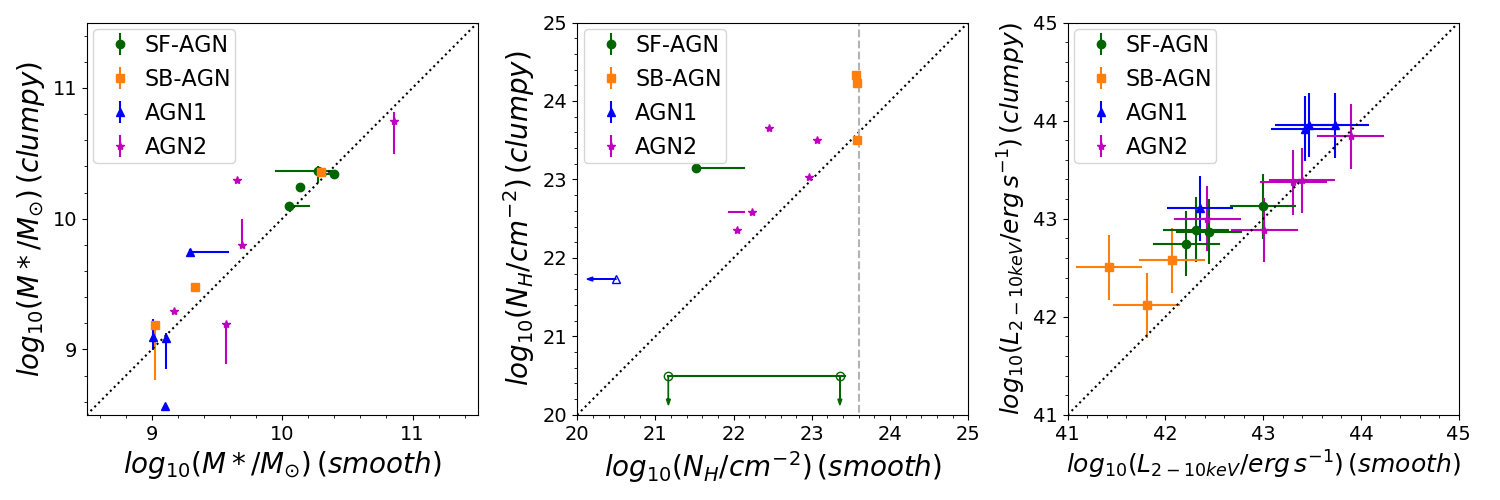}
    \caption{Comparison between the stellar mass (\textit{left}), hydrogen column density (\textit{center}) and intrinsic X-ray luminosity (\textit{right}) derived using {\sc sed3fit} with the clumpy and smooth torus library. All considered SED templates are scaled to have $L_{IR}=10^{11}\,L_{\odot}$. Symbols indicate different galaxy populations: SF-AGN (\textit{dark-green circles}), SB-AGN (\textit{orange squares}), AGN type-1 (\textit{blue triangles}) and AGN type-2 (\textit{magenta stars}). In the central panel, we report with empty symbols and arrows the upper limits to N$_{H}$ for the objects for which N$_{H}=$0 with one torus library, but not with the other one. These points are artificially set to log$_{10}$(N$_{H}/cm^{-2})=$20.5 for visual purposes. We do not show objects for which N$_{H}=$0 using both torus libraries, which are three AGN1 templates and one SF-AGN template. The vertical, grey, dashed line indicates the maximum values of $N_{H}$ of the smooth torus library. Error bars indicate the 16$\%$ and 84$\%$ percentiles and, in the right panel, the scatter of 0.33 dex in the relation to derive X-ray luminosity \citep{Asmus2015}. The black dotted line in each panel is the identity line.}
    \label{fig:ComparisonAGNtorus}
\end{figure*}

The entire list of the model parameters is shown in Table \ref{tab:torus} for both template families. In Figure \ref{fig:ComparisonAGNtorus}, we compare the stellar mass, hydrogen column density and X-ray luminosity (see sec. \ref{sec:Xray} for its derivation) obtained with the two torus libraries after scaling all the templates to have $L_{IR}=10^{11}\,L_{\odot}$. These three derived quantities show a general agreement even if with some scatter. The major differences for the stellar mass correspond to AGN-dominated systems, i.e. AGN1 and AGN2. These discrepancies are not surprising for AGN1, as their stellar mass is difficult to estimate because the AGN light dominates even at optical wavelengths. The maximum value of the hydrogen column density of our library of smooth torus is log$_{10}({\rm N}_{H}/{\rm cm}^{-2})\sim$23.6. The inclusion of the clumpy torus model is therefore fundamental to simulate heavily dust-obscured AGN. In fact, the hydrogen column densities derived considering clumpy torus models are generally larger than those derived with smooth torus models. The two torus models agree on the intrinsic X-ray luminosity of the brightest objects, at fixed IR luminosity, while for the faintest ones the clumpy torus models predict higher X-ray luminosities than the smooth ones. When comparing the two torus libraries, it is important to mention that the intrinsic X-ray luminosity depends not only on the hydrogen column density, but also on the AGN-host galaxy decomposition. Indeed, as visible in the same figure, the intrinsic X-ray luminosities derived with the two torus models differ more for composite systems than for AGN-dominated objects.\par

In the \spr{} simulation we randomly associated a smooth torus model to half of the objects hosting an AGN and a clumpy torus model to the remaining half, but we include in the uncertainties the results derived including separately only one of the two torus models. For more details on the SED fitting code we refer to \citet{Berta2013}, while for the torus models we refer to \citet{Feltre2012}, \citet{Fritz2006} and \citet{Nenkova2008,Nenkova2008b}.

\begin{table}
    \centering
    \caption{Parameters considered for the smooth torus models (\textit{left}) and clumpy torus models (\textit{right}), as described in Sec. \ref{sec:torusmodels}. For more details we refer to \citet{Fritz2006}, \citet{Feltre2012} and \citet{Nenkova2008,Nenkova2008b}.}
    \resizebox{0.49\textwidth}{!}{
    \begin{tabular}{cc|cc}
        \multicolumn{2}{c|}{Smooth torus} & \multicolumn{2}{c}{Clumpy torus}\\
        Parameter  & Values & Parameter  & Values\\
        \hline
        $R_{out}/R_{in}$ & 10,30 & $R_{out}/R_{in}$ & 10,30\\
        $\Theta$ & 60$^{\circ}$,100$^{\circ}$,140$^{\circ}$ & $\sigma$ & 20$^{\circ}$,50$^{\circ}$,70$^{\circ}$\\
        $\tau_{9.7}$ & 0.1,0.3,0.6,1,2,3,6,10 & $\tau_{V}$ & 10,20,60,120,200\\
        $\beta$ & -1,-0.5,0  & q & 0,1,2,3\\
        $\gamma$ & 0,4 & N$_{0}$& 1,5,10,15\\
        $\phi$ & 0$^{\circ}$-90$^{\circ}$  & $\phi$ & 0$^{\circ}$-90$^{\circ}$\\
    \end{tabular}
    \label{tab:torus}}
\end{table}

\subsubsection{X-ray luminosity}\label{sec:Xray}
From the AGN decomposition performed with {\sc sed3fit} we retrieve, for each simulated galaxy, the line of sight optical depth at 9.7 $\mu$m ($\tau_{9.7}$) and the hydrogen column density associated to the torus model. The smooth torus models have values of the hydrogen optical depth below $log_{10}({\rm N_{H}}/{\rm cm}^{-2})<$23.6 and it is therefore not possible to reach the Compton-thick regime (i.e., $log_{10}({\rm N}_{H}/{\rm cm}^{-2})>$24) using such models.
For the clumpy torus models we derive the hydrogen column density of the torus from the optical depth in the V-band ($\tau_{V}$) of the torus, which is among the input parameters of the model, considering that N$_{H}=$1.8$\times 10^{21}\,1.086\,\tau_{V}\,{\rm cm}^{-2}$ \citep{Predehl1995}. 
 The intrinsic hard X-ray luminosity, as derived at 2-10keV, is estimated from the luminosity at 12$\mu$m associated to the AGN component, using the relation derived by \citet{Asmus2015} considering a sample of IR and X-ray selected AGN:
 \begin{equation}
     log_{10}(L_{2-10keV}/ergs^{-1})=43.30+(log_{10}(L_{12\mu m}/ergs^{-1})-43)/0.98.
 \end{equation} \par
\begin{figure}[hbt!]
    \centering
    \includegraphics[width=0.9\linewidth,keepaspectratio]{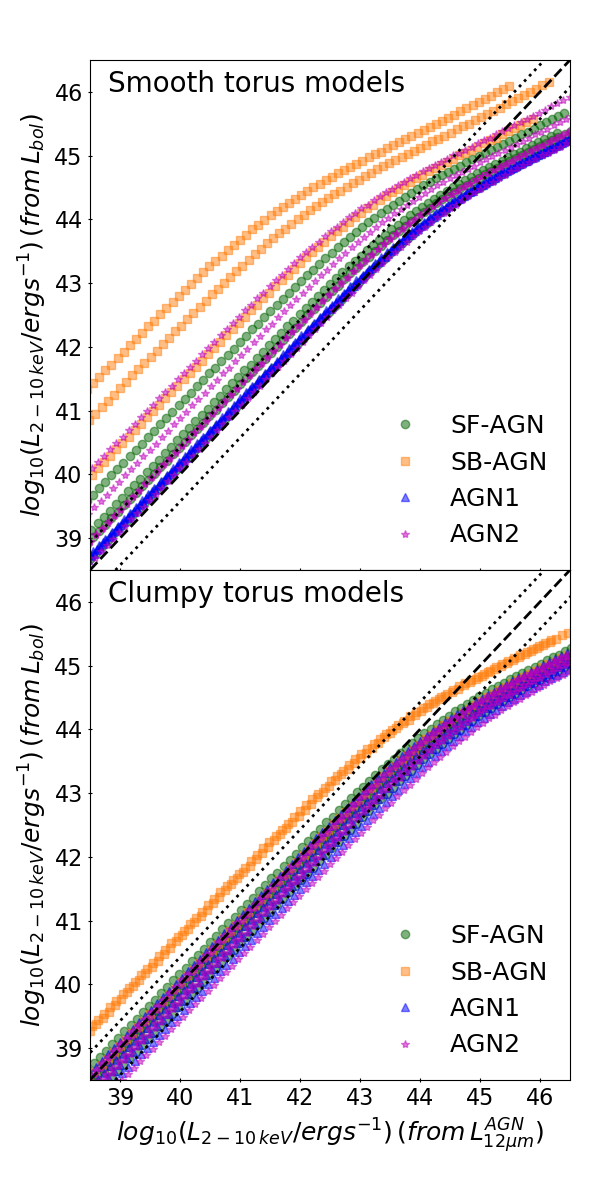}
    \caption{Comparison between the X-ray luminosity obtained from the luminosity at 12$\mu$m \citep{Asmus2015} and the X-ray luminosity obtained from the AGN bolometric luminosity and assuming the correction by \citet{Duras2020}. The comparison is shown separately for the smooth (\textit{top}) and the clumpy torus models (\textit{bottom}). Different colours indicate the galaxy populations with an AGN component (see legend). The X-ray luminosities are compared before applying any dispersion and are shown for both smooth (\textit{top}) and clumpy torus models (\textit{bottom}). The black dashed line is the identity line while the dotted lines show the 1$\sigma$ combined dispersion of the two relations.} 
    \label{fig:Lxcomparison}
\end{figure}

 As mentioned in the same paper, this relation have been tested for AGN with X-ray luminosity below log$_{10}(L_{2-10keV}/ergs^{-1})=$45 and a possible flattening of the relation could be present at higher luminosity. At higher X-ray luminosity, we considered the bolometric correction derived by \citet{Duras2020} using a sample of X-ray selected AGN. The bolometric correction is described as a function of the AGN bolometric luminosity for bright AGN, i.e. $L^{AGN}_{BOL}>10^{11}\,L_{\odot}$, and it is constant otherwise. We derived the AGN bolometric luminosity from the {\sc sed3fit} fit. The comparison between the two intrinsic X-ray luminosity values for all AGN in \spr{} is shown in Figure \ref{fig:Lxcomparison} for both clumpy and smooth torus. The X-ray luminosity derived using the bolometric correction is larger than the other one for the majority of galaxies, except for the brightest AGN. As previously mentioned, the bolometric correction derived by \citet{Duras2020} is a function of the bolometric luminosity for the brightest AGN, while it is constant for fainter AGN. The two X-ray luminosities are quite similar for the AGN1 and AGN2 populations, while they differ for the composite systems, particularly for the SB-AGN population described by a smooth torus. Differences in the two approaches may arise from the different samples of AGN considered to derive the two relations, as may suggest the similarity between the most X-ray luminous AGN (AGN1 and AGN2). On the other hand, it is necessary to consider that there are also uncertainties on the AGN-host galaxy decomposition performed in this work. \par
 
 In the \spr{} simulation, we assigned to each simulated galaxy the X-ray luminosity derived from the luminosity at 12$\mu$m. This choice is motivated by the presence of IR-selected AGN in the sample used by \citet{Asmus2015}, as the main aim of this work is to create realistic IR simulated galaxies. In addition, the peak of the AGN emission at 12 $\mu$m should be less affected by the used torus model than the bolometric AGN luminosity. Last, but not least, the X-ray luminosity derived from the L$_{12\mu m}$ shows a better agreement with the observed X-ray LF than the X-ray luminosity derived with the bolometric correction (see Section \ref{sec:validation}). \par
 We randomly scatter the X-ray luminosity of each simulated galaxy around the value derived from the relation by \citet{Asmus2015}, considering a Gaussian distribution with $\sigma=$0.33, which corresponds to the uncertainty derived in the same work.\par
 
After deriving the intrinsic X-ray luminosity, we retrieved the observed luminosity L$_{2-10keV,obs}$ considering the description by \citet{Lusso2010}:
\begin{equation}
    L_{X,obs}=A\int_{\nu_{1}(1+z)}^{\nu_{2}(1+z)} E^{-\Gamma +1} e^{-E/E_{cut}} e^{-N_{H}\sigma_{E}} dE
\end{equation}
where A is the normalisation derived to match the intrinsic hard X-ray luminosity, E$_{cut}$ is the high-energy cutoff of the primary AGN power-law component, assumed to be 200 keV \citep{Gilli2007} and $\sigma_{E}$ is the effective photoelectric absorption cross-section for hydrogen atoms \citep[Table 2 from][]{Morrison1983}. Finally, $\Gamma$ is the X-ray slope and it assumed to be equal to 2 in the soft X-ray (0.5-2 keV) and equal to 1.7 in the hard X-ray (2-10 keV). The same X-ray spectral energy distribution \citep{Lusso2010} is considered to retrieve the soft and hard X-ray fluxes. In the \spr{} simulation, when deriving the observed X-ray luminosity we apply the scatter associated to both the intrinsic luminosity and the hydrogen column density.
\par
To summarise, using the different empirical relation mentioned above, we assigned to each simulated galaxy with an AGN component an intrinsic and observed luminosity at 2-10 keV, as well as the expected flux in the soft and hard X-ray. The simulation does not include any X-ray emission arising from the host galaxy.

\subsubsection{Radio luminosity}
We derived the expected radio luminosity due to the star-formation activity at 1.4 GHz for each simulated galaxy by considering the redshift evolution of the logarithmic ratio of the 1.4 GHz and total IR luminosity by \citet{Delhaize2017}. In particular, we considered the updated version of this relation, as derived after removing radio-AGN by \citet{Delvecchio2018}:
\begin{equation}\label{eq:1.4GHz}
\begin{aligned} 
    log_{10}\left( \frac{L_{IR,SF}/W}{3.75\times10^{12}\,{\rm Hz}} \right) -& log_{10}\left( \frac{L_{1.4\,GHz}}{{\rm W Hz^{-1}}} \right)= \\
    &=(2.80\pm0.02)(1+z)^{-0.12\pm0.01}.
\end{aligned}
\end{equation}
In the IR luminosity we included only the component due to star-formation. \par
We also include the contribution of radio-loud AGN by combining the estimated fraction of radio-loud AGN by \citet{Best2005} and the pure-luminosity evolution of the 1.4 GHz LF of AGN by \citet{Smolcic2017}.
In particular, \citet{Best2005} found that the local fraction of radio-loud AGN above a certain radio luminosity at 1.4GHz depends on stellar mass as:
\begin{equation}\label{eq:fRAGN}
    f_{RL}=f_{0} \left( \frac{M^{*}}{10^{11}M_{\odot}} \right)^{\alpha}\left[\left(\frac{L}{L_{*}}\right)^{\beta}+\left(\frac{L}{L_{*}}\right)^{\gamma}\right]^{-1}
\end{equation}
where f$_{0}$=0.0055$\pm$0.0004 is the normalisation factor, $\alpha=$2.5$\pm$0.2 accounts for the scaling between the fraction of radio-loud AGN and stellar mass, whereas L$_{*}$, $\beta$ and $\gamma$ are the knee, the faint and the bright-end slope of the LF, respectively. The local LF found by \citet{Best2005} is generally lower than the ones found by other authors \citep{Ceraj2018}; here we decided to consider the results by \citet{Mauch2007}, i.e. L$^{*}=10^{24.59}$ W/Hz, $\beta=$1.27 and $\gamma=$0.49. The same local radio-AGN LF has been considered also by \citet{Smolcic2017}, from which we derived its pure-luminosity evolution with redshift:
\begin{equation}\label{eq:LFRAGN_ev}
    \Phi(L,z,\alpha_{L},\beta_{L})=\Phi_{0} \left[ \frac{L}{(1+z)^{\alpha_{L}+z\,\beta_{L}}} \right]
\end{equation}
where $\alpha_{L}=$2.88$\pm$0.82, $\beta_{L}=$-0.84$\pm$0.34 and $\Phi_{0}$ is the local radio-AGN LF.
As stated also in \citet{Smolcic2017}, considering the pure-density evolution instead of the pure-luminosity evolution brings similar results. We then combined eq. \ref{eq:fRAGN} and \ref{eq:LFRAGN_ev} to estimate the probability for a simulated galaxy in \spr{} to be a radio-loud AGN with a specific 1.4 GHz luminosity, starting from its stellar mass and redshift. We considered a single population of radio-loud AGN, without separating for their accretion mechanisms \citep{Best2005,Tasse2008}. These different radio-loud AGN populations may be considered in future studies.
\par
We also derived for each simulated galaxy the expected luminosity due to the star-formation activity at 150 MHz, following the empirical relation by \citet{Gurkan2018}:
\begin{equation}
\begin{split}
L&_{150MHz}[W\,Hz^{-1}]= \\
    &\Bigg\{  \begin{array}{lr}
    L_{C}\, SFR^{\beta_{low}}(\frac{M_{*}}{10^{10} M_{\odot}})^{\gamma} & ,{\rm if} \,SFR \leq SFR_{break}\\
    L_{C}\,SFR^{\beta_{high}}(\frac{M_{*}}{10^{10} M_{\odot}})^{\gamma}SFR_{break}^{\beta_{low}-\beta_{high}} & , {\rm if}\,SFR > SFR_{break} \\ 
    \end{array} 
\end{split}
\end{equation}
where the normalisation is log$_{10}(L_{C}/ W\,Hz^{-1} )=22.02\pm0.02$, the slopes are $\beta_{low}=0.52\pm0.03$, $\beta_{high}=1.01\pm0.02$ and $\gamma=0.44\pm0.01$, and the position of the break is $log_{10}(SFR_{break} [M_{\odot} yr^{-1}])=0.01\pm0.01$. \par

\subsection{Emission features}\label{sec:lines}
To investigate the spectroscopic capability of future telescopes, we included in the \spr{} simulation a large set of emission features mainly, but not only, in the IR wavelength range.\par
\begin{figure*}[hbt!]
    \centering
    \includegraphics[width=1\linewidth,keepaspectratio]{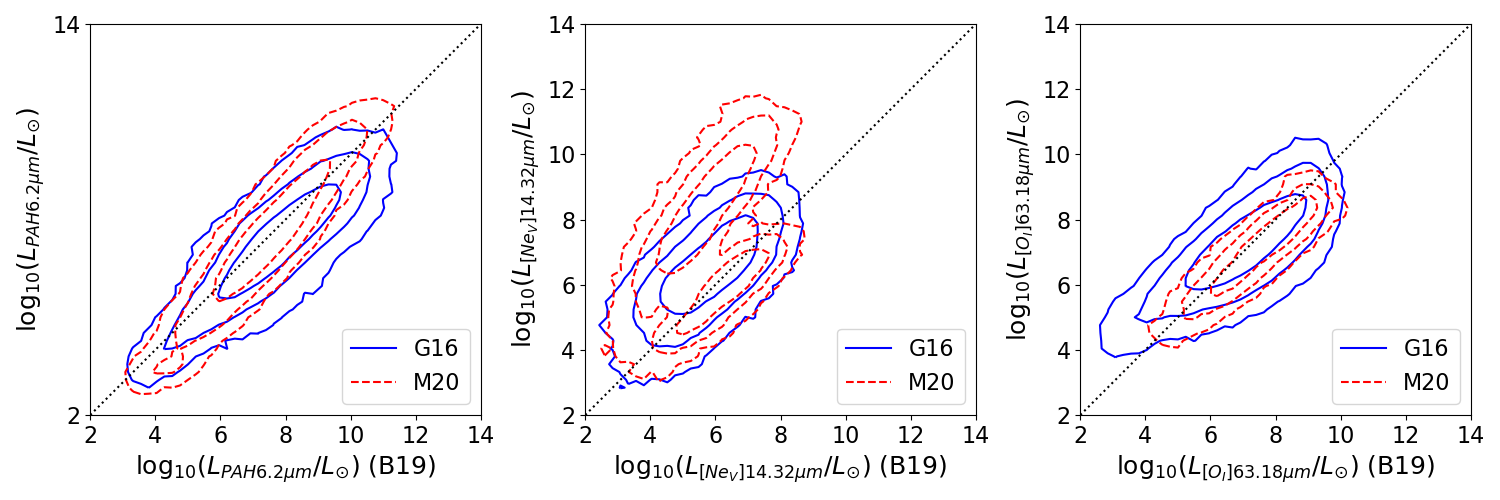}
    \caption{Comparison between the line luminosities derived using the relation by \citet[][B19,x-axis]{Bonato2019} with the ones derived with the relations by \citet[][\textit{blue solid lines}]{Gruppioni2016} and Mordini et al. (in prep., \textit{red dashed lines}). Contour lines indicate the 68$\%$, 95$\%$ and 99.7$\%$ percentiles of the distribution. The three panels show three different lines: PAH at 3.3 $\mu$m (\textit{left}), $[\ion{\rm Ne}{V}]$ 14.32 $\mu$m (\textit{centre}) and $[\ion{\rm O}{I}]$ 63.18 $\mu$m (\textit{right}).  The relations may be different for AGN and star-forming dominated systems, as explained in the text.} 
    \label{fig:lines}
\end{figure*}

From the AGN-galaxy decomposition obtained by {\sc sed3fit} we derived the relative contribution of AGN and star-formation for each galaxy template. We then use several empirical relations between the luminosity of IR emission lines and the IR luminosity, either total or decomposed into SF/AGN contribution. In particular, we considered the empirical relations derived by \citet{Gruppioni2016,Bonato2019} and Mordini et al. (in prep.). For each line, \citet{Gruppioni2016} and Mordini et al. (in prep.) derived two relations, for AGN dominated systems and for galaxies where AGN is a minor component. In the \spr{} simulation this separation is applied to galaxies above or below $f_{AGN}=40\%$, where $f_{AGN}$ is the AGN fraction contributing to the light emitted between 5 and 50 $\mu$m. The full list of IR emission features included in the \spr{} simulation is listed in Table \ref{tab:lines} together with the corresponding reference. In \citet{Bonato2019}, there are also relations linking the L$_{IR}$ to the CO molecular lines, which we also included. All line luminosities, including the luminosity of the different PAH features that are already present in the SED templates associated to each simulated galaxies, are listed separately in the catalogue for a quick use, e.g. to estimate the expected line luminosity of galaxies without fitting the corresponding spectra. In Figure \ref{fig:lines} we show as example the comparison of the line luminosities of the PAH at 3.3 $\mu$m, $[\ion{\rm Ne}{V}]$ 14.32 $\mu$m and $[\ion{\rm O}{I}]$ 63.18 $\mu$m lines, as derived from the corresponding relations.   
\begin{table}
    \caption{List of IR emission features considered in the catalogue with the respective references: B19: \citet{Bonato2019}, G16: \citet{Gruppioni2016}, M20: Mordini et al. (in prep.)}
    \label{tab:lines}
    \centering
    \resizebox{0.49\textwidth}{!}{
    \begin{tabular}{cc||cc}
        Line & Reference & Line & Reference\\
        \hline
        PAH 3.3 $\mu$m & B19 & $[\ion{\rm Ar}{III}]$ 21.82 $\mu$m & B19 \\
        PAH 6.6 $\mu$m & G16,B19,M20 & $[\ion{\rm Fe}{III}]$ 22.90 $\mu$m & B19,M20 \\
        $[\ion{\rm Si}{VII}]$ 6.50 $\mu$m & B19 & $[\ion{\rm Ne}{V}]$ 24.31 $\mu$m & G16,B19,M20 \\
        \ion{\rm H}{2} 6.91 $\mu$m & B19 &  $[\ion{\rm O}{IV}]$ 25.89 $\mu$m & G16,B19,M20 \\
        $[\ion{\rm Ar}{II}]$ 6.98 $\mu$m & B19 & $[\ion{\rm Fe}{II}]$ 25.98 $\mu$m & B19,M20 \\
        $[\ion{\rm Ne}{VI}]$ 7.65 $\mu$m & B19,M20 & $[\ion{\rm S}{III}]$ 33.48 $\mu$m & G16,B19,M20 \\
        PAH 7.7 $\mu$m & B19,M20 & $[\ion{\rm Si}{II}]$ 34.81 $\mu$m & G16,B19,M20 \\
        $[\ion{\rm Ar}{V}]$ 7.90 $\mu$m & B19 & $[\ion{\rm O}{III}]$ 51.81 $\mu$m & B19,M20 \\ 
        PAH 8.6 $\mu$m & B19,M20 & $[\ion{\rm N}{III}]$ 57.32 $\mu$m & G16,B19,M20\\
        $[\ion{\rm Ar}{III}]$ 8.99$\mu$m & B19,M20 & $[\ion{\rm O}{I}]$ 63.18 $\mu$m & G16,B19,M20 \\
        \ion{\rm H}{2} 9.66 $\mu$m & B19 & $[\ion{\rm O}{III}]$ 88.36 $\mu$m & G16,B19,M20 \\
        $[\ion{\rm S}{IV}]$ 10.49 $\mu$m & B19,M20 & $[\ion{\rm N}{II}]$ 121.90 $\mu$m & G16,B19,M20 \\
        PAH 11.3 $\mu$m & G16,B19,M20 & $[\ion{\rm O}{I}]$ 145.52 $\mu$m & G16,B19,M20 \\
        $[\ion{\rm Ca}{V}]$ 11.48 $\mu$m & B19 & $[\ion{\rm C}{II}]$ 157.7 $\mu$m & G16,B19,M20 \\
        \ion{\rm H}{2} 12.28 $\mu$m & B19 & $[\ion{\rm N}{II}]$ 205.18 $\mu$m & B19 \\
        \ion{\rm H}{I} 12.37 $\mu$m & B19 & $[\ion{\rm C}{I}]$ 370.42 $\mu$m & B19 \\
        PAH 12.7 $\mu$m & B19 & $[\ion{\rm C}{I}]$ 609.14 $\mu$m & B19 \\
        $[\ion{\rm Ne}{II}]$ 12.81 $\mu$m & G16,B19,M20 & CO(13-12) 200.27$\mu$m & B19\\
        $[\ion{\rm Ar}{V}]$ 13.09 $\mu$m & B19 & CO(12-11) 216.93$\mu$m& B19\\
        $[\ion{\rm Mg}{V}]$ 13.50 $\mu$m & B19 & CO(11-10) 236.99$\mu$m & B19\\
        PAH 14.2 $\mu$m & M20 & CO(10-9) 268.2$\mu$m & B19\\
        $[\ion{\rm Ne}{V}]$ 14.32 $\mu$m & G16,B19,M20 & CO(9-8) 289.1$\mu$m & B19\\
        $[\ion{\rm Cl}{II}]$ 14.38 $\mu$m & B19,M20  & CO(8-7) 325.2$\mu$m & B19\\
        $[\ion{\rm Ne}{III}]$ 15.55 $\mu$m & G16,B19,M20 & CO(7-6) 371.7$\mu$m & B19\\
        \ion{\rm H}{2} 17.03 $\mu$m & B19,M20 & CO(6-5) 433.6$\mu$m & B19\\
        $[\ion{\rm P}{III}]$ 17.89 $\mu$m & M20 & CO(5-4) 520.2$\mu$m & B19\\
        $[\ion{\rm Fe}{II}]$ 17.93 $\mu$m & B19,M20 & CO(4-3) 650.3$\mu$m & B19\\
        $[\ion{\rm S}{III}]$ 18.71 $\mu$m & G16,B19,M20 & & \\
    \end{tabular}}
\end{table}

In addition, we included in the \spr{} simulation a set of recombination lines of the atomic hydrogen, from the Balmer to the Humphreys series. The H$_{\alpha}$ emission line was derived assuming the relation with the SFR by \citet{Kennicutt1998}. All the other lines were then derived by considering the line intensity ratio by \citet{Hummer1987} and \citet{Osterbrock2006} for the case B recombination with electron temperature T$_{e}=$10000 K in the low density limit, i.e. electron density N$_{e}=$100-10000 cm$^{-3}$. Table \ref{tab:Hlines} contains the complete list of the atomic hydrogen lines considered in the simulation. \par

\begin{table}[hbt!]
     \caption{List of recombination lines of the atomic hydrogen included in the \spr{} simulation with the corresponding vacuum wavelengths \citep{Lang1999,Morton2000}. The relation used to derive each of them is reported in the footnotes.}
    \label{tab:Hlines}
   \centering
    \begin{threeparttable}

    \begin{tabular}{cc||cc}
        line & $\lambda$ [\AA{}] & line & $\lambda$ [\AA{}]\\
        \hline
        H$_{9}$\tnote{1} & 3836.5  &Br$_{9}$\tnote{3} & 18177.0 \\
        H$_{8}$\tnote{1} & 3890.2  &Br$_{\delta}$\tnote{1} & 19451.3 \\
        H$_{\epsilon}$\tnote{1} & 3971.2  &Br$_{\gamma}$\tnote{1} & 21660.9 \\
        H$_{\delta}$\tnote{1} & 4102.9  &Br$_{\beta}$\tnote{1} & 26259.2 \\
        H$_{\gamma}$\tnote{1} & 4341.7  &Br$_{\alpha}$\tnote{1} & 40523.0 \\
        H$_{\beta}$\tnote{1} & 4862.7  &Pf$_{\delta}$\tnote{3} & 32970.0 \\
        H$_{\alpha}$\tnote{2} & 6564.6  &Pf$_{\gamma}$\tnote{3} & 37405.2 \\
         Pa$_{9}$\tnote{3} & 9231.5  &Pf$_{\beta}$\tnote{3} & 46537.7 \\
         Pa$_{8}$\tnote{3} & 9548.6  &Pf$_{\alpha}$\tnote{3} & 74598.3 \\
         Pa$_{\delta}$\tnote{1} & 10052.1  &Hu$_{\gamma}$\tnote{3} & 59082.1 \\
         Pa$_{\gamma}$\tnote{1} & 10941.1  &Hu$_{\beta}$\tnote{3} & 75025.4 \\
         Pa$_{\beta}$\tnote{1} & 12821.6  &Hu$_{\alpha}$\tnote{3} & 123713.7 \\
         Pa$_{\alpha}$\tnote{1} & 18756.1  && \\
    \end{tabular}
  \begin{tablenotes}
    \item[1] \citet{Osterbrock2006}
    \item[2]\citet{Kennicutt1998}
    \item[3] \citet{Hummer1987}
  \end{tablenotes}
    \end{threeparttable}
\end{table}

We also included some of the main nebular emission lines at optical and near-IR wavelengths (Table \ref{tab:opticallines}). In particular, using the relation by \citet{Kennicutt1998}, we derived the luminosity of the $[\ion{\rm O}{II}]$ doublet at 3727/3729 \AA{} from the SFR as:
\begin{equation}
    L([\ion{\rm O}{II}]) [erg/s] = \frac{SFR [M_{\odot} yr^{-1}]}{(1.4\pm0.4)\times 10^{-41}}
\end{equation}
To derive the luminosity of the $[\ion{\rm N}{II}]$ line at 6584 \AA{}, we applied the relation between metallicity and the $[\ion{\rm N}{II}]$/H$_{\alpha}$ by \citet{Pettini2004}:
\begin{equation}
    {\rm log}_{10}\left( \frac{[\ion{\rm N}{II}]}{H_{\alpha}} \right)=\frac{12+{\rm log}_{10}({\rm O/H})-8.90}{0.57} \\
\end{equation}
considering a 1$\sigma$ scatter of 0.18 dex. 
From the metallicity we also derived the [$\ion{\rm Ne}{III}]$ nebular emission line at 3869 \AA{}, using the relation derived at z$\sim$0.8 by \citet{Jones2015}, which is also consistent with local observations:
\begin{equation}
     {\rm log}_{10} \left(  \frac{[\ion{\rm Ne}{III}]}{[\ion{\rm O}{II}]} \right)=16.8974-2.1588\,(12+{\rm log}_{10}({\rm O/H}))
\end{equation}
For this relation we considered an intrinsic scatter of $\sigma=$0.22 dex \citep{Jones2015}.
We also included the expected luminosity of the $[\ion{\rm N}{II}]$ line at 6548 \AA{} considering the theoretical value of one third \citep{Osterbrock2006}. The luminosity of the $[\ion{\rm O}{III}]$ forbidden line at 5007 \AA{} is derived considering the relation derived by \citet{Kewley2013} for galaxies at z$<$3. This relation relates the $[\ion{\rm O}{III}]$/H$_{\beta}$ ratio to the $[\ion{\rm N}{II}]$/H$_{\alpha}$ ratio and includes the evolution with redshift:
\begin{equation}
    {\rm log}_{10} \left(  \frac{[\ion{\rm O}{III}]}{H_{\beta}} \right) =1.1+0.3\,{\rm z} +\frac{0.61}{{\rm log}_{10}([\ion{\rm N}{II}]/H_{\alpha})+0.08-0.1833\,{\rm z}}
\end{equation}
The luminosity of the $[\ion{\rm O}{III}]$ forbidden line at 4959 \AA{} is then assumed to follow the theoretical value and be one third of the luminosity of the [\ion{\rm O}{III}] at 5007 \AA{} \citep{Osterbrock2006}.
The luminosity of the $[\ion{\rm S}{II}]$ doublet at 6717/6731 \AA{} is derived considering the empirical relation between the metallicity, the $[\ion{\rm N}{II}]$ and the H$_{\alpha}$ lines by \citet{Dopita2016}:
\begin{equation}
    12+{\rm log}_{10}({\rm O/H})=8.77+{\rm log}_{10} \left(  \frac{[\ion{\rm N}{II}]}{[\ion{\rm S}{II}]} \right) + 0.264\,{\rm log}_{10} \left(  \frac{[\ion{\rm N}{II}]}{H_{\alpha}} \right)
\end{equation}
To separate the contribution of the two $[\ion{\rm S}{II}]$ doublet lines, which ratio depends on the electron density (n$_{e}$), we first derived the expected electron density from the specific SFR and the stellar mass, following the empirical relation found by \citet{Kashino2019}:

\begin{equation}
\begin{split}
    {\rm log}&_{10}(sSFR/yr^{-1})= -12.661^{+0.124}_{-0.125}+-0.627^{+0.028}_{-0.029}\\
    & \times({\rm log}_{10}(M^{*}/M_{\odot})-10)+1.753^{+0.063}_{-0.064}{\rm log}_{10}(n_{e}/cm^{-3}). 
\end{split}
\end{equation}

We then used the electron density to derive the ratio between the $[\ion{\rm S}{II}]$ doublet lines, assuming an electron temperature of 10$^{4}$ K and the empirical relation by \citet{Proxauf2014}:
\begin{equation}
\begin{split}
    {\rm log}&_{10}(n_{e}/cm^{-3})=0.0543\,{\rm tan}(-3.0553\,R + 2.8506)\\
    & + 6.98 -10.6905\,R + 9.9186\,R^{2} 
\end{split}
\end{equation}

where R is the ratio between the 6716\AA{} and 6731\AA{} $[\ion{\rm S}{II}]$ line flux.
\par

As similar empirical relations are not available for AGN, we considered predictions from photoionization calculations for the same lines listed in Table \ref{tab:Hlines} and \ref{tab:opticallines}. In particular, we incorporated the contribution from the narrow line gas emitting regions (NLR) of AGN to the line emission of AGN-SB, AGN-SF, AGN1 and AGN2. The emission from the broad line gas emitting (BLR) is not included in the current version of \spr{} and will be subject of future works. This impacts only the line emission of the permitted lines in AGN1, whose emission should be considered as a lower limits at present. \par
The model describing the NLR emission is that from \citet{Feltre2016}, based on the phoionization code {\sc CLOUDY}  \citep[v13.03,][]{Ferland2013}. In particular, to derive the desired line luminosities, we considered an ionisation parameter at the Str\"omgren radius between log$_{10}( U_{\rm S})$=-1.5 and -3.5, sub- and super-solar metallicity (0.008, 0.017, 0.03), a dust-to-metal ratio of 0.3, a UV spectral index $\alpha=$-1.4 and internal micro-turbulence velocity v$=$100km/s \citep[see][]{Mignoli2019}. The hydrogen number density is assumed to be 10$^{3}$ cm$^{-3}$ for type-1 AGN, defined as AGN with hydrogen column density log$_{10}(N_{H}/cm^{-2}) <$22. 
This division is obvious for our AGN1 and AGN2 populations and it is mainly necessary to separate SF-AGN and SB-AGN into obscured and non obscured AGN.\par
We apply attenuation from dust to all the lines assuming the \citet{Charlot2000} cloud model in order to be consistent with the dust model used in {\sc Magphys}. In particular, the dust attenuation has a diffuse component, describing the interstellar medium (ISM), and a component concentrated around young stars, representing their birth clouds. As emission lines are predominantly generated by young stellar systems, we assumed that they are affected by both components. The two components of the optical depth depend differently on wavelength, as the ISM component is $\tau^{ISM}_{\lambda}\propto\lambda^{-0.7}$ while the birth clouds component is $\tau^{BC}_{\lambda}\propto\lambda^{-1.3}$. Both the average V-band optical depth and the fraction of the ISM contribution to it are derived from the SED fit performed with {\sc Magphys} or {\sc Sed3fit}. For each line, we include the scatter associated to each considered relation, as written in the respective papers, or, when the relative papers do not quote it, a generic scatter of 0.1 dex.\par
Future releases may include additional lines as well as emissions from the BLR.

\begin{table}[hbt!]
     \caption{List of optical nebular emission lines included in the \spr{} simulation for both star-forming systems and AGN.}
    \label{tab:opticallines}
   \centering
    \begin{tabular}{ccc}
        line & $\lambda$ \\
        \hline       
        $[\ion{\rm O}{II}]$ & 3727/3729 \AA{} \\
        $[\ion{\rm Ne}{III}]$ & 3869 \AA{} \\
        $[\ion{\rm O}{III}]$ & 4959 \AA{} \\
        $[\ion{\rm O}{III}]$ & 5007 \AA{} \\
        $[\ion{\rm N}{II}]$ & 6548 \AA{} \\
        $[\ion{\rm N}{II}]$ & 6584 \AA{} \\
        $[\ion{\rm S}{II}]$ & 6717 \AA{} \\
        $[\ion{\rm S}{II}]$ & 6731 \AA{} \\
    \end{tabular}
\end{table}

\section{Creation of the simulated catalogues}\label{Sec:mocks}
As described in the previous sections, we derived a master catalogue starting from the observed LFs and GSMF of the different galaxy populations. This catalogue contains the number density of each galaxy population in specific L$_{IR}$-z bins, as well as the median values of the physical parameters associated to it. 
From the master catalogue it is possible to derive simulated catalogues corresponding to current and future spectro-photometric surveys, once their area and depth are known.\par
In particular, from the area of the survey we derived the volume of the Universe corresponding to each redshift bin considered in the master catalogue and then the expected number of galaxies N$_{0}(z,L_{IR})$ for each galaxy population at the various redshift. To include the variance on the simulated simulated catalogues, we considered the Poisson errors associated to N$_{0}(z,L_{IR})$ and we generate a number of simulated galaxies equal to N(z,L$_{IR}$)=N$_{0}(z,L_{IR})\pm\sqrt(N_{0}(z,L_{IR}))$. 
To each simulated galaxy we assigned random values of IR luminosity and redshift, considering a flat probability distribution inside the corresponding L$_{IR}$-z bin. We then assigned the various physical properties mentioned in the previous sections, by randomly extracting a value following the corresponding probability distribution function. To summarise, we ended up with a series of simulated galaxies corresponding to different galaxy populations and their number is derived from the corresponding LFs, or GSMF for irregular galaxies, and depends on the area of the simulated survey. It is now necessary to verify which galaxy is expected to be detected in the desired survey by retrieving a set of simulated fluxes and spectra to compare with the spectro-photometric depth of the survey of interest at specific wavelengths. The derivation of simulated fluxes and spectra is described in the following sections.\par

\subsection{Simulated fluxes}
To retrieve the expected flux in different bands for each simulated galaxy, we convolved the SED templates associated to each galaxy population with different filter troughtputs. In particular, we included a set of absolute magnitudes in standard filters: NUV \citep[GALEX,][]{Zamojski2007}, u, r \citep[Sloan Digital Sky Survey,][]{Zehavi2011}, B, V \citep[SUBARU SUPRIME-CAM,][]{Miyazaki2002}, J \citep[UKIRT/WFCAM,][]{Casali2007} and Ks \citep[WIRCam,][]{Puget2004}. We also derived for each simulated galaxy the expected flux for some past, current and future facilities among which the JWST, \textit{Euclid} \citep{Laureijs2010}, \textit{Spitzer} \citep{Werner2004}, the Sloan Digital Sky Survey \citep[SDSS;][]{Gunn1998}, the United Kingdom Infrared Telescope, the \textit{Hubble Space Telescope} (\textit{HST}), \hers{}, the SCUBA-2 instrument on the James Clerk Maxwell Telescope \citep{Holland2013}, the AKARI telescope \citep{Murakami2007}, the Wide-field Infrared Survey Explorer \citep{Wright2010}, the European Extremely Large Telescope \citep[ELT;][]{Gilmozzi2008}, the Vera Rubin Observatory \citep[LSST;][]{Ivezic2008}  and OST. The complete list of filters included in the simulation with the corresponding central wavelengths and references is shown in Appendix \ref{sec:filters}. Additional filters may be included in future updates. In addition, considering the hydrogen column density derived with {\sc sed3fit} for each template and the X-ray spectral energy distribution described in Section \ref{sec:SEDmodels}, we retrieve the expected flux in the soft 0.5-2 keV and hard 2-10 keV X-ray bands. Additional filters may be considered in future updates. \par
To simulate a specific survey we then considered the flux in the filter of interest and we derived the flux error considering the desired observational depth. We applied a random scatter to each flux equal to the desired observational depth.

\subsection{Simulated IR low-resolution spectra}

\begin{table}[]
    \caption{Wavelength coverage and spectral resolution of the three spectrographs considered in \spr{}}
    \centering
    \begin{tabular}{ccc}
        Instrument & wavelength coverage & Spectral resolution \\
        \hline
       JWST/MIRI  & 5-12 $\mu$m & 40-160 \\
       OST/OSS & 25-290 $\mu$m & 300 \\
    \end{tabular}
    \label{tab:instruments}
\end{table}

In the \spr{} simulation we included synthetic spectra derived considering the wavelength coverage, resolution and the sensitivity expected for the low-resolution spectra for the JWST Mid-Infrared Instrument \citep[MIRI,][]{Kendrew2015,Rieke2015,Wright2015} and the low-resolution Origins Survey
Spectrometer \citep[OSS; ][]{Bradford2018} planned for OST (Table \ref{tab:instruments}). Briefly, JWST/MIRI-LR covers between 5 and 12 $\mu$m with resolving power ranging from $\sim$40 at 5$\mu$m to $\sim$160 at 10$\mu$m. OST/OSS is expected to cover between 25 and 590 $\mu$m with spectral resolution R=300 and wide-field survey capability. Additional spectrographs, like other JWST spectroscopic modes, may be considered in future update. \par
For consistency with the photometric predictions, we started from the same set of empirical templates used in the rest of the simulation, i.e 35 templates from \citet{Polletta2007}, \citet{Rieke2009} and \citet{Bianchi2018}. These templates are similar to the starburst templates by \citet{Brandl2006} and \citet{HernanCaballero2011}, in the overlapping galaxy population. The SB-AGN templates show a very deep 9.7$\mu$m silicate absorption feature that is not present in other works. Therefore, for this galaxy population we considered a combination of the continuum of the SED template used for photometry and the features present in the Seyfert2 template by \citet{HernanCaballero2011}. When necessary, we converted all the templates to vacuum wavelengths\footnote{Using the code available in \url{https://github.com/trevisanj/airvacuumvald}} \citep{Morton2000}.\par
We added several nebular emission lines to the spectra considering the expected line luminosity derived using different empirical and theoretical predictions, as explained previously in Section \ref{sec:lines}. We included all the nebular emission lines, but we did not insert PAH features as they are already present in the templates. Since the template spectra included in \spr{} are empirical, their already include some of the optical nebular lines. When this happens, we removed the line from the spectra by linearly interpolating the continuum around the line before adding the new nebular emission line with the desired luminosity. In other words, we assumed a flat continuum around the nebular emission lines, which are generally quite narrow. This assumption is not expected to impact much the results as we aim to simulate low-resolution spectra, however, when necessary, the use of the {\sc Magphys} or {\sc Sed3fit} template can be considered to avoid this issue. For the component due to star-formation, we derived the expected gas velocity dispersion from the stellar mass to take into account the broadening of the nebular emission lines \citep{Bezanson2018}:
\begin{equation}
    {\rm log}_{10}(\sigma_{g,int})=(-1.34\pm0.44)+(0.33\pm0.04){\rm log}_{10}(M^{*}/M_{\odot})
\end{equation}
For nebular emission lines of AGN we assumed full-width-half-maximum (FWHM) between 500 and 800 km/s. We assumed the central wavelengths from \citet{Lang1997} and we converted it to vacuum wavelength \citep{Morton2000}.  
\par
For each simulated galaxy we shifted the corresponding empirical template to the assigned redshift and we applied the wavelength resolution of the desired spectrograph, i.e. JWST/MIRI-LR or OSS/OST. \par
\begin{figure}[hbt!]
    \centering
    \includegraphics[width=1\linewidth,keepaspectratio]{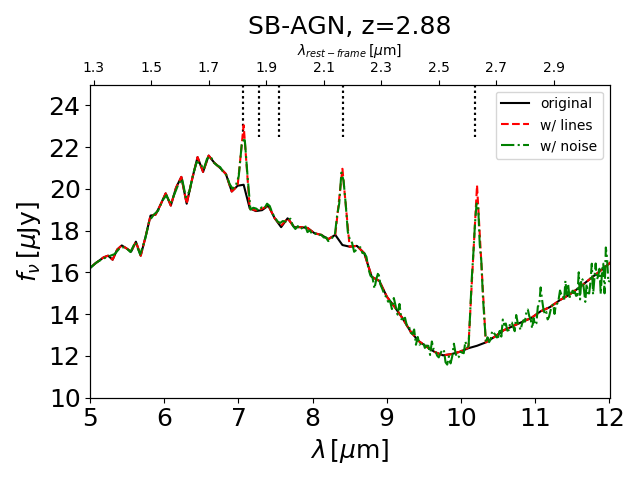}
    \caption{An example of a low-resolution JWST/MIRI spectra of a SB-AGN galaxy at z=2.88. At the top we report the rest-frame wavelengths, while on the bottom we report the observed ones. We show the original spectrum (\textit{solid black line}), the spectrum after the addition of nebular lines (\textit{dashed red line}) and the spectrum after the application of the spectral resolution and noise (\textit{dotted green line}). On the top, we marked with short vertical dotted lines the wavelength position of the nebular emission lines included, despite of their S/N.}
    \label{fig:MIRIspectra_example}
\end{figure}

An example of a JWST/MIRI low-resolution spectra is reported in Figure \ref{fig:MIRIspectra_example}, where we show the simulated spectra of a SB-AGN galaxy at z=2.88, considering the slit observational mode. We considered an exposure time of 10h, reaching a 5$\sigma$ continuum depth of 0.18 and 3.56 $\mu$Jy at 5 and 12 $\mu$m, respectively. The example shows the inclusion of bright hydrogen lines, i.e. Br$_{\alpha}$, Br$_{\beta}$ and Br$_{\gamma}$, with both the contribution of AGN and star-formation. Fainter lines, such as Br$_{\delta}$ and Br$_{9}$, have been included but are below the continuum level. We consider the spectral resolving power and continuum sensitivity reported in \citet{Glasse2015}.

\subsection{Light cone creation}
In order to simulate  galaxies “realistically” distributed in the sky, we made use of the algorithm by \citet{Soneira1978} to distribute galaxies following a two-point correlation function with a specific power-law. The algorithm starts by generating a first layer consisting of $\eta$ random points within a circle of radius R. The second layer consists of $\eta$ points, extracted randomly within a circle of radius R/$\lambda$ centred on each point of the previous level. This procedure is repeated for each level, up to a specific L level, each time reducing the size of the circle by a factor of $\lambda$ and using the points of each level to generate the points of the next one. In this way, the $n$-th level have $\eta^{n}$ points in total and each group of $\eta$ points is generated randomly in a circle, which is centred on the $\eta^{n-1}$ points of the previous level and has a radius of R/$\lambda^{n}$. We repeat this procedure for different redshift bins, from z$=$0 to 10, and we include additional scatter among z bins to avoid sharp features. \par
The two-point correlation function can be roughly described as a power-law in real space $\xi=(r/r_{0})^{-\gamma}$ and, using the approximation by \citet{Limber1953}, the angular correlation function is $w(\theta)=A_{w}\theta^{1-\gamma}$. In our simulation we assumed a power-law slope for the angular correlation function $\delta$=$\gamma$-1=0.7, as suggested by observations \citep[e.g.,][]{Wang2013}. In the algorithm by \citet{Soneira1978}, the power-law slope of the angular correlation function is related to the $\eta$ and $\lambda$ parameters as:
\begin{equation}
    \delta=M-\frac{log_{10}(\eta)}{log_{10}(\lambda)}, \mbox{with } \frac{R}{\lambda^{L-1}}<r<R;
\end{equation}
where M is the number of dimensions and R, L and $\lambda$ are in arc-second. We therefore assumed $\eta$=6 and $\lambda$=4. The starting radius R is chosen to cover the desired survey area and the parameter L is fine-tuned to obtained a number of positions equal or larger than the number of simulated galaxies.\par
The clustering of galaxies shows a dependence with stellar mass, as more massive galaxies tend to live in denser environments \citep[e.g.,][]{Li2006,Meneux2008,Wake2011,Marulli2013}. To reproduce this effect we derived the dependence of the spatial correlation length r$_{0}$ on stellar mass limit using data from \citet{Wake2011} and \citet{Hatfield2016}, which consider the same power-law slope considered in this work. 
Literature works have showed that r$_{0}$ has little or no evolution with redshift \citep{Bethermin2015,Schreiber2015}, therefore we considered only its dependence on the stellar-mass limit. Briefly, we described the r$_{0}$ dependence on stellar-mass limit as a broken power-law:
\begin{equation}
\begin{array}{l}
 r_{0} \propto \begin{cases} M^{k_{M,1}}, & \mbox{if } M^{*}<=M^{*}_{break} \\ M^{k_{M,2}}, & \mbox{if } M^{*}>M^{*}_{break}  \end{cases} \\
\end{array}
\end{equation}
with the stellar mass break equal to log$_{10}({\rm M}^{*}_{break}/{\rm M}_{\odot})=$10.6. The resulting best fit corresponds to a low-mass slope of k$_{M,1}=$0.0959$\pm$0.0003 and a high-mass slope of k$_{M,2}=$0.181$\pm$0.006. From the spatial correlation length r$_{0}$, it is then possible to derive the normalisation of the angular correlation function A$_{w}$ for different redshifts and stellar-mass limits. This can be done deriving first the projected correlation function:
\begin{equation}
    w_{p}(r_{p})=r_{p} \left( \frac{r_{0}}{r_{p}} \right) ^{-\gamma}\frac{\Gamma(1/2)\Gamma(\frac{\gamma-1}{2})}{\Gamma(\gamma/2)}
\end{equation}
where r$_{p}$ is the projected radius in h$^{-1}$Mpc in which the correlation function is calculated, and $\Gamma$ is the gamma function. For each redshift in the simulation, r$_{p}$ is then converted to angular scale using the angular distance. \par
Once we derived the normalisation A$_{w}$ of the angular correlation function for each redshift and stellar mass limit, we fine tuned the positions obtained using the algorithm by \citet{Soneira1978} to match the desired normalisation, similarly to what has been proposed in \citet{Schreiber2017}. In particular, we derived the angular two point correlation function of the position obtained with the original algorithm by \citet{Soneira1978}, using the \citet{Landy1993} method, and we fit a power-law with $\delta=$0.7 to derive its normalisation. We then compared the derived normalisation with the normalisation expected for the considered stellar mass limit and redshift. In case the algorithm predicts a larger normalisation, we decreased it by substituting a fraction $f$ of the positions with completely random ones. We then derived the normalisation of the new positions and repeated the procedure iteratively until we reached the desired A$_{w}$, considering a tolerance of 5$\%$. The fraction $f$ is a function of both the redshift and the stellar mass limits. \par
\begin{figure}
    \centering
    \includegraphics[width=0.7\linewidth,keepaspectratio]{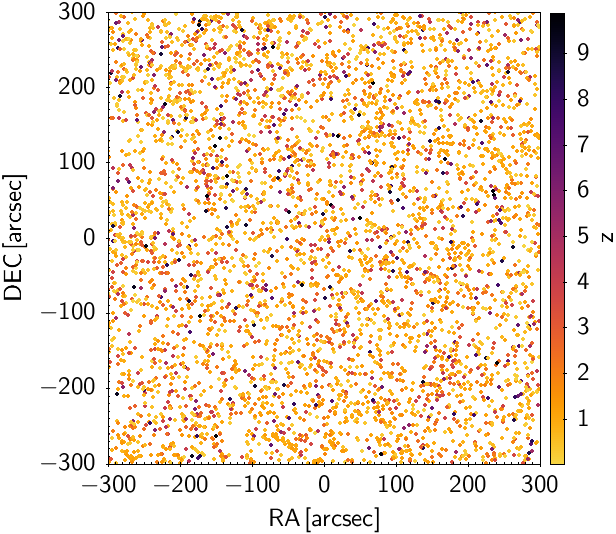}
    \caption{Example of a simulated field of 10$^{\prime}\times10^{\prime}$. Each point represents a simulated galaxy and it is colour coded depending on its redshift.}
    \label{fig:example_field}
\end{figure}

Once we obtained the positions corresponding to the desired angular two-point correlation function, considering both the slope and the normalisation, we associated a position to each simulated galaxy, considering its redshift and stellar-mass. In Figure \ref{fig:example_field} we show an example of a "realistic" sky distribution over a field of 10$^{\prime}\times10^{\prime}$.

\section{Validation and discussion}\label{sec:validation}
In this section we perform some analysis to validate the simulated catalogues obtained using \spr{}. In all the figures we show as reference the simulation with $k_{\Phi}=$-1 (see eq. \ref{eq:zevol}), but we include in the uncertainties the results considering the other high-z extrapolations. \par
It is worth mentioning that the conversion between the L$_{IR}$ and any other quantity is fixed to a specific value that depends on the template associated to each simulated galaxy. As the number of considered templates is limited, i.e. 35 templates from \citet{Polletta2007}, \citet{Rieke2009} and \citet{Bianchi2018}, the ratio between L$_{IR}$ and other quantities, such as stellar mass and luminosities at specific wavelengths, is discrete. To limit such issue, we considered the observational scatter associated to any assumed empirical relation as well as the probability distribution of the different parameters derived by {\sc Magphys} or {\sc Sed3fit}. \par

\begin{figure}[hbt!]
    \centering
    \includegraphics[width=1\linewidth,keepaspectratio]{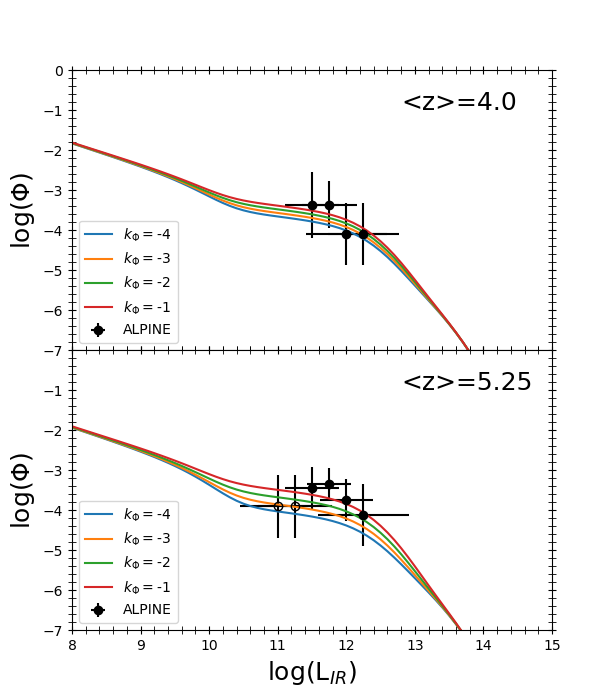}
    \caption{The observed IR LFs derived in the ALPINE survey \citep[\textit{black circles},][]{Gruppioni2020} at z=3.5-4.5 (\textit{top}) and z$=$4.5-6 (\textit{bottom}). \textit{Empty circles} are below the completeness limits. We compare the ALMA IR LFs with the four extrapolations of the total \hers{} LFs used in this work to extract simulated galaxies at z$>$3 (\textit{coloured solid lines}).}
    \label{fig:LF_ALPINE}
\end{figure}

 As investigated by \citet{Gruppioni2019}, the \hers{} IR LF considered in this work is in perfect agreement with results from independent works at various wavelengths, including IR \citep{Rodighiero2010,Lapi2011,Marchetti2016,Rowanrobinson2016}, radio \citep{Novak2017} and CO observations \citep{Vallini2016,Riechers2019}. In addition, the extrapolations done at z$>$3, in particular the extrapolation with $k_{\Phi}$=-1, are consistent with the constraints obtained from the continuum luminosity function up to z$\sim$6 of ALMA serendipitous detections in the ALPINE survey \citep{Gruppioni2020}, as shown in Figure \ref{fig:LF_ALPINE}. \par 
In the next paragraphs, we will compare the derived number counts in different bands, the stellar mass function, the SFR-stellar mass relation, the X-ray, FUV and K-band luminosity functions and an AGN diagnostic diagram \citep[i.e BPT diagram;][]{BPT1981} with various observational results from previous works. 

\subsection{Number counts}\label{sec:Ncounts}
\begin{figure*}
    \centering
    \includegraphics[width=1\linewidth,keepaspectratio]{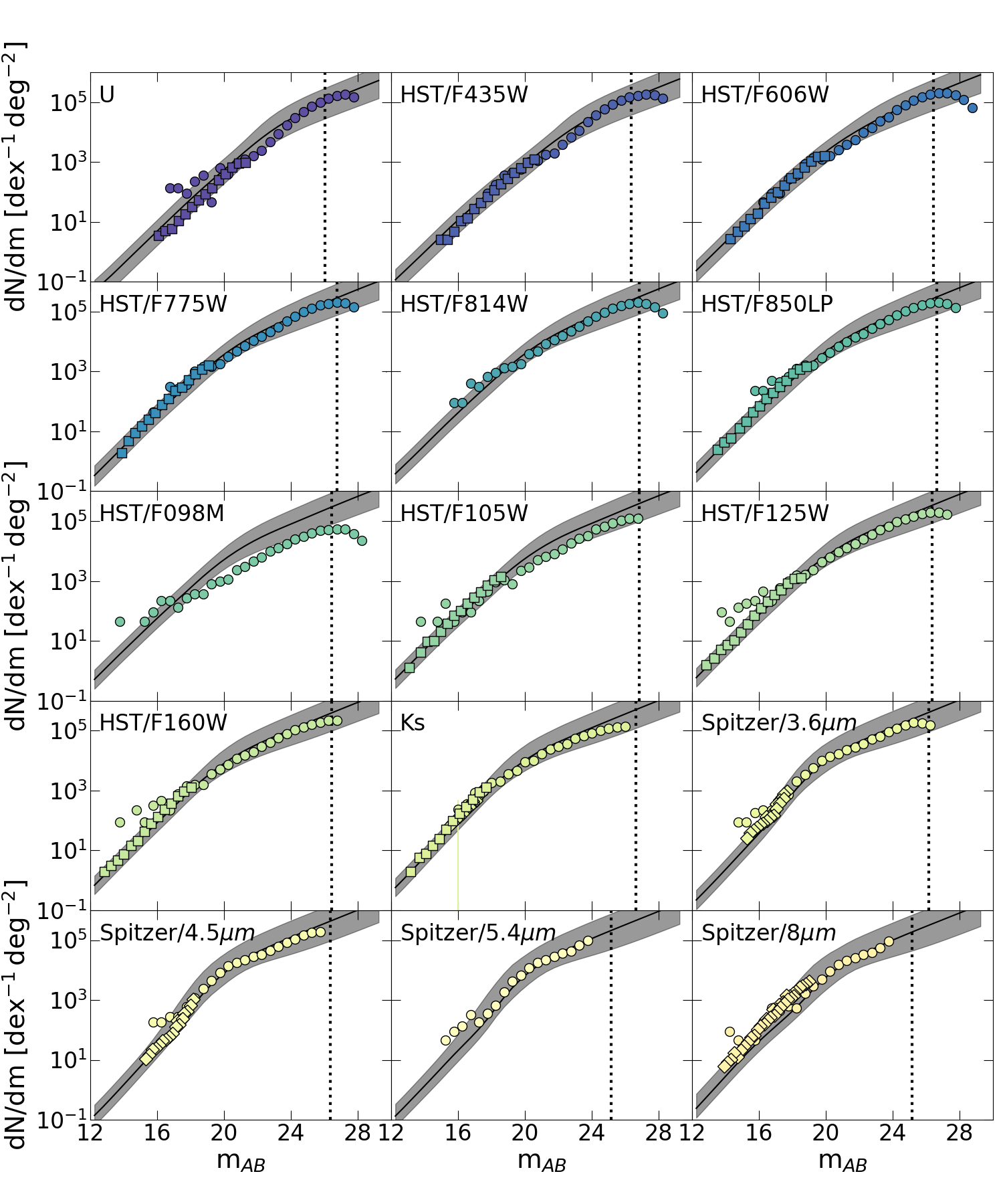}
    \caption{Differential number counts of the \spr{} simulated galaxies from the U-band to the near-IR, with high-redshift extrapolation $k_{\Phi}=$-1 (eq. \ref{eq:zevol}, \textit{solid black line}). The shaded regions show the uncertainties due to the the high-z extrapolation and the 1$\sigma$ errors of the LFs and GSMF used to generate simulated galaxies. We include also observations in the same bands of the CANDELS GOODS-S survey \citep[\textit{filled circles},][]{Guo2013} as well as data from the GAMA survey \citep[\textit{filled squares},][]{Hill2011} and from the AKARI telescope \citep[\textit{filled diamonds},][]{Murata2014}, both corrected for the differences in the broad-band filters. For clarity, only one every four data points of the GAMA survey are shown. Vertical black dotted lines show a rough estimate of the 50$\%$ completeness level for CANDELS data.}
    \label{fig:Ncounts}
\end{figure*}

We start our tests by comparing the differential number counts of the \spr{} simulation with the ones of the Cosmic Assembly Near-infrared Deep Extragalactic Legacy Survey \citep[CANDELS;][]{Grogin2011} in the GOODS-S field \citep{Guo2013}. Broad-band filters cover from the U-band up to the 8$\mu$m \textit{Spitzer}-IRAC band. The comparison is shown in Figure \ref{fig:Ncounts}, where we included also data from the Galaxy And Mass Assembly survey \citep[GAMA;][]{Hill2011} and the AKARI telescope \citep{Murata2014}. We corrected these observations for the difference in the filters by computing the median ratio of the fluxes derived considering the CANDELS filters and the closest GAMA or AKARI ones, both included in the \spr{} simulation. To avoid spurious sources we reported only CANDELS observations above the 5$\sigma$ detection limit reported by \citet{Guo2013}. Neither the GAMA, AKARI or the CANDELS data are corrected for completeness, but we report only AKARI data above 90$\%$ completeness, as derived in their paper. Completeness levels are difficult to estimate for the CANDELS GOODS-S survey, given its non-homogeneous observational depth \citep[see ][]{Guo2013}. We anyway obtained an approximate estimation of the 50$\%$ completeness level by deriving the magnitude at which the number counts deviate by more than a factor of two from a power-law. A similar method was tested by \citet{Guo2013} on the Ks band. We report the estimated 50$\%$ completeness in Figure \ref{fig:Ncounts}, but the estimates at 5.6 and 8 $\mu$m may be underestimated as the number counts at these wavelengths deviate significantly from a power-law.  \par
The \spr{} differential number counts are in agreement with observations over a large range in magnitudes. CANDELS observations show some bias at bright magnitudes due to the small area covered by the survey in GOODS-S, as visible when both GAMA or AKARI data in similar filters are available. Some inconsistencies are present for the \textit{HST}/F098M filter, but they are not present in adjacent bands suggesting that there may be some specific and relative narrow features affecting only data in that band. The considered filter is included only in the CANDELS survey and not in the GAMA or AKARI ones, therefore we can not check the presence of some observational biases. In some filters, number counts are slightly overestimated in the \spr{} simulation at intermediate magnitudes and, given the just mentioned completeness level, this can not be explained fully with observational incompleteness. Except fot the \textit{HST}/F098M filter, all estimates are overall within the model uncertainties.
\par

\begin{figure*}
    \centering
    \includegraphics[width=1\linewidth,keepaspectratio]{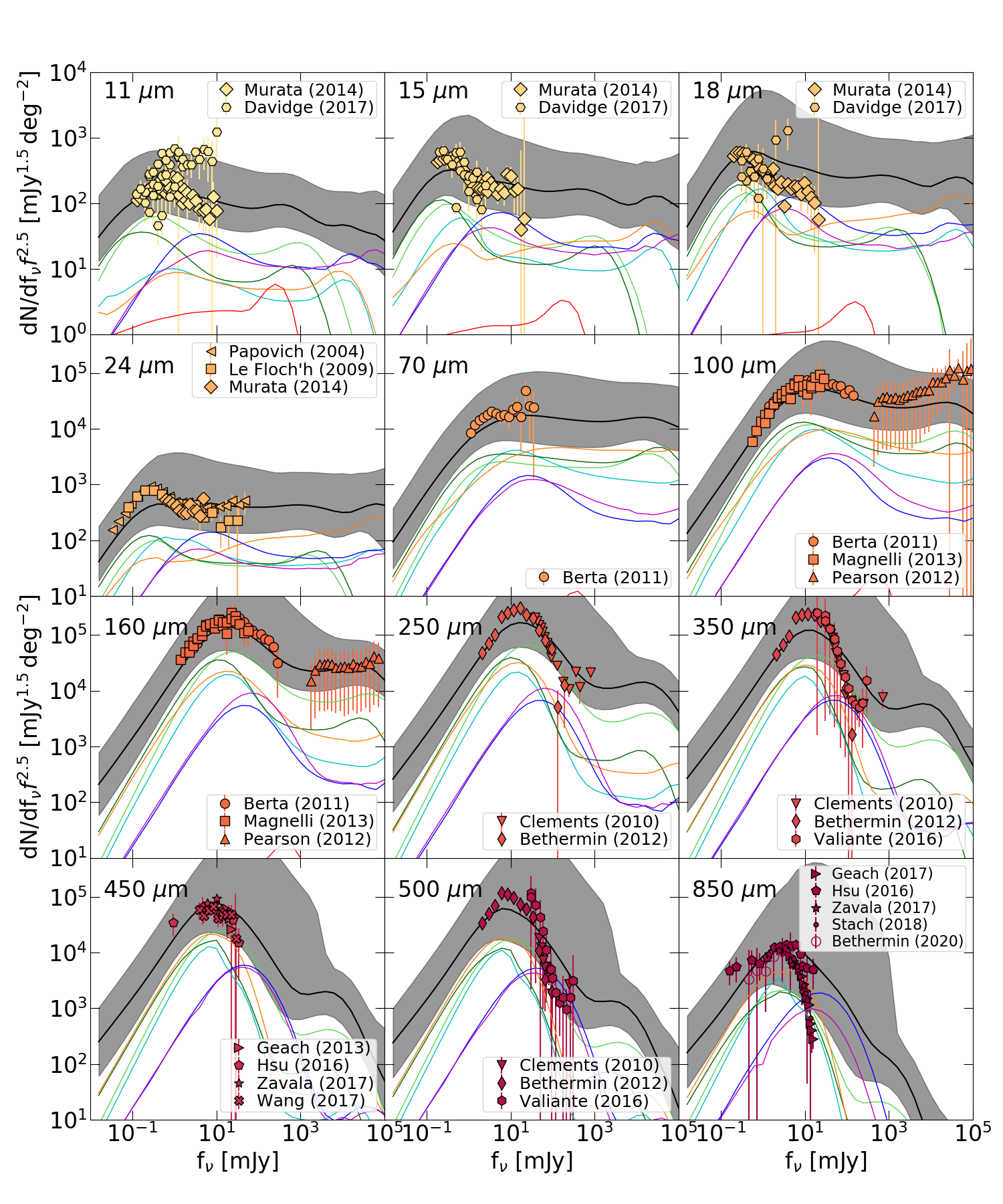}
    \caption{Differential number counts normalised to the Euclidean slope of the \spr{} simulated galaxies in the mid and far-IR with high-redshift extrapolation $k_{\Phi}=$-1 (eq. \ref{eq:zevol}, \textit{solid black line}). The shaded regions show the uncertainties due to the the high-z extrapolation and the 1$\sigma$ errors of the LFs and GSMF used to generate simulated galaxies. We also include several observed values present in the literature \citep{Papovich2004,LeFloch2009,Clements2010,Berta2011,Bethermin2012,Pearson2012,Geach2013,Magnelli2013,Murata2014,Hsu2016,Valiante2016,Davidge2017,Geach2017,Wang2017,Zavala2017,Stach2018,Bethermin2020}. Thin coloured lines are the normalised number counts for the different galaxy populations considered in \spr{}: spirals (\textit{light green line}), starbursts (\textit{cyan line}), SF-AGN (\textit{dark-green line}), SB-AGN (\textit{orange line}), AGN type-1 (\textit{blue line}), AGN type-2 (\textit{magenta line}), elliptical galaxies (\textit{red line}) and irregular galaxies (\textit{brown line}).}
    \label{fig:NcountsIR}
\end{figure*}

We compare the differential number counts, normalised to the Euclidean slope, of \spr{} galaxies in the mid and far-IR together with results from the literature (Fig. \ref{fig:NcountsIR}). In particular, we consider differential normalised number counts derived using \textit{Spitzer} \citep{Papovich2004,LeFloch2009}, \hers{} \citep{Berta2011,Magnelli2013,Clements2010,Bethermin2012,Valiante2016}, AKARI \citep{Pearson2012,Murata2014,Davidge2017}, SCUBA-2 \citep{Geach2013,Geach2017,Hsu2016,Wang2017,Zavala2017} and ALMA observations \citep{Stach2018,Bethermin2020}. The \spr{} differential normalised number counts are in overall agreement with observations over a large range of wavelengths and fluxes. There are however some discrepancies, in particular at 18$\mu$m, where number counts are overestimated, and at 24$\mu$m where the peak in the normalised number counts is slightly underestimated. At 100 $\mu$m some discrepancies are present with observations from the AKARI All-Sky Survey \citep{Pearson2012} for sources brighter than 10$^{4}$ mJy, which mainly correspond to low-z spirals. However, these are also galaxies from which beam corrections are expected to be large in the AKARI observations, as they are generally extended \citet{Clements2019}. This effect would make some AKARI fluxes underestimated and would move the tension at larger fluxes or, in case only some galaxies are affected by this issue, would flatten the observed normalised number counts, reducing the tension with the simulation. At longer wavelengths, $\lambda>500\mu$m, some discrepancies are present in the bright-end regime, i.e. f$_{\nu}>10^{2}$ mJy, mainly due to the AGN1 and AGN2 populations at z$=$1 to 4. This may indicate the necessity to include an AGN template with a less prominent cold-dust component. 

\subsection{Stellar mass function}\label{sec:MF}
\begin{figure*}[hbt!]
    \centering
    \includegraphics[width=1\linewidth,keepaspectratio]{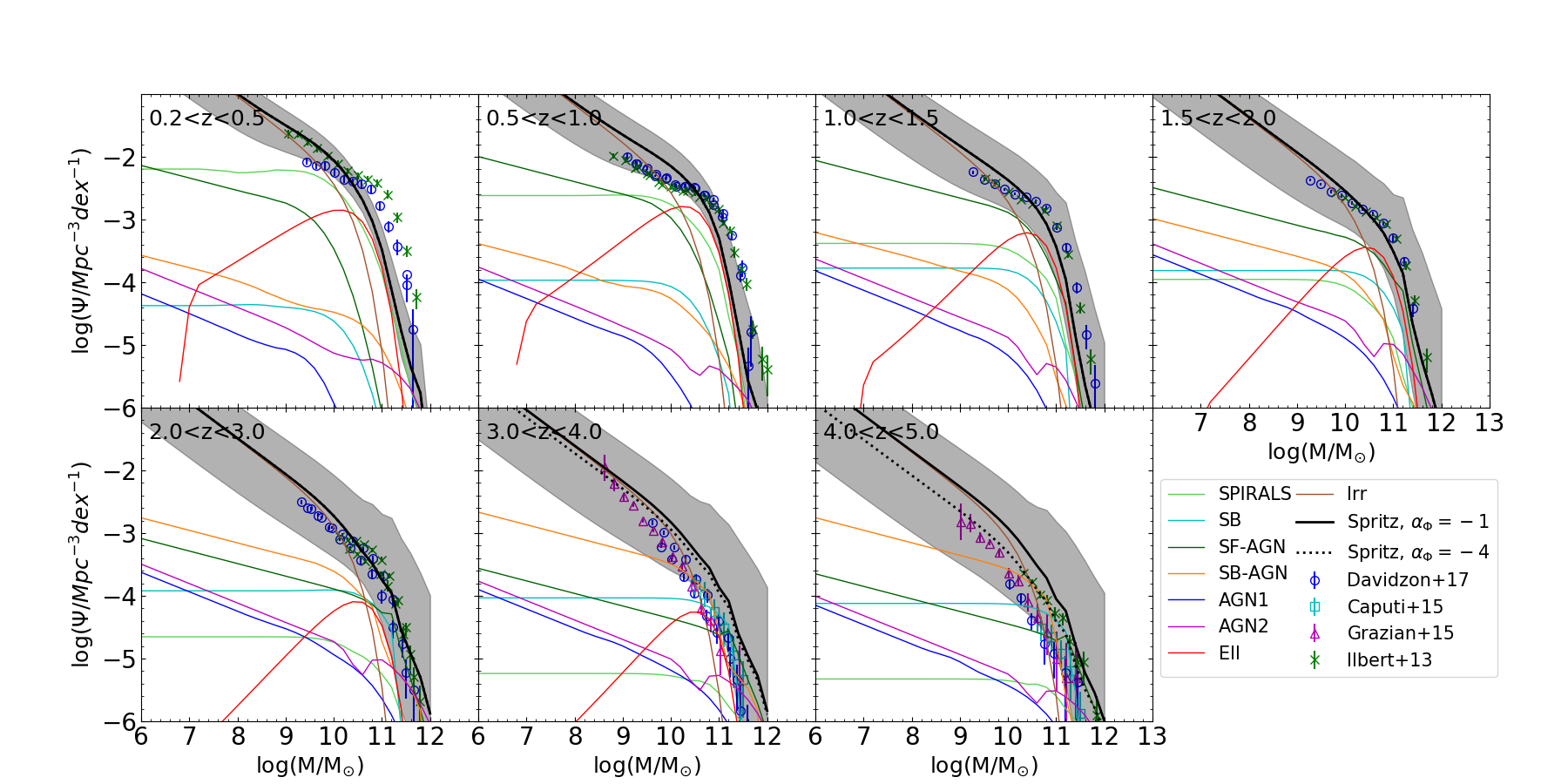}
    \caption{Stellar mass function of the \spr{} simulation at different redshifts (\textit{black thick solid line}), with high-redshift extrapolation $k_{\Phi}=$-1 (eq. \ref{eq:zevol}) and -4 (\textit{black thick dotted line}). The grey shaded area shows the uncertainties arising from the high-z extrapolation, the different torus library considered, the 1$\sigma$ value derived from the probability distribution of the stellar mass of each SED template and the 1$\sigma$ errors of the LFs and GSMF used to derive simulated galaxies. We report some of the values present in the literature: \citet[][\textit{green crosses}]{Ilbert2013}, \citet[][\textit{blue circles}]{Davidzon2017}, \citet[][\textit{cyan squares}]{Caputi2015} and \citet[][\textit{magenta triangles}]{Grazian2015}. Thin coloured lines are the stellar mass functions for the different galaxy populations considered in \spr{}: spirals (\textit{light green line}), starbursts (\textit{cyan line}), SF-AGN (\textit{dark-green line}), SB-AGN (\textit{orange line}), AGN type-1 (\textit{blue line}), AGN type-2 (\textit{magenta line}), elliptical galaxies (\textit{red line}) and irregular galaxies (\textit{brown line}).}
    \label{fig:MF}
\end{figure*}

In Figure \ref{fig:MF} we report the comparison between the stellar mass-function of \spr{} galaxies and the data available in the literature \citep{Ilbert2013,Caputi2015,Grazian2015,Davidzon2017}. Galaxies are selected in the Ks band in \citet{Ilbert2013}, at 4.5$\mu$m in \citet{Caputi2015}, in the HST/F160W at 1.6$\mu$m in \citet{Grazian2015} and in the stacked zYJHK image in \citet{Davidzon2017}. These results are corrected for completeness, as mentioned in each corresponding paper. The results from the \spr{} simulations are shown for the master catalogue, before applying any observational selection to mimic any spectro-photometric survey. We converted the observed stellar mass function to the same IMF \citep{Chabrier2003}, and cosmology used in this work, when necessary. \par
The \spr{} stellar-mass function is overall consistent with the observed values, both in the high-mass regime and in the low-mass end. The high-mass end is dominated by elliptical galaxies up to z$=$2 and by SB, SB-AGN and SF-AGN at higher redshifts. This change in the bright-end, from dominated by elliptical to dominated by SB-AGN, suggests that SB-AGN may be indicated as the progenitors of elliptical galaxies. The low-mass end is instead, not surprisingly, dominated by dwarf irregular galaxies. At z$>$3 the simulation with $k_{\Phi}=$-1 (eq. \ref{eq:zevol}) brings to an overestimation of the faint-end of the stellar mass function, while the simulation with $k_{\Phi}=$-4 is more in agreement with observations. Therefore, while the IR LF favours the $k_{\Phi}=$-1 high-z extrapolation for \hers{} galaxies (see \ref{fig:LF_ALPINE}), the stellar mass function favours a steeper evolution, i.e. $k_{\Phi}=$-4 (eq. \ref{eq:zevol}), for Irregular galaxies. It is however necessary to take into account that large uncertainties on the predictions are present at low-masses at z$<$3 end at all masses at higher redshifts. These are mainly resulting from the extrapolations performed in the simulation, as \hers{} data probed z$<$3 and relatively bright galaxies.  Future IR observations reaching much fainter fluxes may help on anchoring both the high-z and low-mass extrapolations and reducing the scatter on the predictions.  

\subsection{SFR and stellar-mass plane}\label{sec:M_SFR}
The majority of star-forming galaxies show a tight relation between their current SFR and their stellar mass, historically called the main-sequence of star-forming galaxies\citep{Brinchmann2004,Noeske2007}. This relation has been interpreted as the result of the interplay of secular and stochastic effects which boost or suppress the SFR \citep{Peng2014,Tacchella2016}. Galaxies with suppressed star-formation are below this almost-linear relation while starburst galaxies, featuring a temporary boost on their SFR, lay above the main-sequence. The number of starburst galaxies is quite low in the local Universe \citep{Rodighiero2011,Sargent2012} with a possible strong increase of their fraction with redshift \citep{Caputi2017,Bisigello2018}\par
\begin{figure}
    \centering
    \includegraphics[width=1\linewidth,keepaspectratio]{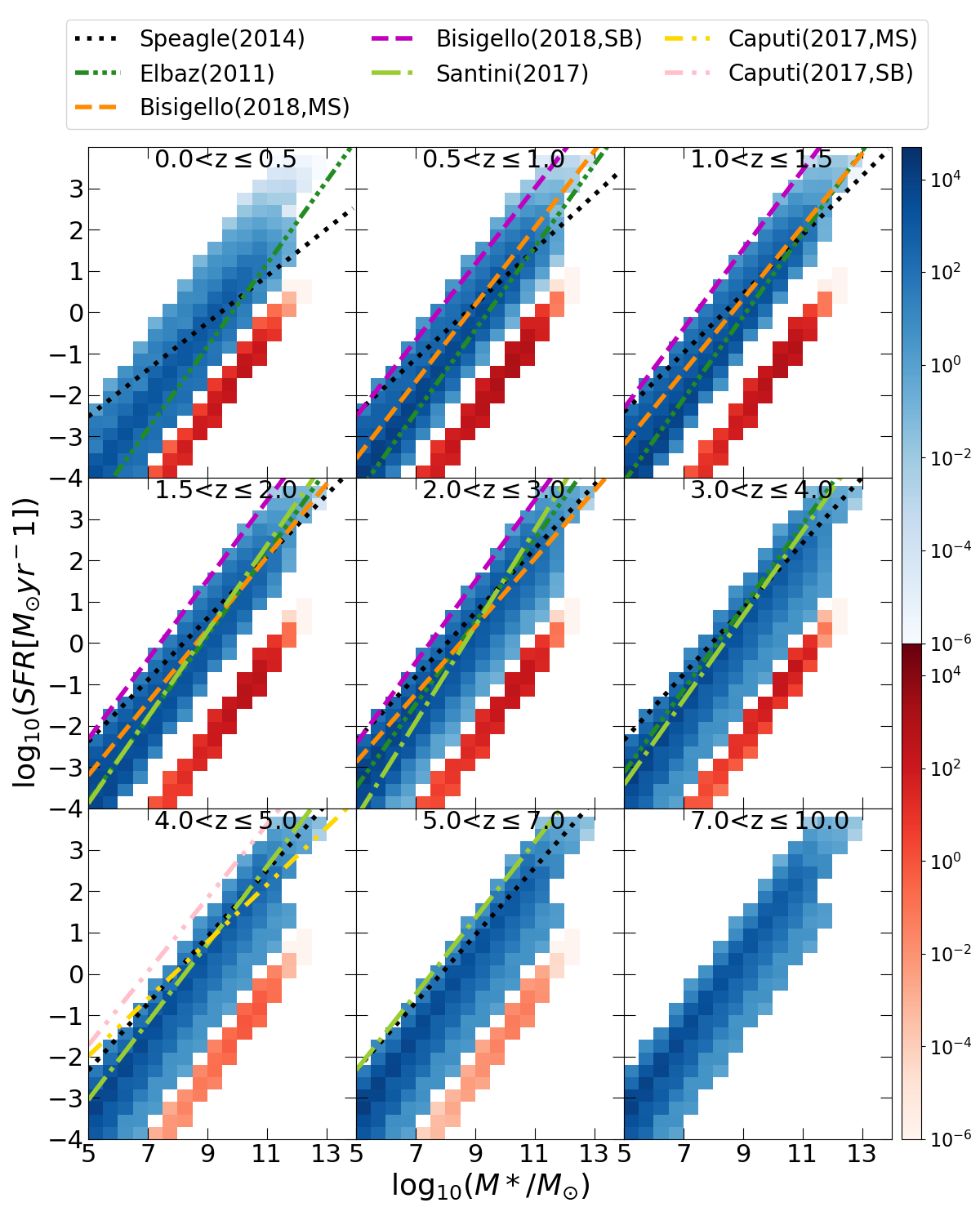}
    \caption{Stellar mass vs. SFR at different redshifts for simulated \spr{} galaxies, with high-redshift extrapolation $k_{\Phi}=$-1 (eq. \ref{eq:zevol}). The blue colour shows the distribution of galaxies derived from the IR LF or the irregular GSFM, while the red colour is the distribution of elliptical galaxies derived from the K-band LF. The colour gradients show the number of galaxy in 1deg$^{2}$ and they are in logarithmic scale, as shown on the right for star-forming galaxies (\textit{top color bar}) and ellipticals (\textit{bottom color bar}). Different lines show the observed position of normal star-forming galaxies as derived by \citet[][\textit{dash-dot-dot-dotted green line}]{Elbaz2011}, \citet[][\textit{black dotted line}]{Speagle2014}, \citet[][\textit{dash-dot-dotted yellow line}]{Caputi2017}, \citet[][\textit{dash-dotted light green line}]{Santini2017} and \citet[][\textit{dashed orange line}]{Bisigello2018}, as well as the relation derived for starbursts by \citet[][\textit{pink dash-dot-dotted lines}]{Caputi2017} and  \citet[][\textit{magenta dashed lines}]{Bisigello2018}.}
    \label{fig:MSFR}
\end{figure}

In Figure \ref{fig:MSFR} we show the distribution of galaxies simulated with \spr{} in the SFR-stellar mass plane. No scatter has been considered for the  SFR and stellar mass associated to each simulated galaxy. The results are shown for the master catalogue, before any flux selection. For comparison, we report the sequence of normal
star-forming galaxies derived by \citet{Elbaz2011}, \citet{Speagle2014}, \citet{Caputi2017}, \citet{Santini2017} and \citet{Bisigello2018} and the one occupied by starburst derived by \citet{Caputi2017} and \citet{Bisigello2018}. 
The sample of \citet{Elbaz2011} comprises galaxies detected in the far-IR and their SFR is derived directly from the total IR luminosity. \citet{Speagle2014} derived the sequence using a large set of results available in the literature, from both UV or IR data, carefully taking into account possible biases. In \citet{Bisigello2018}, the sequence is derived considering a sample of optically selected galaxies with multi-wavelength observations and the SFRs were derived from the dust-corrected UV fluxes complemented, when possible, by IR observations. The work by \citet{Caputi2017} considers instead galaxies detected at rest-frame optical wavelengths with multi-wavelength observations and SFRs are derived from the dust-corrected H$_{\alpha}$ fluxes, as obtained from the observed flux excess in the available broad-band filters. Finally, \citet{Santini2017} considered a sample of optically selected galaxies, including also gravitational lenses to push the analysis to low-mass and high-z systems, and SFRs are derived from the dust-corrected UV fluxes. We chose to compare with works which apply different techniques and span different redshifts and stellar masses to avoid large biases in the comparison with our simulation.
We remind the reader that the SFR in \spr{} includes both the 'unobscured' component, as derived from the IR, and the obscured one from the UV, both computed after removing the AGN contribution. \par
Elliptical galaxies have stellar masses and SFR consistent with the quiescent population and they occupy the area $\sim$2 dex below the sequence of normal star-forming galaxies. All the other simulated galaxies occupy the area expected for star-forming and starburst galaxies at the different redshifts. Simulated galaxies form a sequence that at low-z is steeper than the one observed by \citet{Speagle2014}. However, \citet{Bisigello2018} has shown that the slope estimated considering only relative massive galaxies, i.e. log$_{10}(M^{*}/{\rm M}_{\odot})>$9.5, and a classical $\chi^{2}$ fitting method, as done by \citet{Speagle2014}, can produce a flattening of the derived main-sequence. For comparison, at the same redshift, \citet{Elbaz2011} derived a steeper main-sequence and a smaller normalisation than \citet{Speagle2014}. This relation shows a SFR $\sim$0.05 (0.5) dex below the average SFR of the star-forming galaxies in our sample at log(M/M$_{\odot})\sim$11 (8). At z$\sim$4.5, there are few galaxies with SFR as high as observed by \citet{Caputi2017}, showing that more extreme starburst templates may be necessary to reproduce these observations. Indeed, SB galaxies in the \spr{} simulation are above the MS at z$<$2, but they have sSFR consistent with the MS at z$>$2. At even higher redshifts, i.e. z$=$5-7, simulated star-forming galaxies are in agreement with predictions from \citet{Speagle2014} at stellar masses above 10$^{10}{\rm M}_{\odot}$, while they are generally below the expected main-sequence at lower stellar masses. Discrepancies are larger when considering observations by \citet{Santini2017}, which data pointed out a larger normalisation for the main-sequence at such high-z. It is however necessary to take into account that the work by \citet{Speagle2014} is limited to the brightest galaxies, while the work by \citet{Santini2017} is limited in statistics at this very high-z. Future observations over large areas of galaxies at z$>$5 will help to anchor the main-sequence and the starburst locus.\par 
The linear relation among SFR and stellar mass observed in our simulation is not surprising, as the ratio between stellar mass and IR/UV luminosity is constant for each template and all considered templates, excluding the elliptical ones, have SFR and stellar masses consistent with normal star-forming systems or starbursts. However, the evolution with redshift and the exact slope of the relation is dictated by the evolution of the considered IR LF and the fraction of the templates used for describing each galaxy population at different redshift and L$_{IR}$. Therefore, as the evolution of the LF is consistent with the observed evolution of the SFR-stellar mass relation, the evolution of the main-sequence of star-formation is simply dictated by the luminosity ($\sim$SFR) evolution.  

\subsection{Luminosity functions}
In the next sections we compare the \spr{} LF with observations available in the literature, focusing on some representative wavelengths spanning from the X-ray to the radio.

\subsubsection{8$\mu$m}
We start the comparison from the 8$\mu$m luminosity function. In particular, we compare the \spr{} simulation with results from \citet{Caputi2007}, \citet{Rodighiero2010} and \citet{Magnelli2011}, all obtained from a sample of galaxies observed with Spitzer at 24$\mu$m. We complement this comparison with the results by \citet{Goto2015,Goto2019} using AKARI observations.  We do not show higher redshifts as there are no available observations. Both the observations and the \spr{} simulation show a large scatter, but they generally agree with each other at any redshifts and luminosities. A light overestimation may be present in the \spr{} simulation in the faint-end at z$\sim$0.5, but few observations are available in this luminosity range, i.e. L$_{8\mu m}<10^{10}\,L_{\odot}$.

\begin{figure}[hbt!]
    \centering
    \includegraphics[width=1\linewidth,keepaspectratio]{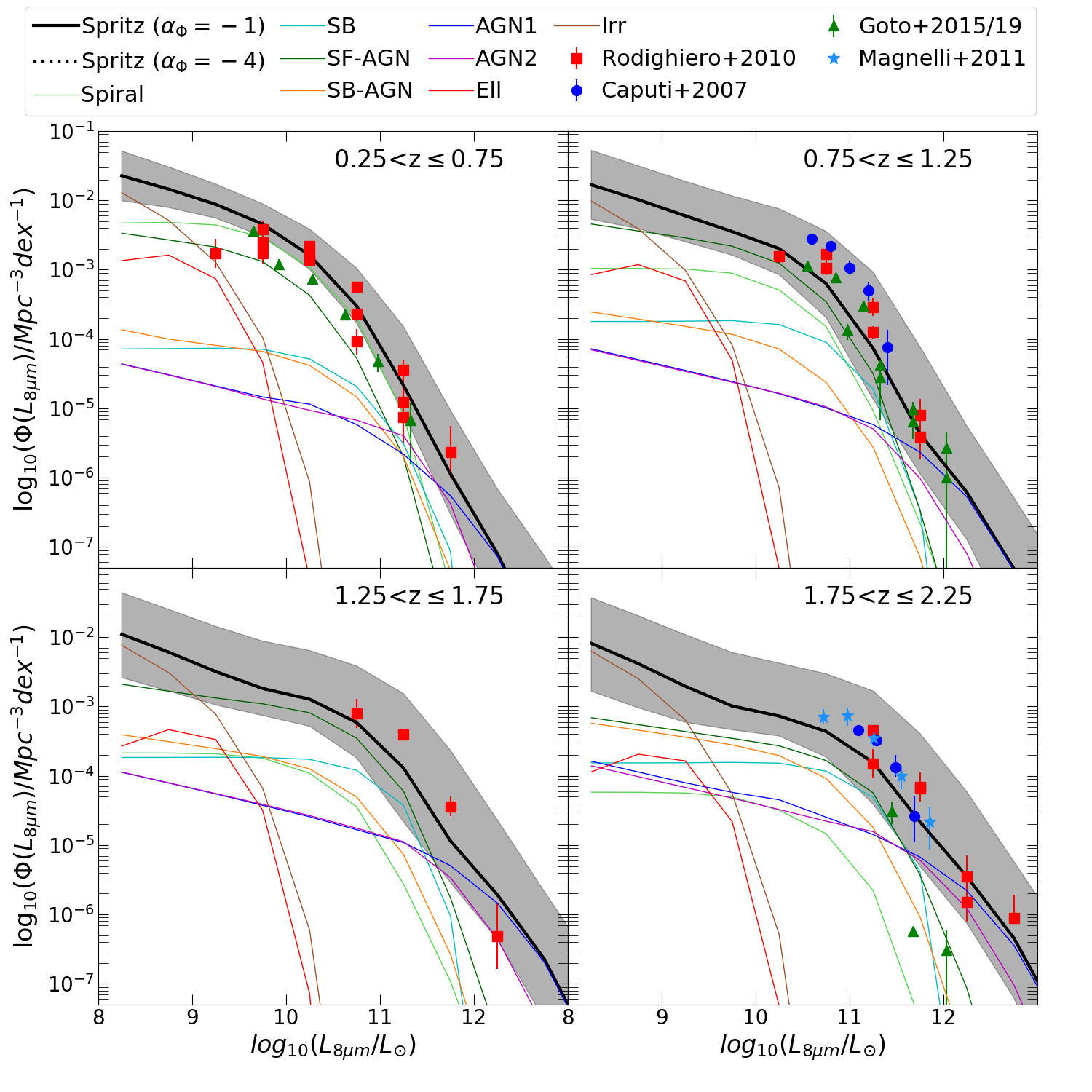}
    \caption{Luminosity function at 8 $\mu$m for all galaxies simulated with \spr{} (\textit{black tick solid line}), with high-redshift extrapolation $k_{\Phi}=$-1 (eq. \ref{eq:zevol}). The grey area shows the uncertainties due to the high-z extrapolation and to the 1$\sigma$ errors of the GSFM, IR and K-band LFs used to generate simulated galaxies. The luminosity function for the different sub-populations is also shown: spirals (\textit{green line}), starbursts (\textit{cyan line}), the two composite populations SF-AGN (\textit{dark-green line}) and SB-AGN (\textit{orange line}), AGN type-1 (\textit{blue line}), AGN type-2 (\textit{magenta line}), elliptical galaxies (\textit{red line})  and irregular galaxies (\textit{brown line}). We also report the observed 8$\mu$m luminosity functions by \citet[][\textit{red squares}]{Rodighiero2010}, \citet[][\textit{blue circles}]{Caputi2007}, \citet[][\textit{green triangles}]{Goto2015,Goto2019} and \citet[][\textit{light blue stars}]{Magnelli2011}.}
    \label{fig:LF8}
\end{figure}

\subsubsection{K-band}
\begin{figure}[hbt!]
    \centering
    \includegraphics[width=1\linewidth,keepaspectratio]{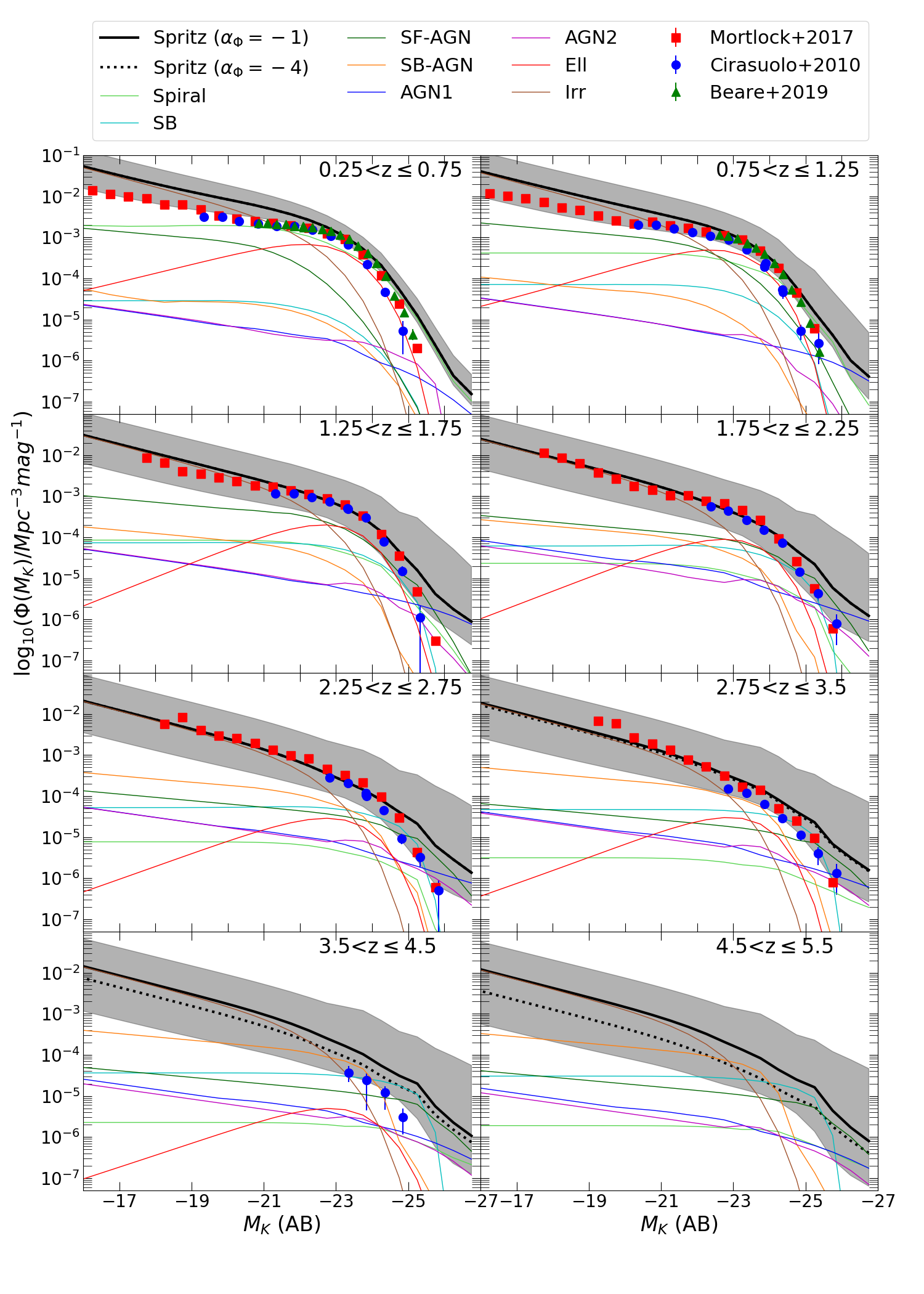}
    \caption{The same as Fig. \ref{fig:LF8}, but for the K-band LF. We also report the observed K-band LFs by \citet[][\textit{red squares}]{Mortlock2017}, \citet[][\textit{blue circles}]{Cirasuolo2007} and \citet[][\textit{green triangles}]{Beare2019} in the overlapping redshift bins.}
    \label{fig:LFK}
\end{figure}

Another test we perform to validate our results consists in reproducing the observed K-band LF with our simulated catalogue. We show the K-band LF of \spr{} galaxies in Figure \ref{fig:LFK} together with the observed LF by \citet{Mortlock2017}, \citet{Cirasuolo2007} and \citet{Beare2019}. The LF is generally well reproduced, the knee of the K-band LF of \spr{} galaxies is consistent with observations up to z$\sim$3.5 while small deviations from observations are present in the bright and faint-end regime and at z>3.5. Below z<1.25 the bright-end regime is dominated by elliptical galaxies with spirals and irregulars dominating at faint magnitudes (M$_{Ks}>$-20). Spiral galaxies and AGN-1 are also present at extremely bright magnitudes, i.e. M$_{Ks}<$-25, where the number densities are generally below the ones probed by observations. At higher redshift, composite systems and AGN dominate the bright-end, with a decreasing contribution of elliptical galaxies. Some differences in the bright-end at z>1.25 may arise from the selection done in the literature, as bright X-ray sources have been removed before deriving the observed K-band LF.  \par
The agreement between the simulated and the observed K-band LF is in line with the consistency between the \spr{} and the observed stellar mass function analysed in Section \ref{sec:MF}, considering that the K-band is dominated by the light of the old stellar populations and it is therefore a relative good tracer of the stellar mass. We can conclude that the K-band luminosity function of \spr{} simulated galaxies slightly overestimates the observed K-band LF at z$>$3.5, where however few observations are available, while it is generally in agreement with observations at other redshifts. The Spritz LF is dominated by AGN in the bright-end regime at high-z and the absence of AGN in the literature derivations could be responsible for this inconsistency.
\par

\subsubsection{B-band}
\begin{figure}
    \centering
    \includegraphics[width=1\linewidth,keepaspectratio]{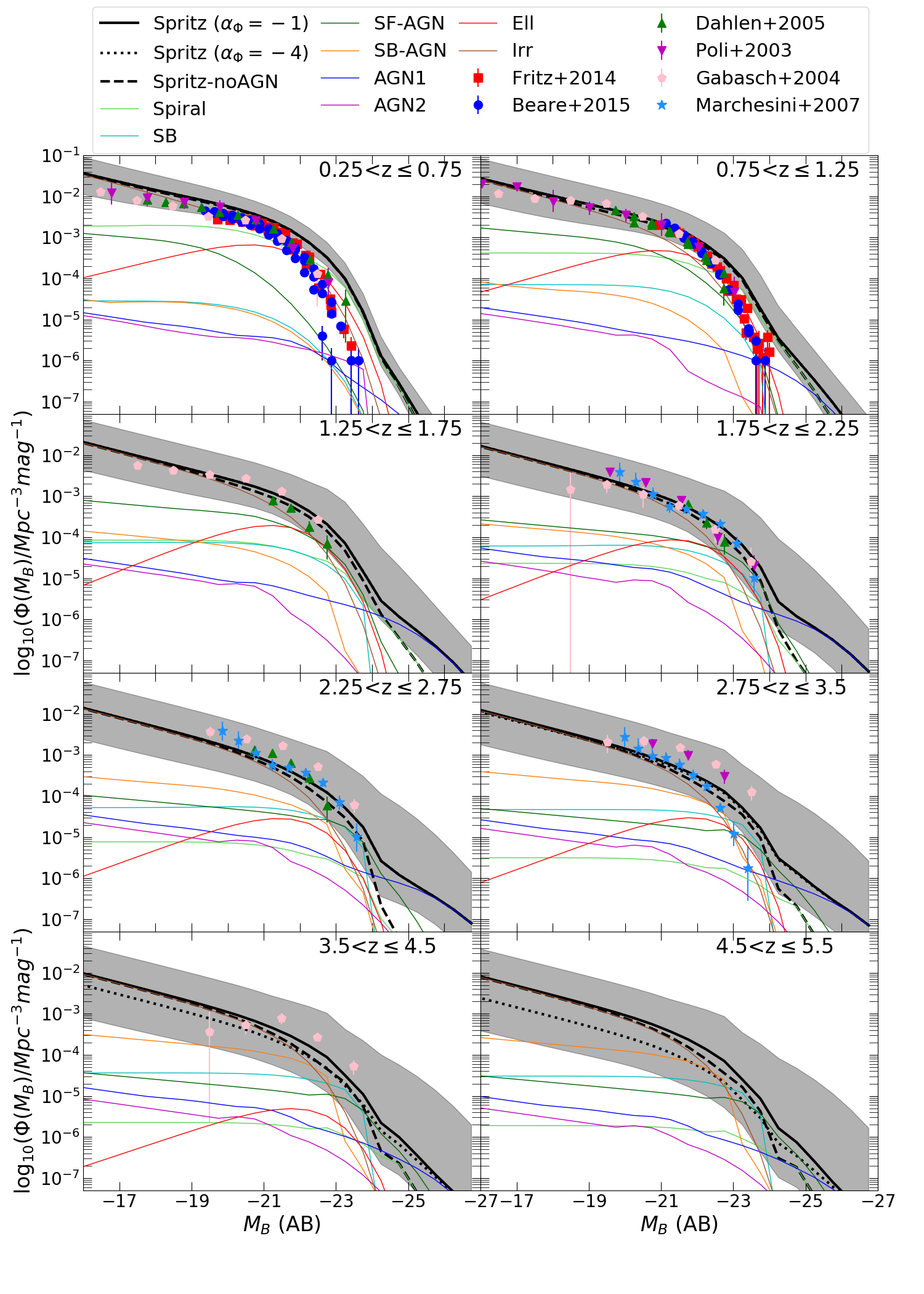}
    \caption{The same as Fig. \ref{fig:LF8}, but for the B-band LF. We also report the observed B-band LFs by \citet[][\textit{red squares}]{Fritz2014}, \citet[][\textit{blue circles}]{Beare2015}, \citet[][\textit{green triangles}]{Dahlen2005}, \citet[][\textit{magenta upside-down triangles}]{Poli2003}, \citet[][\textit{pink pentagon}]{Gabasch2004} and \citet[][\textit{light blue stars}]{Marchesini2007ApJ}  in the overlapping redshift bins.}
    \label{fig:LFB}
\end{figure}

In this section we compare the rest-frame B-band LF of the \spr{} simulation with observations available in the literature. In particular, we considered the B-band LF by \citet{Beare2015}, \citet{Fritz2014} and \citet{Gabasch2004}, whose samples are selected in the observed i-band, the work by \citet{Dahlen2005} using R- and K-band selected galaxies, results by \citet{Poli2003} using i- and K-band selected galaxies and the LF by \citet{Marchesini2007ApJ} derived from a sample of K-selected galaxies. The comparison is shown in Figure \ref{fig:LFB}. \par
At z$<$1.25, the bright-end of the B-band LF is generally overestimated in the \spr{} simulation similarly to the K-band LF. On the other hand, the faint-end of the B-band LF is remarkably similar to observations, at least at z$>$0.75. This reassures on the validity of the approach considered to introduce irregular galaxies, which dominate the faint-end. At z$>$1.25 the \spr{} B-band LF is generally in agreement with observations, at least in the magnitude range covered by observations. However, few studies are present in the literature showing a large scatter (see z$\sim$3). In the plot we also report the \spr{} B-band LF without considering AGNs, as their are at least partially removed in some of the considered studies, i.e. in \citet{Fritz2014} and \citet{Beare2015}. However, the two LF do not differ significantly, as AGNs never dominate the B-band LF except in the very bright-end, which has not been covered by the considered observations. 

\subsubsection{FUV}
\begin{figure}[hbt!]
    \centering
    \includegraphics[width=1\linewidth,keepaspectratio]{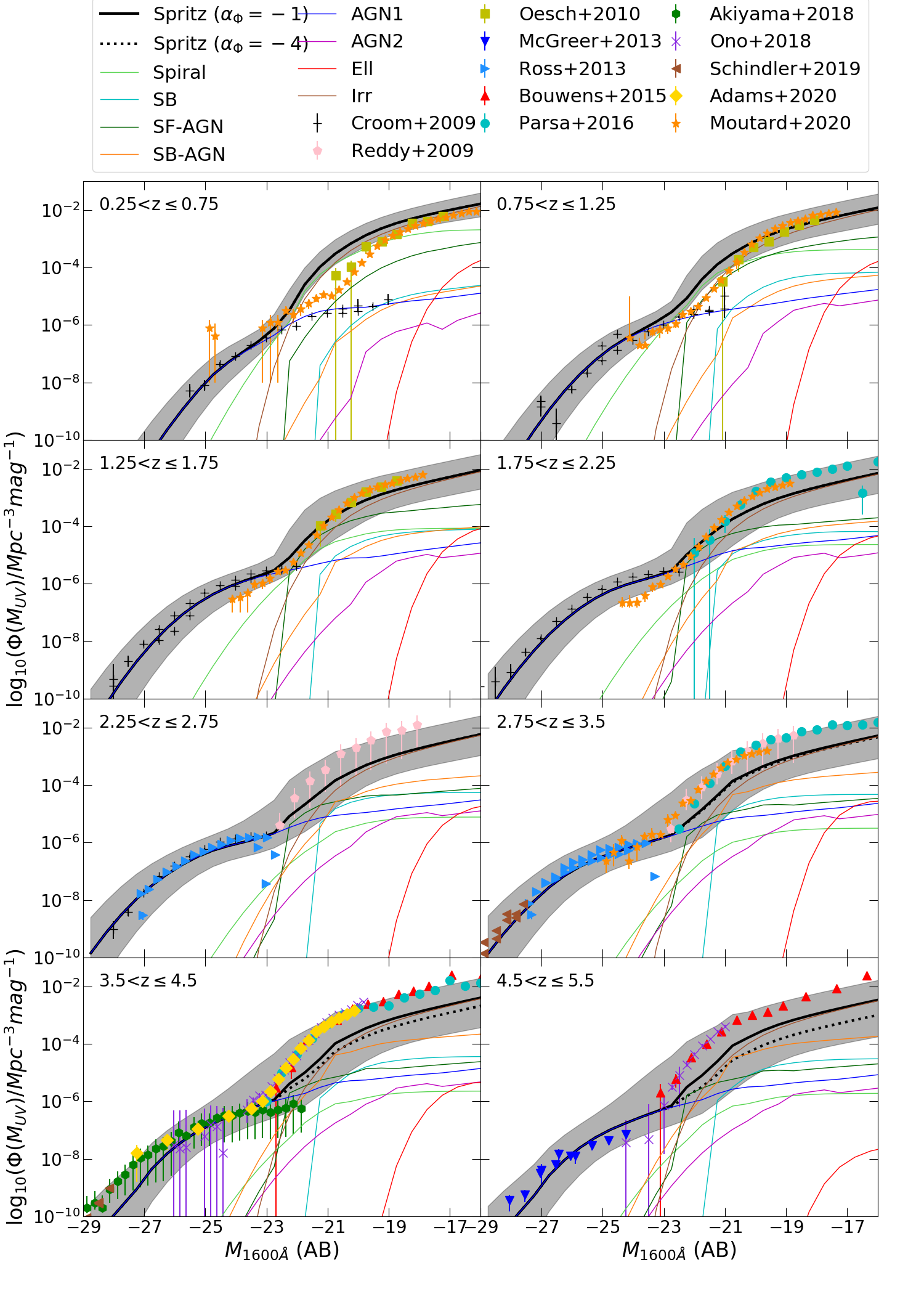}
    \caption{The same as Fig. \ref{fig:LF8}, but for the LF at 1600\AA{}. Our simulation is compared with other data from the literature (see legend).}
    \label{fig:LFUV}
\end{figure}

 We perform the next comparison of \spr{} simulation with the observed luminosity functions by analysing the far-ultraviolet regime. Figure \ref{fig:LFUV} shows the LF derived at 1600\AA{} for \spr{} simulated galaxies and the observed LF for galaxies and QSO by \citet{McGreer2013}, \citet{Akiyama2018} and \citet{Schindler2019} at 1450\AA{}, by \citet{Moutard2020} at 1546\AA{}, by \citet{Oesch2010}, \citet{Parsa2016} and \citet{Adams2020} at 1500\AA{}, by \citet{Bouwens2015} at 1600\AA{} and by \citet{Reddy2009} at 1700\AA{}. Observations by \citet{Ono2018} correspond to rest-frame between 1230 to 1600\AA{}, depending on redshift. Data by \citet{Croom2009} and \citet{Ross2013} have been converted to 1450\AA{}. \par
The bright-end of the UV LF is dominated by unobscured QSO, which are equivalent to the AGN1 population considered in the simulation. We took advantage of the separation of the LF of this galaxy population to improve the IR LF of AGN1 and AGN2 in the \spr{} simulation (see Sec. \ref{sec:AGN1} and Appendix \ref{sec:MCMC}), therefore by construction the bright-end UV LF of the \spr{} simulation agrees with observations up to high-z. The part dominated by galaxies agrees with observations at all redshifts, except for galaxies at z<0.75 in the bright-end regime, i.e. M$_{UV}<$-20. At z$>$2, the galaxy LF is consistent with observations, but tends to slightly underestimate the faint-end. This light discrepancy in the faint-end slope may be explained with a dust-poor population of galaxies missed by the \hers{} observations and not completely described by the irregular population. Indeed, it has been shown by observations that both the fraction of dusty objects and the amount of dust extinction decreases towards faint FUV magnitudes \citep{Bouwens2016}. However these two quantities seem to have little or no evolution at high-z, which is in line with our underestimation of the faint-end slope that remains similar between z$=$2 and 5. On the other hand, it is also possible that the SED templates considered for irregular galaxies, which dominate the galaxy population with M$_{1600\AA}>$-21, are not representative for all redshifts. Indeed, for all the galaxy population derived from the \hers{} LF we considered different probabilities for each template based on observations and depending on redshift and L$_{IR}$. On the other hand, we considered a constant probability for the three SED templates of irregular galaxies. More observations are necessary to verify the validity of this assumption. However, despite the large distance in wavelength between the basis of the SPRITZ simulations (e.g., mid-/far-IR) and the UV regime, the overall agreement between data and model is very good.\par

\subsubsection{X-ray}
\begin{figure}
    \centering
    \includegraphics[width=1\linewidth,keepaspectratio]{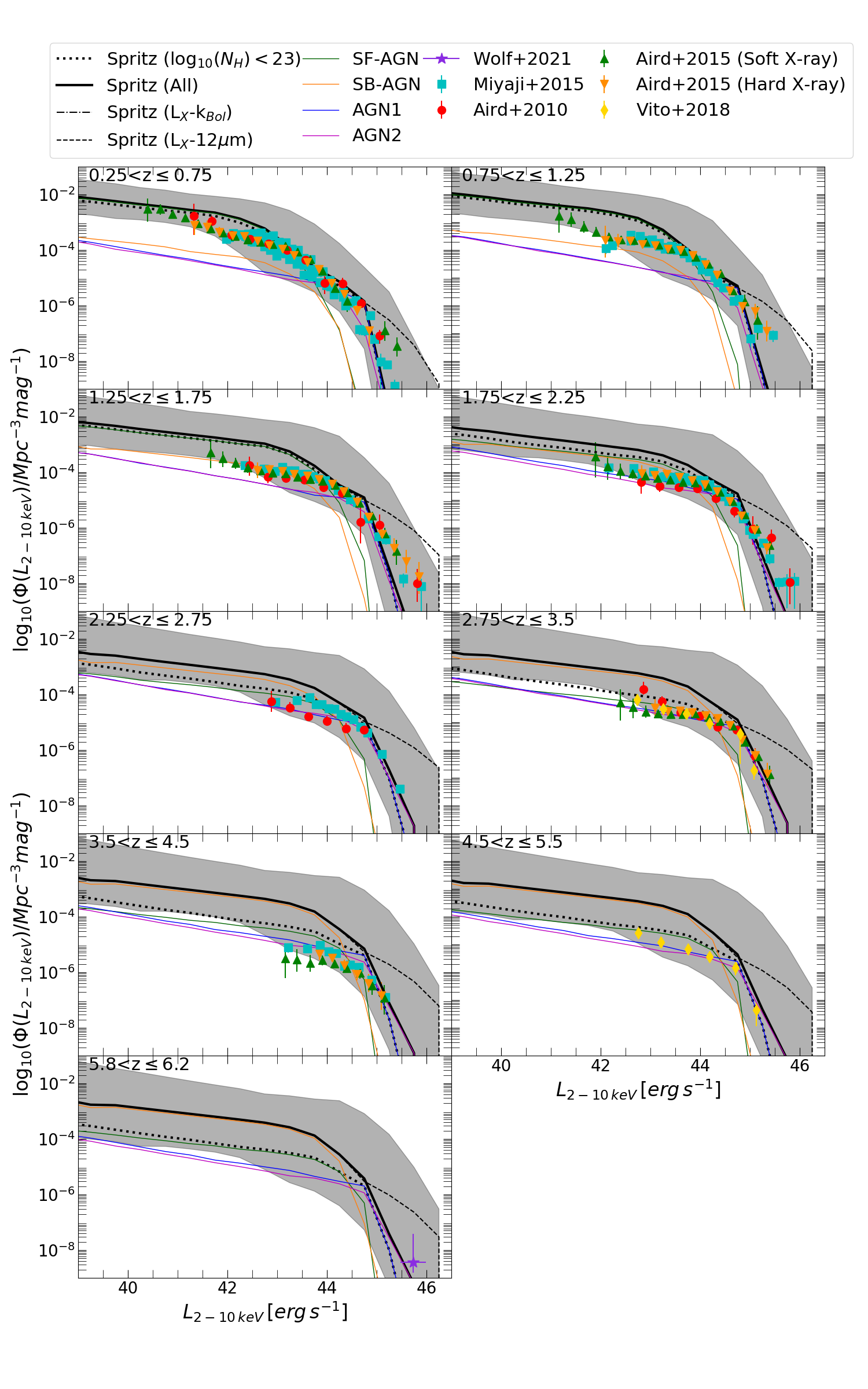}
    \caption{Luminosity function at 2-10 keV for all AGN simulated with \spr{}  (\textit{black solid line}) with high-redshift extrapolation $k_{\Phi}=$-1 (eq. \ref{eq:zevol}) and for the sub-sample of AGN with log$_{10}(N_{H}/cm^{-2})<23$ (\textit{black dotted line}). The grey area shows the uncertainties due to the conversion from the L$_{12\mu m}$ to L$_{2-10keV}$, to the use of different torus models and the 1$\sigma$ errors of the IR LFs used to derive the simulated galaxies hosting an AGN. The LF for each different sub-population is also shown (see Fig. \ref{fig:LF8}). We reported the observed X-ray luminosity functions by \citet[][\textit{purple star}]{Wolf2021}, \citet[][\textit{cyan squares}]{Miyaji2015}, \citet[][\textit{red circles}]{Aird2010}, \citet[][\textit{green and orange triangles}]{Aird2015} and \citet[][\textit{yellow diamonds}]{Vito2018}.}
    \label{fig:LFX}
\end{figure}

In Figure \ref{fig:LFX} we compare the hard X-ray LF obtained with the \spr{} simulation with various X-ray observations \citep{Aird2010,Aird2015,Miyaji2015,Vito2018,Wolf2021}. At low-z the \spr{} LF is consistent with observations, but at increasing redshift the faint-end starts to be overestimated. This discrepancy is reduced if we exclude from the comparison heavily obscured AGN, i.e. $log_{10}(N_{H}/cm^{-2})>$23, that can be missed by X-ray observations or whose intrinsic X-ray luminosity may be underestimated. The bright-end of the LF is consistent with observations within one $\sigma$ even if it is generally overestimated. This happens even at z$\sim$6, taking into account that the X-ray LF by \citet{Wolf2021} is a conservative estimation, but it is anyway derived from a single object. \par
It is necessary to take into account that the conversion to X-ray luminosity has more uncertainties that the UV or K-band luminosity analysed before. In the \spr{} simulation the latter two mainly depend on the shape and evolution of the IR LF and on the templates associated to galaxies in different luminosity regimes. The X-ray luminosity in addition depends on the AGN-host galaxy decomposition and, finally, on the conversion from the luminosity at 12$\mu$m to the X-ray luminosity. Moreover, we considered a single X-ray spectrum for all simulated AGN, as described in Section \ref{sec:Xray}. All these effects contribute to generate the large uncertainties on the predictions as well as the discrepancy between the simulated and the observed X-ray LF. From the point of view of observations, heavily obscured AGN can explain a possible underestimation of the observed faint-end slope. Indeed, if we remove these sources from the \spr{} simulation, the faint-end of the new X-ray LF is in agreement with observations at redshift between 2 and 3.5, while at other redshifts the tension with observations is anyway reduced. An additional possible source of error can be the mis-classification of low-luminosity AGN as galaxies in X-ray surveys, due to their faint X-ray luminosity. \par
Overall, the \spr{} simulation can reproduce the shape and normalisation of the X-ray LF even if with large uncertainties. 

\subsubsection{1.4 GHz}
In this section we compare the rest-frame 1.4 GHz luminosity function in the \spr{} simulation with results from the literature. In particular, we consider the luminosity function at 1.4 GHz derived by \citet{Smolcic2009,Novak2017,Upjohn2019} and \citet{Bonato2021} considering only the star-formation activity, i.e. excluding radio-loud AGN. The 1.4 GHz luminosity is derived in \spr{} from the IR luminosity function using eq. \ref{eq:1.4GHz} and considering the component due to star-formation of the IR luminosity. The \spr{} simulation is overall in good agreement with observations at all redshifts and luminosities. \par
In Figure \ref{fig:L1.4GHz_AGN} we show the comparison between the rest-frame 1.4 GHz LF of radio-loud AGN in \spr{} and the the same LF derived by \citet{Smolcic2017,Ceraj2018,Butler2019} and \citet{Bonato2021}. The agreement between \spr{} and observations is remarkable, considering the different selection criteria for radio-loud AGN applied in the literature and the uncertainties present in the simulation to select radio-loud AGN. Consistently with observations \citep[e.g.,][]{Matthews1964}, radio-loud AGN in \spr{} are preferably hosted by elliptical galaxies, at least up to z$=$1.75. This happens because elliptical galaxies dominate the low-z stellar-mass function and the fraction of radio-loud AGN depends on stellar mass (see eq. \ref{eq:fRAGN}). 

\begin{figure}[hbt!]
    \centering
    \includegraphics[width=1\linewidth,keepaspectratio]{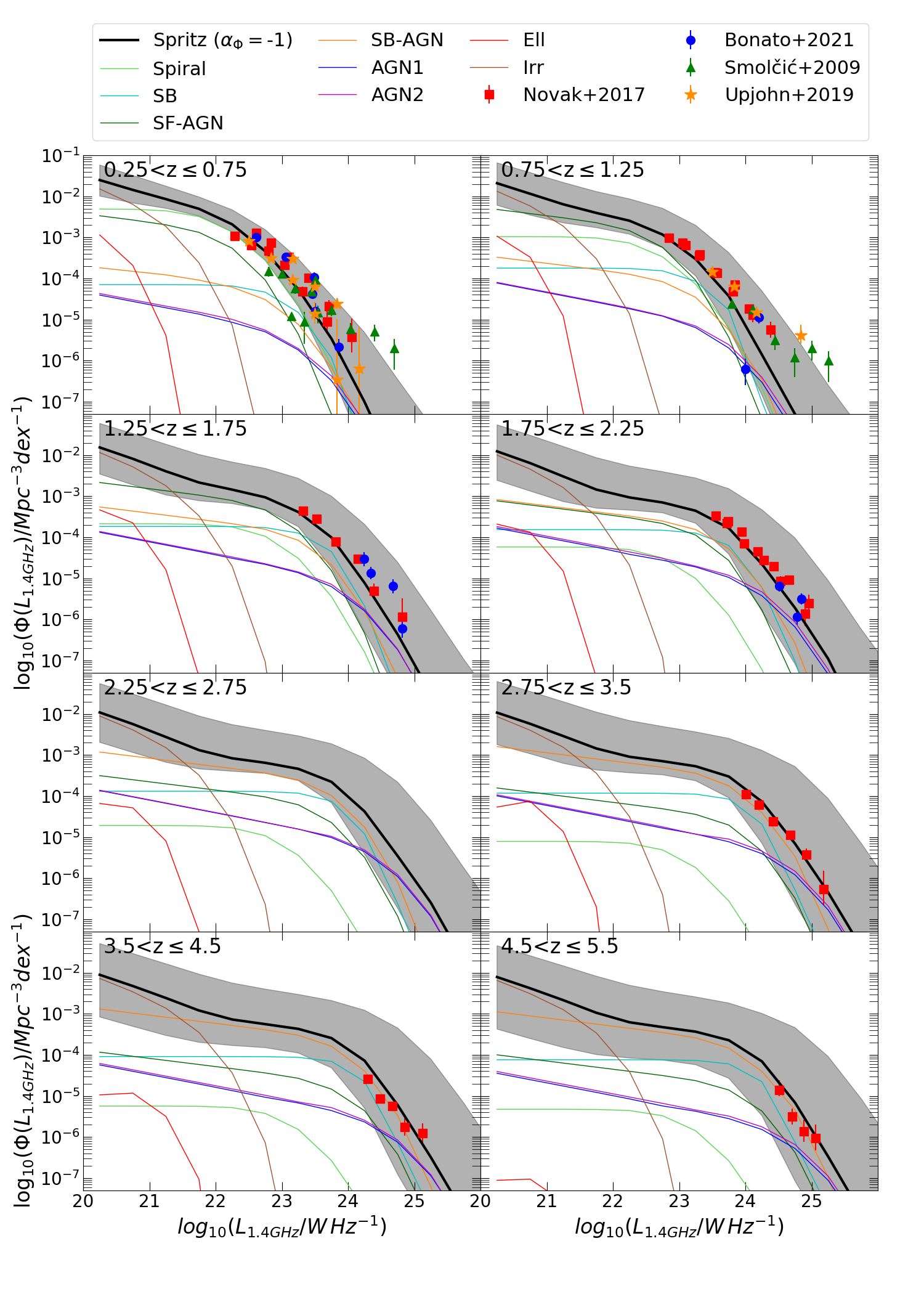}
    \caption{Rest-frame luminosity function at 1.4 GHz for all galaxies simulated with \spr{} (\textit{black tick solid line}) with high-redshift extrapolation $k_{\Phi}=$-1 (eq. \ref{eq:zevol}), taking into account only the luminosity originated from the star-formation activity. The grey areas indicate the uncertainties due to the high-z extrapolation, to the different torus models included and the 1$\sigma$ errors of the GSMF, IR and K-band LFs used to generate simulated galaxies. The luminosity function for each different sub-population is also shown (see Fig. \ref{fig:LF8}). We reported the observed 1.4 GHz luminosity functions, limited to the star-formation activity, by \citet[][\textit{red squares}]{Novak2017}, \citet[][\textit{blue circles}]{Bonato2021}, \citet[][\textit{green trangles}]{Smolcic2009} and \citet[][\textit{orange stars}]{Upjohn2019}.}
    \label{fig:L1.4GHz}
\end{figure}

\begin{figure}[hbt!]
    \centering
    \includegraphics[width=1\linewidth,keepaspectratio]{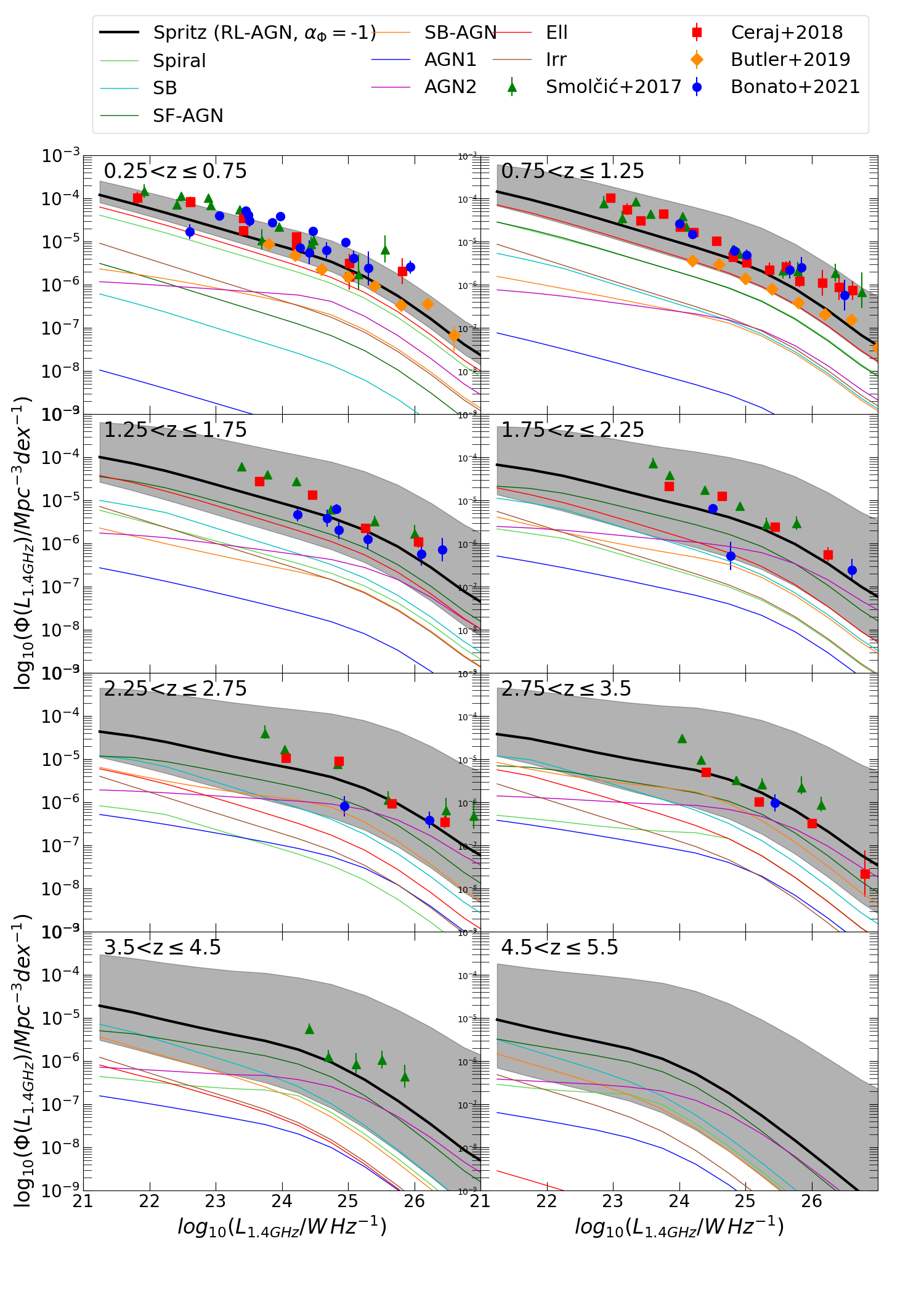}
    \caption{The same as Fig. \ref{fig:L1.4GHz}, but for radio-loud AGN. We reported the observed 1.4 GHz luminosity functions of radio-loud AGN by \citet[][\textit{red squares}]{Ceraj2018}, \citet[][\textit{blue circles}]{Bonato2021}, \citet[][\textit{green trangles}]{Smolcic2017} and \citet[][\textit{orange diamonds}]{Butler2019}.}
    \label{fig:L1.4GHz_AGN}
\end{figure}

\subsection{AGN line diagnostics}
\begin{figure}[htb!]
    \centering
    \includegraphics[width=1\linewidth,keepaspectratio]{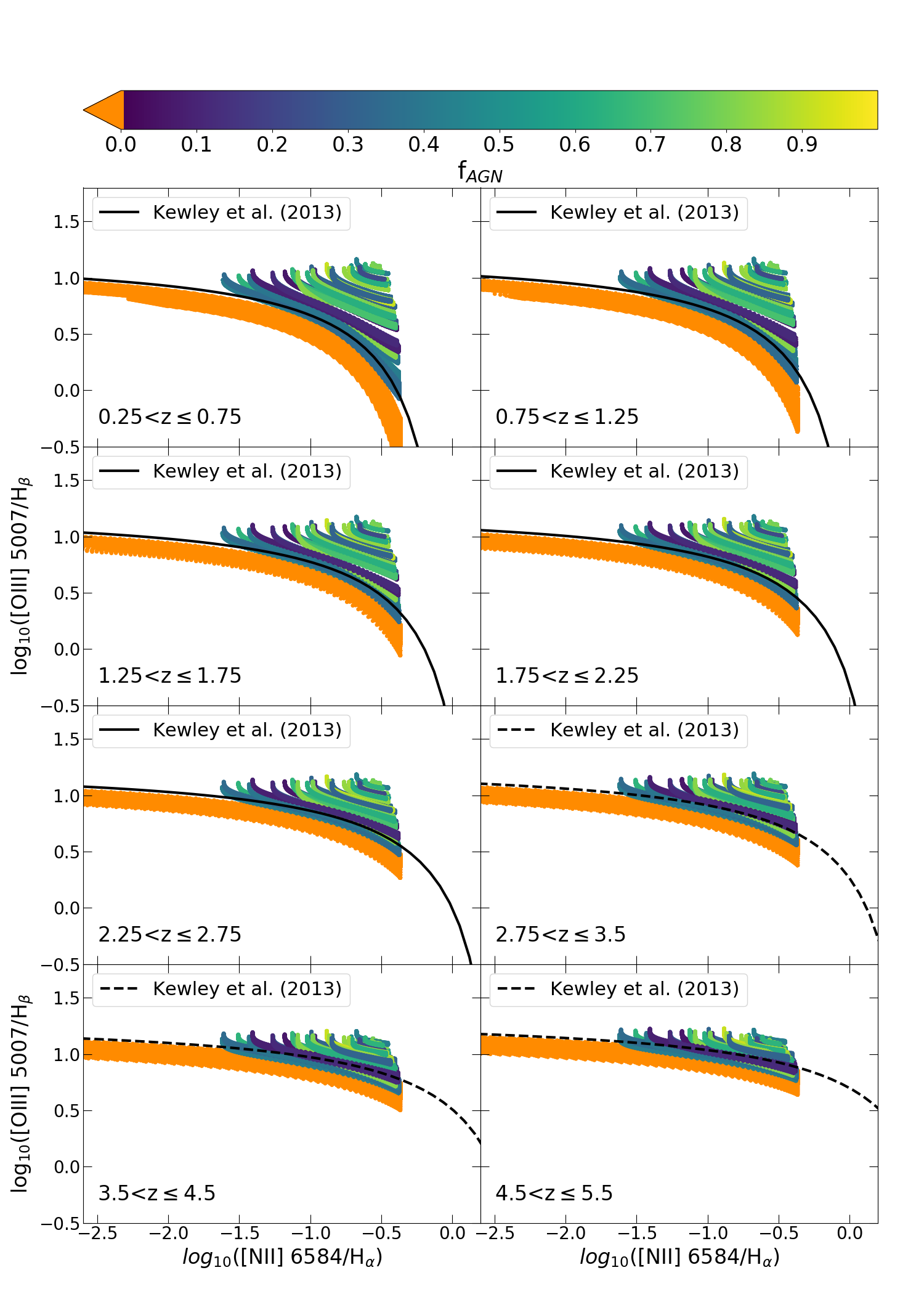}
    \caption{BPT diagram for simulated galaxies. Data points are colour coded depending on the fraction of AGN measured in the IR. The black solid line shows the separation between AGN and $\ion{H}{II}$ region, as derived by \citet{Kewley2013b} at z$\leq$2.5. The black dashed line indicates the extrapolation of the same separation criteria at high reshifts.}
    \label{fig:BPT}
\end{figure}

The ratio between some nebular emission lines can give indication on the nature of the source powering them, allowing us to separate between star-forming systems and AGN. The most used diagnostic, which allows us to separate AGN from $\ion{H}{II}$ regions, is the BPT diagram \citep{BPT1981} which includes the ratio of the optical lines $[\ion{O}{III}]$ 5007\AA/H$_{\beta}$ and $[\ion{N}{II}]$ 6584\AA/H$_{\alpha}$. We made use of such diagram to verify some nebular emission line flux implemented in the simulation. In Fig. \ref{fig:BPT}, we show the position of the \spr{} simulated galaxies in the BPT diagram, as derived from the master catalogue. 
We incorporated the nebular fluxes in the master catalog before applying any observational scatter. The nebular line fluxes of the AGN component of the master catalog has been obtained by averaging the predictions of the models spanning the range of parameters described in Sec. \ref{sec:lines}. This flux is then normalised to the AGN accretion disk luminosity of each mock galaxy. When specific simulated galaxies are derived, considering the full probability distribution of physical parameters of the AGN models and the photometric errors, the $[\ion{N}{II}]$ 6584\AA/H$_{\alpha}$ ratio of AGN are not limited below log$_{10}([\ion{N}{II}]6584\AA/H_{\alpha})<$-0.37.
The sharp limit at high $[\ion{N}{II}]$ 6584\AA/H$_{\alpha}$ values for galaxies without AGN is instead dictated by the asymptotic metallicity value of the considered mass metallicity relation (see eq. \ref{eq:met}). \par
In the same Figure we also report the semi-empirical line, evolving with redshift, that separates AGN from star-forming systems \citep{Kewley2013b}. AGN dominated systems lay above the separation line at all redshifts, while some of the galaxies with a fraction of AGN below 50$\%$, as derived in the IR, are below the separation line at high-z. First, it is necessary to consider that these objects are the most uncertain ones, as the AGN only marginally contribute to the nebular emission lines considered in this work. Second, the separation line by \citet{Kewley2013b} has been derived from data collected up to z$\sim$2.5, therefore its extrapolation at higher-z needs to be further verified.\par

\section{Application: predictions for JWST and OST}\label{sec:App}
In this section we provide some examples on the application of \spr{} to the construction of simulated catalogues, useful mostly for future space programs. In particular, we simulated the following observations: 
\begin{itemize}
    \item JWST NIRCam observations in 9 filters (F090W, F115W, F150W, F200W, F277W, F335M, F356W, F410M and F444W) with 5$\sigma$ depths between 29.5 and 30.6 AB magnitude and covering an area of 46 square arcmin. This follows the planned JADES-Deep survey (JWST Advanced Deep Extragalactic Survey), as prepared by the NIRSpec and NIRCam Guaranteed Time Observers \citep{Rieke2019}.
    \item a far-IR survey with OST over 500 deg$^{2}$ area\footnote{Predictions for this area are derived using the master catalogue, while a complete simulated catalogue is avavilable only for 5 deg$^{2}$.} and with 5$\sigma$ depths equal to 40$\mu$Jy and 1 mJy at 50 and 250 $\mu$m.
\end{itemize} \par

\subsection{JWST}
\begin{figure}
    \centering
    \includegraphics[width=1\linewidth,keepaspectratio]{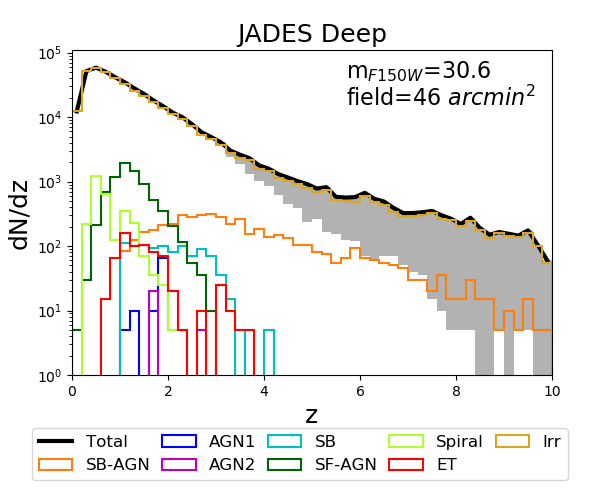}
    \includegraphics[width=1\linewidth,keepaspectratio]{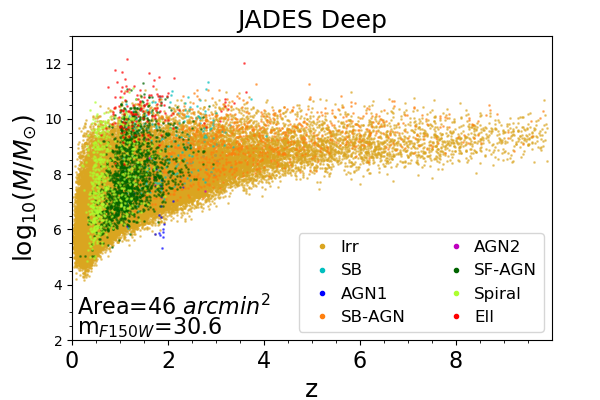}
    \caption{Redshift distribution (\textit{top}) and stellar-mass vs. redshift (\textit{bottom}) of the simulated galaxies obtained with the \spr{} simulation for the JWST JADES-Deep survey, considering only galaxies brighter than $m_{AB,F150W}=$30.6. Different colours indicate different galaxy populations, as listed in the legend. On the top, the grey shaded area indicates uncertainties due to the high-z extrapolation of the different LFs. }
    \label{fig:JADES}
\end{figure}

Figure \ref{fig:JADES} shows the number-redshift distribution and the stellar mass distributions at different redshifts for the JWST JADES-Deep simulated catalogue, considering only galaxies with m$_{AB,F150W}<$30.6. Irregular galaxies are expected to dominate the number counts of the survey, which is not surprising given that the flux cut corresponds in general to stellar masses as low as M$^{*}\sim10^{8}$ M$_{\odot}$ even at the highest redshift. However, these galaxies are expected to be detected on 7 out of the 9 considered filters, on average, with a median S/N$\sim$12. Excluding irregulars, at z$>$4 we expect around 260 galaxies, all with an AGN contribution, but this number decreases to 29 when considering the most extreme redshift extrapolation, i.e. $k_{\Phi}=$-4. It is however necessary to point-out that the redshift extrapolation with $k_{\Phi}=$-4 slightly underestimates the ALPINE IR LF at z$\sim$5 (see Fig. \ref{fig:LF_ALPINE}), so it is to be regarded as conservative prediction. The simulated catalogue also contains almost no AGN-dominated objects (AGN1 and AGN2), due to a combination of depth and area. This shows the necessity of larger areas or pointed observations to study the AGN population with JWST. The number of irregular galaxies at z$>$4 varies from 4$\times$10$^2$ to 1.7$\times$10$^3$, showing the large uncertainties linked to this galaxy population. \par
\begin{figure*}
    \centering
    \includegraphics[width=1\linewidth,keepaspectratio]{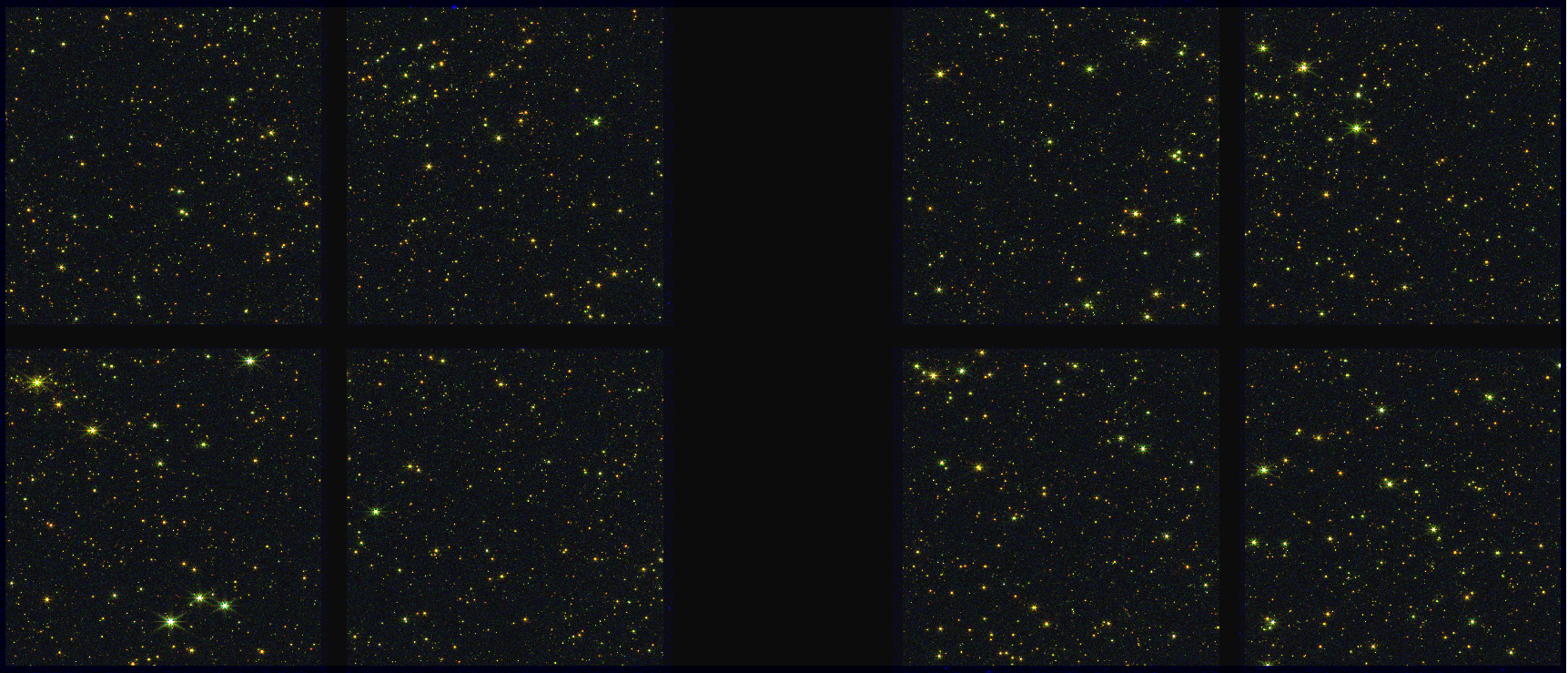}
    \caption{Example of a NIRCam pointing in three filters: F090W (\textit{blue}), F277W (\textit{green}) and F44W (\textit{red}).}
    \label{fig:NIRCam_pointing}
\end{figure*}

To further show the wealth of possible applications of the \spr{} simulation, we generate simulated NIRCam images in different filters. In particular, we considered a single realisation of NIRCam point-spread-function, as simulated by WebbPSF \citep{Perrin2012,Perrin2014} considering the "requirements" optical path difference map\footnote{\label{JWSTsite}\url{https://jwst-docs.stsci.edu}}. Each simulated galaxy is considered as unresolved, but future developments may include morphological information. To generate the simulated images, we considered a gap of 43" (47") between the two NIRCam modules in the blue (red) filters, i.e. bands at wavelengths shorter (longer) than 2.3$\mu$m (2.4$\mu$m), and gaps of 4" or 5" between the 4 quadrants of each module, depending on the filter wavelengths. It is however necessary to consider that these distances are approximated, as they may change after the launch\footref{JWSTsite}. The simulated image is shown in figure \ref{fig:NIRCam_pointing} combining three different filters, each shown with different colours: i.e. F090W (\textit{blue}), F277W (\textit{green}) and F44W (\textit{red}). Simulated images in other filters will be made publicly available\footref{mywebsite}. \par

\subsection{OST}
\begin{figure*}
    \centering
    \includegraphics[width=1\linewidth,keepaspectratio]{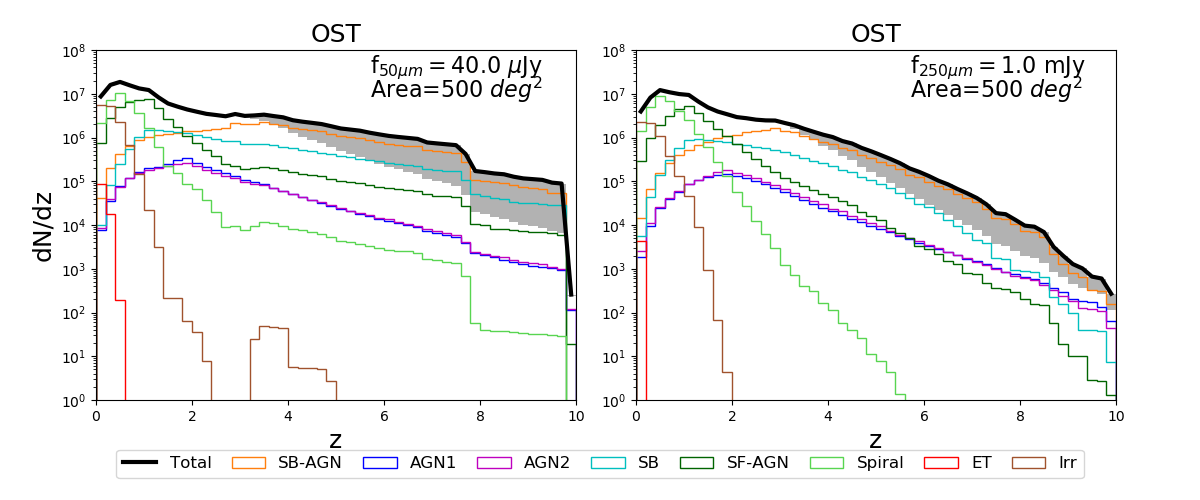}
    \includegraphics[width=1\linewidth,keepaspectratio]{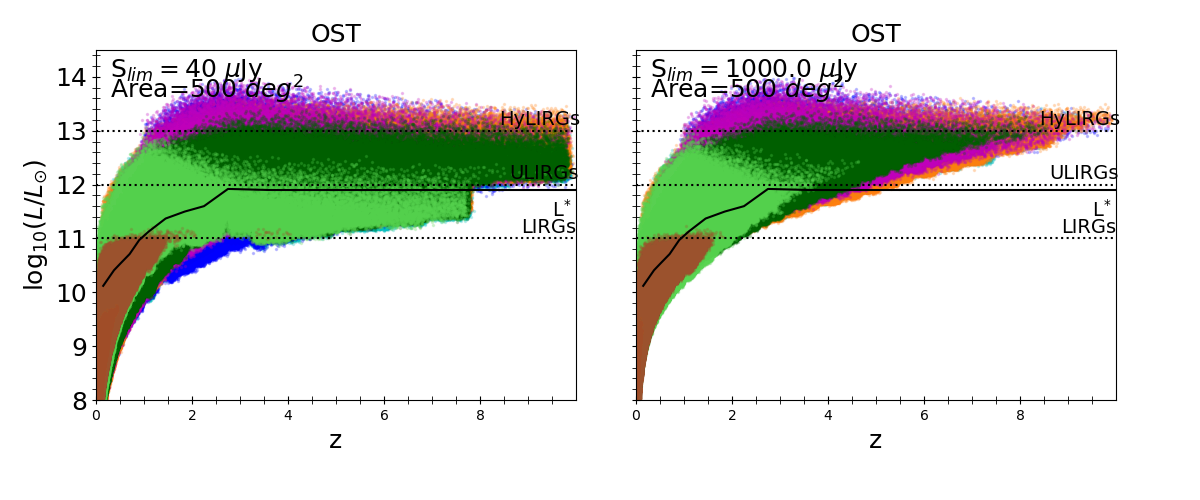}
    \caption{Redshift distribution (\textit{top}) and L$_{IR}$ vs. redshift (\textit{bottom}) of the OST simulated catalogues covering 500 deg$^{2}$. We considered galaxies detected in both photometric filters, i.e. the band centred at 50$\mu$m (\textit{left}) and the one centred at 250$\mu$m(\textit{right}). Different colours indicate different galaxy populations, as listed in the legend. On the top, the grey shaded area indicates uncertainties due to the high-z extrapolation of the different LFs. }
    \label{fig:z_OST}
\end{figure*}

In Figure \ref{fig:z_OST} we show our predictions for the redshift distribution and the L$_{IR}$ at different redshift of galaxies detected with OST. This mission is planned to have a collecting area of 5.9-m primary mirror, at the moment of writing, and photometric filters observing at 50 and 250 $\mu$m. In addition, the great mapping capability will allow the observation of a sky area larger than JWST. For this reason, the planned OST Medium Survey is expected to cover 500 deg$^{2}$ with a 5$\sigma$ depth of 40$\mu$Jy and 1 mJy at 50 and 250$\mu$m. Both photometric filters are, at the moment, described as simple boxes and this is the reasons behind some sharp features in the Figure. At both 50 and 250$\mu$m we predict that some irregular galaxies should be observable up to z$\sim$2, while elliptical galaxies will be limited to very low redshifts. On the other hand, spirals galaxies are expected to be observed at all redshifts at 50 $\mu$m and up to z$=$5 at 250$\mu$m. The SB-AGN population dominates the redshift distributions at both wavelengths at z$>$2, while the remaining populations, i.e. SB, AGN1, AGN2 and SF-AGN, should anyway be observable up to z$=$10 at both wavelengths. Looking at the IR luminosity of the observed populations, we expect to detect ULIRG (i.e. $L_{IR}>10^{12}L_{\odot}$) at all redshifts at 50$\mu$m and at least up to z$=$5 in the other filter. Observations at more moderate IR luminosities, i.e. $L_{IR}=10^{11}L_{\odot}$, will instead be limited to z$<$4 (z$<$2) at 50 (250) $\mu$m. \par

These simulated catalogues were chosen to give an example of the possible application of the \spr{} simulation and they are all publicly available\footref{mywebsite}. Additional simulated catalogues involving these or other missions may be included in future releases or may be produced on demand.
\section{Summary and conclusions}\label{sec:conclusions}
In this paper we presented the new \spr{} simulation to construct simulated catalogues for current and future facilities, such as \textit{JWST} and OST. The need for this new simulation is motivated by the difficulty of current semi-analytical and hydro-dynamical simulations to reproduce the observed number density of the most luminous IR galaxies. \par
The \spr{} simulation is semi-empirical and starts from the IR LF based on \hers{} sources up to z$\sim$3 \citep{Gruppioni2013}, complemented by the K-band LF of elliptical galaxies \citep{Arnouts2007,Cirasuolo2007,Beare2019} and the GSMF of irregular galaxies by \citet{Huertas-Company2016}. Both irregular and elliptical galaxies were not observed in sufficient number by \hers{} and, therefore, it was necessary to include them separately as a simple extrapolation of the \hers{}  IR LF would not be sufficient to include them. We updated the \hers{} IR LF of both "unobscured" and "obscured" AGN by performing a combined fit of IR and FUV observations. We considered a series of empirical relations to assign several physical parameters as well as optical and IR emission features to each simulated galaxy, and we performed a host-AGN decomposition to estimate the contribution of AGN. Among the considered physical parameters there are redshift, stellar mass, IR luminosity, SFR, hydrogen column density, the X-ray and the 1.4 GHz luminosity. We created a simulated light cone by assigning a sky position to each simulated galaxies following a two-point angular correlation function, whose normalisation varies with stellar mass. 
For all these simulated galaxies we also derived the expected fluxes in different filters, from current and future facilities, among which SDSS, \textit{JWST}, \textit{Euclid} and OST, as well as some low-resolution spectra, such as JWST MIRI-LR and OST/OSS. \par
In this work we have shown that the \spr{} simulation well reproduces a large number of observables, among which the number counts of galaxies from the U-band to the far-IR, the stellar mass function, SFR-stellar mass plane, the IR LF at z$>$3\footnote{below z$=$3 the IR LF function is reproduced by construction.}, the K-band, FUV and X-ray LFs as well as the BPT diagram. 
In the future, the inclusion of additional observations and an increase of the number of considered SED templates may help to improve the simulation, reducing the few discrepancies as well as the uncertainties in the predictions. \par
We have presented the use of \spr{} to generate simulated catalogues, considering a JWST survey mimicking the planned JADES Deep and an OST survey of 500 deg$^{2}$. Future releases may include additional surveys, not only for the considered missions but also for additional ones (e.g. Euclid, Athena).\par
To conclude, the \spr{} simulation is suitable for predictions for a broad set of future facilities operating at, but not limited to, IR wavelengths. This is possible by simulating realistic spectro-photometric data spanning a wide range in wavelengths.    

\begin{acknowledgements}
LB acknowledge financial support by the Agenzia Spaziale Italiana (ASI) under the research contract 2018-31-HH.0. CG, AF and FC acknowledge the support from grant PRIN MIUR 2017 20173ML3WW.
\end{acknowledgements}

%
\bibliographystyle{aa} 
\bibliography{main} 
%
\begin{appendix} 
\section{luminosity function of AGN1 derived with an MCMC}\label{sec:MCMC}
In this section we report in detail the MCMC applied to derive the new IR LF for AGN1 and AGN2, using simultaneously IR and FUV observations. \par
In particular, we considered the same IR \hers{} data used by \citetalias{Gruppioni2013} to derive the IR LF. Considering the unification theory of AGN, AGN1 and AGN2 are the same population and differences are due to orientation effects. We therefore expected AGN1 and AGN2 to have similar IR LFs, with some possible differences in the normalisation due to different orientations, i.e. line-of-sight that intercepts or not the dusty torus. \hers{} observations do not show any significant difference between the IR LF of AGN1 and AGN2, therefore we decided to consider both populations together to have a better sample of the IR LF. \par
In addition, we considered observations at rest-frame 1450 \AA{} by \citet{McGreer2013}, \citet{Akiyama2018} and \citet{Schindler2019}, with the addition of the observations by \citet{Croom2009} and \citet{Ross2013} converted to 1450 \AA{} using the correction reported in the last mentioned work. 
From the \hers{} observations we have also the number of times from which each SED template for AGN1 has been chosen to model observations. From the same SED templates we also derived the conversion between L$_{IR}$ and 1450 \AA{} luminosity, for AGN1 only. In this way we have the possibility to convert every possible AGN1 IR LF to a specific FUV LF and therefore use at the same time the IR and FUV observations to derive the IR AGN1 and AGN2 LFs. As our interested is mainly in reproducing IR observations and we want to compensate for the difference in number between FUV and IR observations, we arbitrary multiply the observational errors of FUV data to have an average value equal to twice the average value of IR errors. In addition, we increase the minimum FUV observational errors to match the minimum IR one. \par
To derive the AGN1 IR LF we adopted a Bayesan approach where the probability $\mathcal{P}$ of the model described by the parameters $\overrightarrow{x}$ given the data $\mathcal{D}$, which contains both IR and FUV, described as:
\begin{equation}
    \mathcal{P}(\overrightarrow{x},\mathcal{D})=\mathcal{P}(\mathcal{D},\overrightarrow{x})\mathcal{P}(\overrightarrow{x})
\end{equation}
The list of prior $\mathcal{P}(\overrightarrow{x})$ is listed in Table \ref{tab:Prioirs}.  The likelihood $\mathcal{P}(\mathcal{D},\overrightarrow{x})$ is a combination of the IR and FUV likelihoods, weighted for the number of available points:
\begin{multline}
    \mathcal{P}(\mathcal{D}_{\rm IR},\mathcal{D}_{\rm UV}|\overrightarrow{x}_{\rm IR},\overrightarrow{x}_{\rm FUV})\propto \\
    \prod\exp \left( -w_{\rm IR}\frac{(\mathcal{D}_{\rm IR}-\overrightarrow{x}_{\rm IR})^{2}}{\sigma_{IR}^{2}}-w_{\rm FUV}\frac{(\mathcal{D}_{\rm FUV}-\overrightarrow{x}_{\rm FUV})^{2}}{\sigma_{FUV}^{2}} \right)
\end{multline}
The weights are derive to compensate the different number of data points available in FUV and IR, i.e. $w_{\rm IR}=\frac{N_{\rm FUV}+N_{\rm IR}}{N_{\rm IR}}$ and $w_{\rm FUV}=\frac{N_{\rm FUV}+N_{\rm IR}}{N_{\rm FUV}}$. \par
We considered the modified Schechter function evolving with redshift, as reported in equation \ref{eq:LF} and \ref{eq:zevol}, but we removed the extrapolation at z$>$3, as FUV observations are available at higher redshifts. The \citetalias{Gruppioni2013} LF for AGN1 is considered as starting point for the MCMC chains. \par

\begin{table}[]
    \centering
    \caption{Summary of the parameters of the evolving, modified Schechter function and their priors.}
    \label{tab:Prioirs}
    \begin{tabular}{cc|cc}
        Parameter & Prior range &  Parameter & Prior range  \\
        \hline
        log$_{10}(L^{*}_{0}/{\rm L}_{\odot})$ & ]8,12.5[ &         k$_{\rho,2}$ & ]-12,12[ \\
        log$_{10}(\Phi^{*}_{0}/{\rm Mpc}^{-3})$ & ]-9,-1[ & $\alpha$ & ]1,1.6[ \\
        k$_{L,1}$ & ]-10,10[ &   $\sigma$ & ]0.1,1[ \\
        k$_{L,2}$ & ]-10,10[ &   z$_{L}$ & ]0,5[ \\
        k$_{\rho,1}$ & ]-12,12[ & z$_{\rho}$ & ]0,5[ \\
    \end{tabular}
\end{table}

\begin{figure*}[hbt!]
    \centering
    \includegraphics[width=1\linewidth,keepaspectratio]{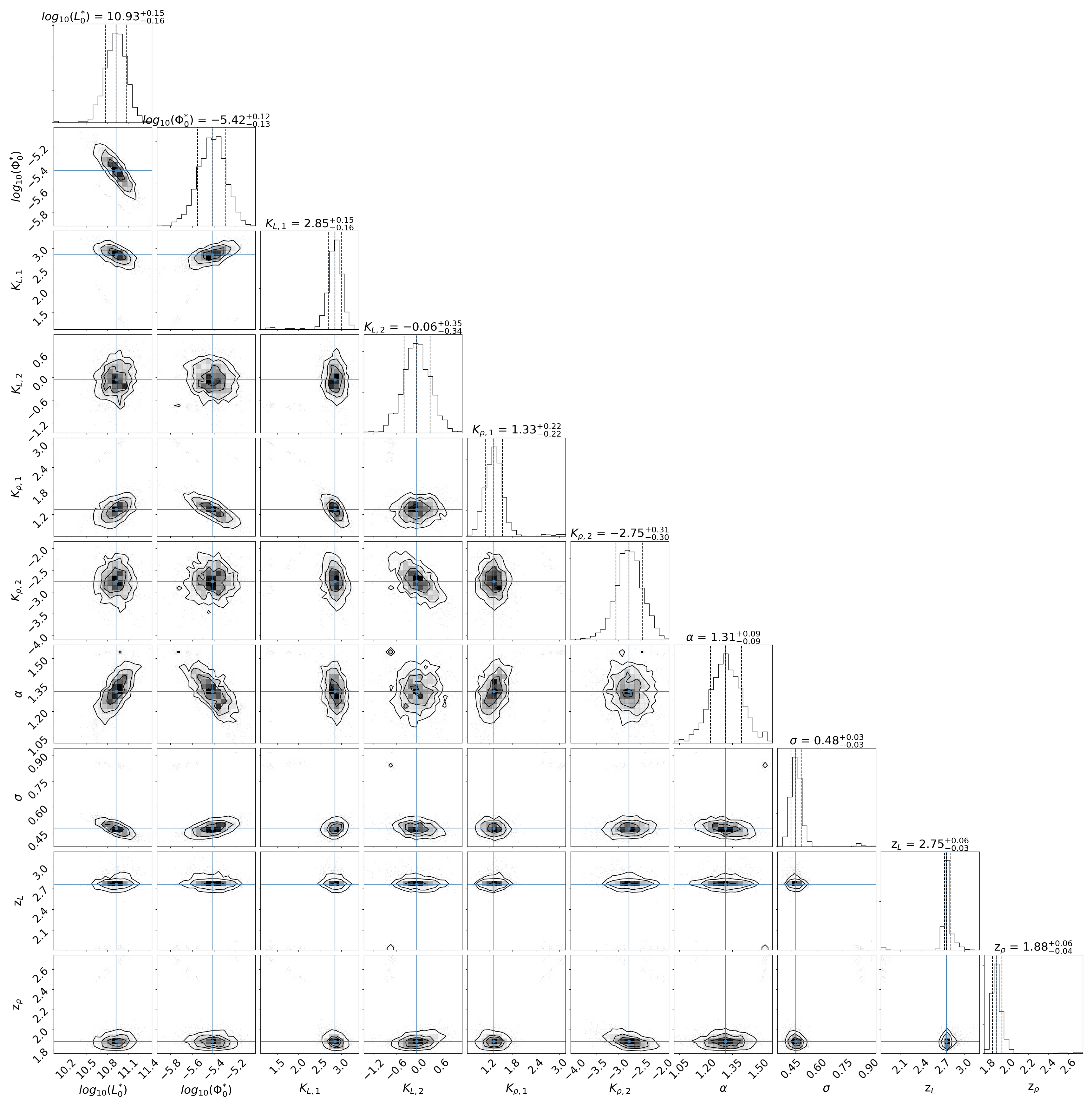}
    \caption{Corner plot of the MCMC performed considering simultaneously the FUV and IR data, contour lines are shown at arbitrary iso-density levels. Marginalised histograms are shown at the top of each column for each parameter. This plot has been derived using the \citet{corner} code.}
    \label{fig:cornerplot}
\end{figure*}

In Figure \ref{fig:cornerplot} we report the corner plots, showing the correlation between the different parameters, as well as the marginalised histogram of each parameter. All parameters show well defined solutions and we report them in Table \ref{tab:IRLF}.\par

\begin{figure}[hbt!]
    \centering
    \includegraphics[width=1\linewidth,keepaspectratio]{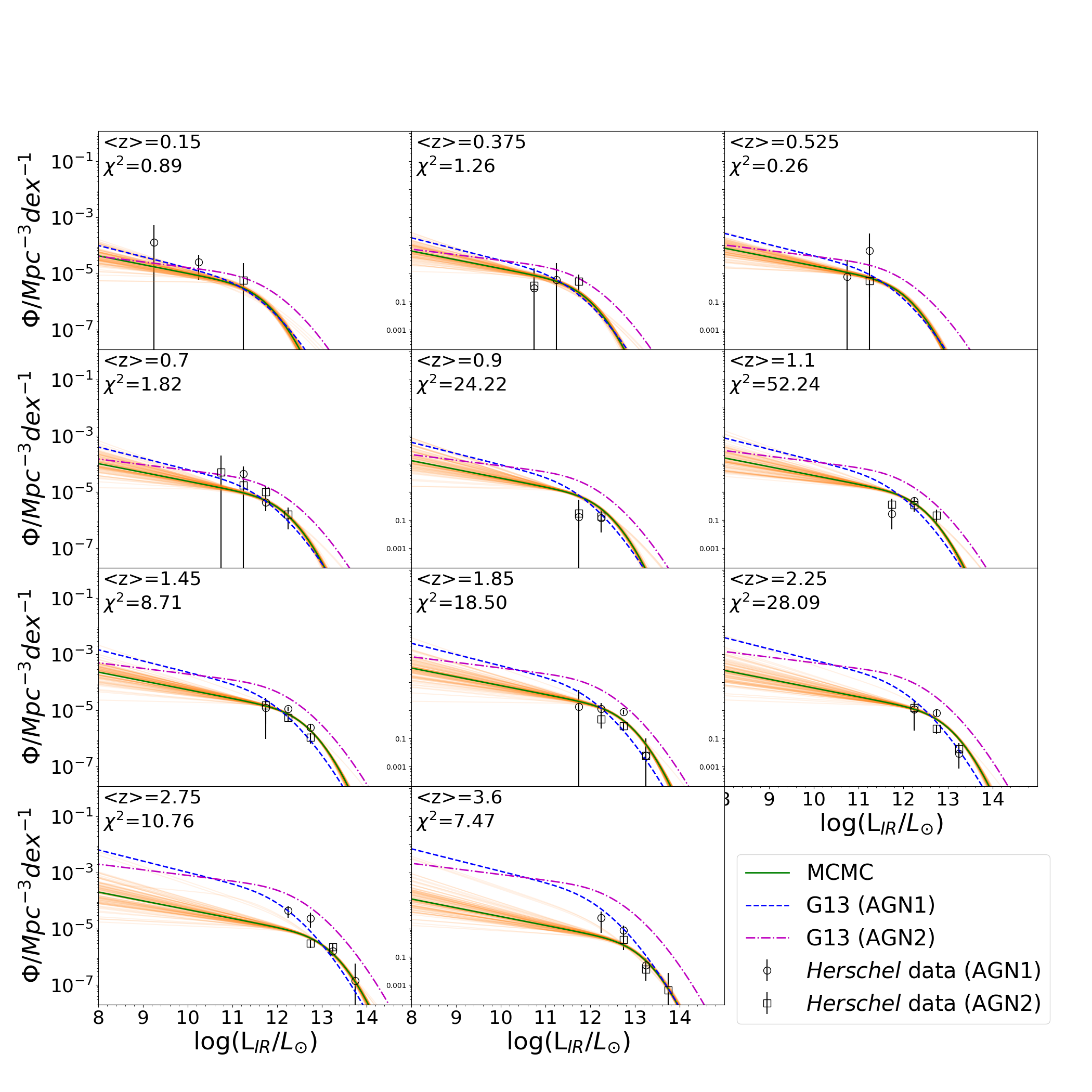}
    \caption{MCMC best result (\textit{green solid line}) compared with AGN1 (\textit{black circles}) and AGN2 (\textit{black squares}) IR \hers{} data and the AGN1 (\textit{blue dashed line}) and AGN2 (\textit{magenta dot-dashed line}) IR LF by \citetalias{Gruppioni2013}. \textit{Orange lines} are random extractions within 1$\sigma$ uncertainty of the MCMC results. On the top left of each panel we report the average redshift and the $\chi^{2}$. }
    \label{fig:IRLF_MCMC}
\end{figure}

\begin{figure}[hbt!]
    \centering
    \includegraphics[width=1\linewidth,keepaspectratio]{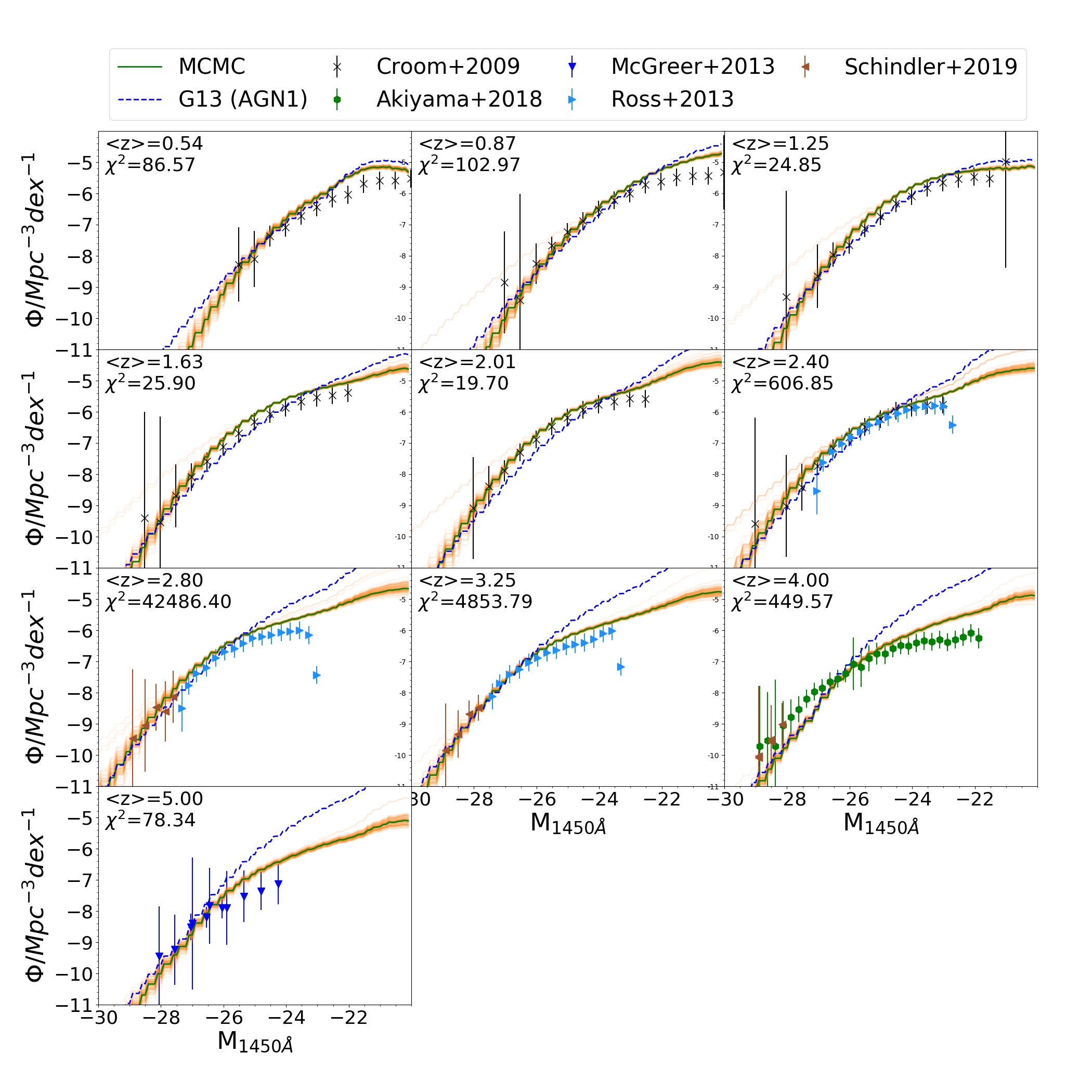}
    \caption{MCMC best result (\textit{green solid line}) compared with the converted AGN1 IR LF by \citetalias{Gruppioni2013} (\textit{blue dashed line}) and FUV observations (see legend). \textit{Orange lines} are random extractions within 1$\sigma$ uncertainty of the MCMC results. On the top left of each panel we report the average redshift and the $\chi^{2}$.}
    \label{fig:FUVLF_MCMC}
\end{figure}

In Figure \ref{fig:IRLF_MCMC} and \ref{fig:FUVLF_MCMC}, we show the result of the MCMC compared with the IR and FUV observations and the \citetalias{Gruppioni2013} IR LF for AGN2 and AGN1. The previous one well describes the IR LF of AGN1, but overestimates the FUV LF of the same galaxy population. The new IR LF still describes the IR data, of both AGN1 and AGN2, but it also well represents the FUV LF of AGN1, with some discrepancies around z$\sim$2. This discrepancy may be due to the low number of SED templates associated in this work to the AGN1 population. In the future, increasing the number of SED templates may allow to better sample the $L_{\rm IR}-{\rm to}-L_{\rm FUV}$ conversion.

\section{Filters included in \spr{}}\label{sec:filters}
In this Appendix we report the complete list of filters included in the \spr{} simulation. In particular, in Table \ref{tab:filters1} we report the filter for which we derived absolute magnitudes. In the same table we also report the corresponding reference and the central wavelength of each filter, derived as follows:
\begin{equation}
    \lambda_{cen}=\frac{\int \lambda T(\lambda) d\lambda}{\int T(\lambda) d\lambda}
\end{equation}
In Table \ref{tab:filters2} we instead report the list of filters included in the \spr{} simulation to derive the observed fluxes with their central wavelengths and the reference of each instrument. 

\begin{table}[]
    \centering
    \caption{Filters included in the \spr{} simulation to derive absolute magnitudes, their central wavelength and the corresponding reference.}
    \label{tab:filters1}
    \begin{tabular}{ccc}
    Filter name & $\lambda_{cen}$ & reference\\
    \hline
    GALEX/FUV & 1538.62 \AA & \citet{Zamojski2007}\\ 
    GALEX/NUV & 2315.66 \AA & \citet{Zamojski2007} \\ 
    SDSS/u & 3573.89 \AA & \citet{Gunn1998}\\ 
    SUBARU/B & 4458.32 \AA & \citet{Miyazaki2002}\\    
    SUBARU/V & 5477.83 \AA & \citet{Miyazaki2002}\\
    SDSS/r & 6202.46 \AA & \citet{Gunn1998} \\
    UKIRT/J & 1.25 $\mu$m & \citet{Casali2007}\\
    WIRCAM/Ks & 2.16 $\mu$m & \citet{Puget2004}\\
    Spitzer/IRAC/Ch4 & 8 $\mu$m & \citet{Fazio2004}\\ 
    \end{tabular}
\end{table}

\onecolumn
\begin{center}
{\small  
\begin{longtable}{ccc|ccc}
\caption{Filters included in the \spr{} simulation to derive observed fluxes, their central wavelength and the corresponding reference.} \label{tab:filters2} \\
    Filter name & $\lambda_{cen}$ & Reference & Filter name & $\lambda_{cen}$ & Reference \\
    \hline
    \endfirsthead
    \multicolumn{6}{c}%
    {\bfseries \tablename\ \thetable{} -- continued from previous page} \\
    \hline  
    \endhead

    \multicolumn{6}{r}{{Continued on next page}} \\
    \endfoot

    \endlastfoot

    SDSS/u & 3561.79 \AA & \citet{Gunn1998} &      Euclid/H & 1.78 $\mu$m & \citet{Laureijs2010}\\
    ELT/MICADO/U & 3605.07 \AA & \citet{Leschinski2016} & ELT/MICADO/Spec-HK & 1.86 $\mu$m & \citet{Leschinski2016}\\ 
    LSST/u & 3684.83 $\AA$ & \citet{Ivezic2008} & JWST/NIRCam/F200W & 1.99 $\mu$m & \citet{Rieke2008}\\  
    VIMOS/U & 3720.47 \AA & \citet{Mieske2007} &  ELT/MICADO/K-short & 2.06 $\mu$m & \citet{Leschinski2016}\\ 
    SUBARU/Suprime-Cam/IB427 & 4263.45 $\AA$ & \citet{Miyazaki2002} & ELT/MICADO/xK1 & 2.06 $\mu$m & \citet{Leschinski2016}\\
    HST/ACS/F435W & 4331.7 \AA & \citet{Sirianni2005} & ELT/MICADO/He-I & 2.06 $\mu$m & \citet{Leschinski2016}\\
    ELT/MICADO/B & 4412.92 \AA & \citet{Leschinski2016} & ELT/MICADO/K-mid & 2.10 $\mu$m &\citet{Leschinski2016} \\
    SUBARU/Suprime-Cam/B & 4458.32 $\AA$ & \citet{Miyazaki2002} & ELT/MICADO/H2-1-0S1 & 2.13 $\mu$m &\citet{Leschinski2016}\\
    SUBARU/Suprime-Cam/IB464 & 4635.13 $\AA$ & \citet{Miyazaki2002} & ELT/MICADO/Ks & 2.14 $\mu$m &\citet{Leschinski2016}\\
    SDSS/g & 4718.87 \AA  & \citet{Gunn1998}&  ELT/MICADO/Ks2 & 2.14 $\mu$m &\citet{Leschinski2016}\\
    SUBARU/Suprime-Cam/g & 4777.07 $\AA$ & \citet{Miyazaki2002} & WIRCAM/Ks & 2.16 $\mu$m & \citet{Puget2004}\\
    LSST/g & 4802.0 $\AA$ & \citet{Ivezic2008} & ELT/MICADO/Br$_{\gamma}$ & 2.17 $\mu$m& \citet{Leschinski2016}\\
    SUBARU/Suprime-Cam/IB484 & 4849.2 $\AA$ & \citet{Miyazaki2002} & ELT/MICADO/K-cont & 2.20 $\mu$m & \citet{Leschinski2016}\\ 
    SUBARU/Suprime-Cam/IB505 & 5062.51 $\AA$ & \citet{Miyazaki2002} & UKIRT/K & 2.20 $\mu$m & \citet{Casali2007} \\ 
    SUBARU/Suprime-Cam/IB527 & 5261.13 $\AA$ & \citet{Miyazaki2002} & ELT/MICADO/xK2 & 2.22 $\mu$m & \citet{Leschinski2016}\\
    SUBARU/Suprime-Cam/V & 5477.83 \AA & \citet{Miyazaki2002} & ELT/MICADO/K-long & 2.31 $\mu$m & \citet{Leschinski2016}\\ 
    ELT/MICADO/V & 5512.0 \AA & \citet{Leschinski2016} & AKARI/IRC/N2 & 2.41 $\mu$m & \citet{Murakami2007}\\  
    SUBARU/Suprime-Cam/IB574 & 5764.76 $\AA$ & \citet{Miyazaki2002} & JWST/NIRCam/F277W & 2.78 $\mu$m & \citet{Rieke2008} \\
    HST/ACS/F606W & 5956.83 \AA &  \citet{Sirianni2005} &  ELT/METIS/H2Oice & 3.10 $\mu$m & \citet{Leschinski2016}\\ 
    SDSS/r & 6185.19 \AA  & \citet{Gunn1998} & AKARI/IRC/N3 & 3.28 $\mu$m  & \citet{Murakami2007}\\
    LSST/r & 6231.2 $\AA$ & \citet{Ivezic2008} & ELT/METIS/shortL & 3.30 $\mu$m & \citet{Leschinski2016}\\ 
    SUBARU/Suprime-Cam/IB624 & 6233.09 $\AA$ & \citet{Miyazaki2002} & ELT/METIS/PAH3.3 & 3.30 $\mu$m &\citet{Leschinski2016} \\ 
    SUBARU/Suprime-Cam/r & 6288.71 $\AA$ & \citet{Miyazaki2002} & WISE/WISE/W1 & 3.40 $\mu$m & \citet{Wright2010} \\ 
    ELT/MICADO/R & 6592.93 \AA & \citet{Leschinski2016} &  Spitzer/IRAC/Ch1 & 3.56 $\mu$m & \citet{Fazio2004}\\
    SUBARU/Suprime-Cam/IB679 & 6781.13 $\AA$ & \citet{Miyazaki2002} & JWST/NIRCam/F356W & 3.57 $\mu$m & \citet{Rieke2008}\\ 
    JWST/NIRCam/F070W & 7006.11 \AA & \citet{Rieke2008} & ELT/METIS/HCI-L-short & 3.60 $\mu$m & \citet{Leschinski2016}\\
    SUBARU/Suprime-Cam/IB709 & 7073.63 $\AA$ & \citet{Miyazaki2002} & ELT/METIS/Lp & 3.81 $\mu$m & \citet{Leschinski2016}\\
    SUBARU/Suprime-Cam/NB711 & 7119.88 $\AA$ & \citet{Miyazaki2002} & ELT/METIS/HCI-L-long & 3.82 $\mu$m & \citet{Leschinski2016}\\
    Euclid/VIS & 7156.46 \AA & \citet{Laureijs2010} & ELT/METIS/Bra & 4.05 $\mu$m & \citet{Leschinski2016}\\ 
    SUBARU/Suprime-Cam/IB738 & 7361.56 $\AA$ & \citet{Miyazaki2002} & JWST/NIRCam/F444W & 4.41 $\mu$m & \citet{Rieke2008}\\
    SDSS/i & 7499.7 \AA  & \citet{Gunn1998} & AKARI/IRC/N4 & 4.47 $\mu$m  & \citet{Murakami2007}\\ 
    LSST/i & 7541.69 $\AA$ & \citet{Ivezic2008} &Spitzer/IRAC/Ch2 & 4.51 $\mu$m & \citet{Fazio2004} \\
    SUBARU/Suprime-Cam/i & 7683.88 $\AA$ & \citet{Miyazaki2002} &ELT/METIS/CO01ice & 4.65 $\mu$m & \citet{Leschinski2016}\\
    SUBARU/Suprime-Cam/IB767 & 7684.89 $\AA$ & \citet{Miyazaki2002} &  WISE/WISE/W2 & 4.65 $\mu$m & \citet{Wright2010}\\ 
    HST/ACS/F775W & 7712.58 \AA &  \citet{Sirianni2005} &   ELT/METIS/Mp & 4.78 $\mu$m & \citet{Leschinski2016}\\
    HST/ACS/F814W & 8012.07 \AA &  \citet{Sirianni2005} & JWST/MIRI/F560W & 5.65 $\mu$m & \citet{Wright2015} \\ 
    ELT/MICADO/I & 8059.97 \AA & \citet{Leschinski2016} & Spitzer/IRAC/Ch3 & 5.76 $\mu$m & \citet{Fazio2004}\\
    SUBARU/Suprime-Cam/NB816 & 8149.39 $\AA$ & \citet{Miyazaki2002} &AKARI/IRC/S7 & 7.31 $\mu$m  & \citet{Murakami2007}\\ 
    SUBARU/Suprime-Cam/IB827 & 8244.52 $\AA$ & \citet{Miyazaki2002} & JWST/MIRI/F770W & 7.66 $\mu$m & \citet{Wright2015} \\ 
    ELT/MICADO/xI1 & 8374.01 \AA & \citet{Leschinski2016} &   Spitzer/IRAC/Ch4 & 7.96 $\mu$m & \citet{Fazio2004}\\ 
    ELT/MICADO/I-long & 8680.73 \AA & \citet{Leschinski2016} & ELT/METIS/PAH8.6 & 8.60 $\mu$m & \citet{Leschinski2016}\\ 
    LSST/z & 8690.47 $\AA$ & \citet{Ivezic2008} & ELT/METIS/N1 & 8.65 $\mu$m & \citet{Leschinski2016}\\ 
    SDSS/z & 8961.49 \AA  & \citet{Gunn1998} &  AKARI/IRC/S9W & 9.22 $\mu$m  & \citet{Murakami2007}\\ 
    ELT/MICADO/xI2 & 9022.12 \AA & \citet{Leschinski2016} &   JWST/MIRI/F1000W & 9.94 $\mu$m & \citet{Wright2015} \\ 
    SUBARU/Suprime-Cam/z & 9036.88 $\AA$ & \citet{Miyazaki2002} & ELT/METIS/SIV & 10.50 $\mu$m &\citet{Leschinski2016} \\ 
    HST/ACS/F850LP & 9043.26 \AA &  \citet{Sirianni2005} & AKARI/IRC/S11 & 10.95 $\mu$m  & \citet{Murakami2007}\\ 
    JWST/NIRCam/F090W & 9045.79 \AA & \citet{Rieke2008} &  ELT/METIS/PAH11.5 & 11.20 $\mu$m & \citet{Leschinski2016}\\
    LSST/y & 9736.41 $\AA$ & \citet{Ivezic2008} & ELT/METIS/N2 & 11.63 $\mu$m & \citet{Leschinski2016}\\ 
    HST/WFC3/F098M & 9875.26 \AA & {\tiny \url{www.stsci.edu/hst/wfc3}} & WISE/WISE/W3 & 12.81 $\mu$m  & \citet{Wright2010}\\
    ELT/MICADO/xY1 & 1.00 $\mu$m & \citet{Leschinski2016} &  ELT/METIS/NeII & 12.82 $\mu$m & \citet{Leschinski2016}\\ 
    UKIRT/Y & 1.02 $\mu$m & \citet{Casali2007} &  AKARI/IRC/L15 & 16.16 $\mu$m  & \citet{Murakami2007}\\
    ELT/MICADO/Y & 1.04 $\mu$m & \citet{Leschinski2016} &  AKARI/IRC/L18W & 19.81 $\mu$m  & \citet{Murakami2007}\\ 
    HST/WFC3/F105W & 1.06 $\mu$m & {\tiny \url{www.stsci.edu/hst/wfc3}}& WISE/WISE/W4 & 22.38 $\mu$m  & \citet{Wright2010}\\
    ELT/MICADO/xY2 & 1.08 $\mu$m & \citet{Leschinski2016} & AKARI/IRC/L24 & 23.35 $\mu$m  & \citet{Murakami2007}\\
    Euclid/Y & 1.08 $\mu$m & \citet{Laureijs2010} & Spitzer/MIPS/24mu & 23.84 $\mu$m & \citet{Rieke2004}\\
    ELT/MICADO/Spec-IJ & 1.11 $\mu$m & \citet{Leschinski2016} & OST/OSS/Ch1 & 34.50 $\mu$m & \citet{Bradford2018}\\
    JWST/NIRCam/F115W & 1.16 $\mu$m & \citet{Rieke2008} & OST/FIP/50 & 50.00 $\mu$m & \citet{Staguhn2018}\\
    ELT/MICADO/J-short & 1.19 $\mu$m & \citet{Leschinski2016} & OST/OSS/Ch2 & 58.00 $\mu$m & \citet{Bradford2018}\\
    ELT/MICADO/xJ1 & 1.20 $\mu$m & \citet{Leschinski2016} & AKARI/FIS/N60 & 66.69 $\mu$m  & \citet{Murakami2007}\\
    UKIRT/J & 1.25 $\mu$m & \citet{Casali2007} &  Herschel/PACS/70 & 71.93 $\mu$m & \citet{Poglitsch2010}\\
    ELT/MICADO/J & 1.25 $\mu$m & \citet{Leschinski2016} & Spitzer/MIPS/70 & 72.56 $\mu$m & \citet{Rieke2004}\\
    HST/WFC3/F125W & 1.25 $\mu$m & {\tiny \url{www.stsci.edu/hst/wfc3}} & AKARI/FIS/WIDE-S & 89.20 $\mu$m  & \citet{Murakami2007}\\
    ELT/MICADO/J-long & 1.27 $\mu$m & \citet{Leschinski2016} & OST/OSS/Ch3 & 97.50 $\mu$m & \citet{Bradford2018}\\
    ELT/MICADO/Pa$_{\beta}$ & 1.29 $\mu$m & \citet{Leschinski2016} &  Herschel/PACS/100 & 102.62 $\mu$m & \citet{Poglitsch2010}\\
    ELT/MICADO/xJ2 & 1.30 $\mu$m & \citet{Leschinski2016} & AKARI/FIS/WIDE-L & 149.94 $\mu$m & \citet{Murakami2007}\\
    Euclid/J & 1.37 $\mu$m & \citet{Laureijs2010} & Spitzer/MIPS/160mu & 156.96 $\mu$m & \citet{Rieke2004}\\
    JWST/NIRCam/F150W & 1.50 $\mu$m & \citet{Rieke2008} &   AKARI/FIS/N160 & 163.07 $\mu$m  & \citet{Murakami2007}\\
    HST/WFC3/F160W & 1.54 $\mu$m & {\tiny \url{www.stsci.edu/hst/wfc3}} & OST/OSS/Ch4 & 163.50 $\mu$m & \citet{Bradford2018}\\
    ELT/MICADO/xH1 & 1.55 $\mu$m & \citet{Leschinski2016} & Herschel/PACS/160 & 167.14 $\mu$m & \citet{Poglitsch2010}\\
    ELT/MICADO/H-cont & 1.57 $\mu$m & \citet{Leschinski2016} & OST/FIP/250um & 250.00 $\mu$m & \citet{Staguhn2018}\\
    ELT/MICADO/H-short & 1.58 $\mu$m & \citet{Leschinski2016} &  Herschel/SPIRE/250 & 251.50 $\mu$m &  \citet{Griffin2010}\\
    UKIRT/H & 1.63 $\mu$m & \citet{Casali2007} & OST/OSS/Ch5 & 275.00 $\mu$m & \citet{Bradford2018}\\
    ELT/MICADO/H & 1.64 $\mu$m & \citet{Leschinski2016} & Herschel/SPIRE/350 & 352.82 $\mu$m &  \citet{Griffin2010}\\
    ELT/MICADO/FeII & 1.65 $\mu$m & \citet{Leschinski2016} & JCMT/SCUBA2/450 & 450.12 $\mu$m & \citet{Holland2013} \\
    ELT/MICADO/H-long & 1.69 $\mu$m & \citet{Leschinski2016} &  OST/OSS/Ch6 & 462.50 $\mu$m & \citet{Bradford2018}\\
    ELT/MICADO/xH2 & 1.70 $\mu$m & \citet{Leschinski2016} & Herschel/SPIRE/500 & 511.50 $\mu$m & \citet{Griffin2010}\\
    
\end{longtable} 
}
\end{center}

\end{appendix}
\end{document}